\documentclass[12pt]{article}
 
\usepackage{amsmath,amssymb}
\usepackage{bm}
\usepackage{graphicx, color}
\usepackage{wrapfig}
\usepackage{here}
\usepackage{cite}
\begin{document}
\title{
\begin{flushright}
\ \\*[-80pt] 
\begin{minipage}{0.22\linewidth}
\normalsize
%arXiv:YYMM.NNNN \\
APCTP Pre2020-024
 \\*[50pt]
\end{minipage}
\end{flushright}
{\Large \bf 
  \ Modular invariant flavor model of $\rm A_4$ and \\
   hierarchical structures   at nearby fixed points
\\*[20pt]}}

\author{ 
\centerline{
Hiroshi Okada $^{a,b}\footnote{E-mail address: hiroshi.okada@apctp.org}$~ and 
~~Morimitsu Tanimoto $^{c}\footnote{E-mail address: tanimoto@muse.sc.niigata-u.ac.jp}$} \\*[5pt]
\centerline{
\begin{minipage}{\linewidth}
\begin{center}
$^a${\it \normalsize
Asia Pacific Center for Theoretical Physics, Pohang 37673, Republic of Korea} \\*[5pt]
$^b${\it \normalsize
Department of Physics, Pohang University of Science and Technology, Pohang 37673,\\Republic of Korea} \\*[5pt]
$^c${\it \normalsize
Department of Physics, Niigata University, Niigata 950-2181, Japan}
\end{center}
\end{minipage}}
\\*[50pt]}

\date{
\centerline{\small \bf Abstract}
\begin{minipage}{0.9\linewidth}
\medskip 
\medskip 
\small 
In the  modular invariant flavor model of $\rm A_4$, we study the hierarchical structure of lepton/quark flavors
at nearby fixed points of $\tau=i$ and $\tau=\omega$ of the  modulus, which are in the fundamental domain of $\rm PSL(2,\mathbb{Z})$.
  These fixed points  correspond to   the  residual symmetries 
   $\mathbb{Z}_2^{S}=\{I, S \}$
   and   $\mathbb{Z}_3^{ST}=\{ I, ST, (ST)^2 \}$ of  $\rm A_4$,
   where $S$ and $T$ are  generators of the $A_4$ group.   
   The infinite  $\tau= i \infty$
    also preserves the residual symmetry of the subgroup 
    $\mathbb{Z}^T_3=\{ I,T,T^2 \}$ of $\rm A_4$.
  We study typical two-type mass matrices for  charged leptons and quarks
   in terms of  modular forms of weights $2$, $4$ and $6$ while
    the neutrino mass matrix  with the modular forms of weight $4$ 
     through the Weinberg operator. 
    Linear  modular forms are obtained 
    approximately by performing Taylor expansion of modular forms around fixed points. By using them, the flavor structure of the  lepton and quark mass matrices are examined  at nearby fixed points.
    The hierarchical structure of these mass matrices is clearly shown in the diagonal base of $S$, $T$ and $ST$.
 The observed PMNS and CKM mixing matrices can be  reproduced  
  at  nearby fixed points in some cases of mass matrices.
   By scanning model parameters numerically at nearby fixed points,
    our discussion are confirmed
    for both the normal hierarchy and inverted one of neutrino masses. 
    Predictions are given for the sum of neutrino masses and
    the CP violating Dirac phase of leptons at each nearby fixed point.  
%%%%%%%%%%%%%%%%%%%%%%%%%%%%%%%%%%%%%%%%%%%%%%%%%%%
\end{minipage}
}

\begin{titlepage}
\maketitle
\thispagestyle{empty}
\end{titlepage}

%%%%%%%%%%%%%%%%%%%%%%%%%%%%%%%%%%%%%%%%%%%%%%%%%%%%%%%%%%%%%%%%%%
%%%%%%%%%%%%%%%%%%%%%%%%%%  Introduction %%%%%%%%%%%%%%%%%%%%%%%%% %%%%%%%%%%%%%%%%%%%%%%%%%%%%%%%%%%%%%%%%%%%%%%%%%%%%%%%%%%%%%%%%%%
\section{Introduction}
%%%%%%%%%%%%%%%%%%%%%%%%%%%%%%%%%%%%%%%%%%%%%%%%%%%
In spite of the remarkable success of the standard model (SM), the origin of the 
flavor of quarks and leptons is still a challenging issue. 
Indeed, a lot of  works have been presented by using the  discrete  groups
for flavors to understand the flavor structures of quarks and leptons.
In  the early models of quark masses and mixing angles, 
the  $\rm S_3$ symmetry was used 
\cite{Wilczek:1977uh,Pakvasa:1977in}.
It was also discussed  to understand the large mixing angle
\cite{Fukugita:1998vn} in the oscillation of atmospheric neutrinos \cite{Fukuda:1998mi}. 
For the last twenty years, the  discrete symmetries of flavors have been developed, that is
motivated by the precise observation of  flavor mixing angles of  leptons
\cite{Altarelli:2010gt,Ishimori:2010au,Ishimori:2012zz,Hernandez:2012ra,King:2013eh,King:2014nza,Tanimoto:2015nfa,King:2017guk,Petcov:2017ggy,Feruglio:2019ktm}.

Many models have been proposed by using 
the non-Abelian discrete groups  $\rm S_3$, $\rm A_4$, $\rm S_4$, $\rm A_5$ and other groups with larger orders to explain the large neutrino mixing angles.
Among them, the $\rm A_4$ flavor model is attractive one 
because the $\rm A_4$ group is the minimal one including a triplet 
irreducible representation, 
which allows for a natural explanation of the  
existence of  three families of leptons 
\cite{Ma:2001dn,Babu:2002dz,Altarelli:2005yp,Altarelli:2005yx,
	Shimizu:2011xg,Petcov:2018snn,Kang:2018txu}.
However, variety of models is so wide that it is difficult to show 
 clear evidences of the $\rm A_4$ flavor symmetry.

Recently, a new  approach to the lepton flavor problem appeared
based on the invariance of the modular group \cite{Feruglio:2017spp}, 
where the model of the finite
modular group  $\rm \Gamma_3 \simeq A_4$ has been presented.
This work  inspired further studies of the modular invariance 
to the lepton flavor problem. 
%It should be emphasized that there is a significant difference between the 
%models based on the $\rm A_4$ modular symmetry and those based on the usual 
%non-Abelian discrete $\rm A_4$ flavor symmetry.
%Yukawa couplings transform non-trivially under the modular group
%and are written in terms of modular forms which are  
%holomorphic functions of a complex parameter,  the modulus  $\tau$.
The finite groups $\rm S_3$, $\rm A_4$, $\rm S_4$ and $\rm A_5$ are
  formed as the quotient groups of the modular group and its principal congruence subgroups \cite{deAdelhartToorop:2011re}.
Therefore, an interesting framework for the construction of flavor models
has been put forward based on  the $\rm \Gamma_3 \simeq A_4$ modular group \cite{Feruglio:2017spp}, and further, based on $\rm \Gamma_2 \simeq S_3$ \cite{Kobayashi:2018vbk}.
The  flavor models have been proposed by using modular symmetries  
$\rm \Gamma_4 \simeq S_4$ \cite{Penedo:2018nmg} 
and  $\rm \Gamma_5 \simeq A_5$ \cite{Novichkov:2018nkm}. 
Phenomenological discussions of the neutrino flavor mixing have been done
based on  $\rm A_4$ \cite{Criado:2018thu,Kobayashi:2018scp,Ding:2019zxk}, $\rm S_4$ \cite{Novichkov:2018ovf,Kobayashi:2019mna,Wang:2019ovr} and 
$\rm A_5$ \cite{Ding:2019xna}.
A clear prediction of the neutrino mixing angles and the CP violating phase was presented in  the  simple lepton mass matrices  with $\rm A_4$ modular symmetry \cite{Kobayashi:2018scp}.
The  Double Covering groups  $\rm T'$~\cite{Liu:2019khw,Chen:2020udk}
and $\rm S_4'$ \cite{Novichkov:2020eep,Liu:2020akv} have also
obtained from the modular symmetry.

The $\rm A_4$ modular symmetry has been also applied to
the leptogenesis \cite{Asaka:2019vev,Behera:2020sfe,Mishra:2020gxg},
 on the other hand,
it is discussed  in the SU$(5)$ grand
unified theory (GUT) of  quarks and leptons  \cite{deAnda:2018ecu,Kobayashi:2019rzp}.
The residual symmetry of the $\rm A_4$ modular symmetry has presented the interesting phenomenology \cite{Novichkov:2018yse}.
Furthermore, modular forms for $\Delta(96)$ and $\Delta(384)$ were constructed \cite{Kobayashi:2018bff},
and the extension of the traditional flavor group  is discussed with modular symmetries \cite{Baur:2019kwi}.
The level $7$ finite modular group $\rm \Gamma_7\simeq PSL(2,Z_7)$
is also presented for the lepton mixing \cite{Ding:2020msi}.
Moreover, multiple modular symmetries are proposed as the origin of flavor\cite{deMedeirosVarzielas:2019cyj}.
The modular invariance has been also studied combining with the CP symmetries for theories of flavors \cite{Novichkov:2019sqv,Kobayashi:2019uyt}.
The quark mass matrix  has been discussed in the $\rm S_3$ and $\rm A_4$ modular symmetries as well \cite{Kobayashi:2018wkl,Okada:2018yrn,Okada:2019uoy}.
%%%%%%%%%
Besides mass matrices of quarks and leptons,
related topics have been discussed 
in the baryon number violation  \cite{Kobayashi:2018wkl}, 
the dark matter \cite{Nomura:2019jxj, Okada:2019xqk}
and the modular symmetry anomaly  \cite{Kariyazono:2019ehj}.
Further phenomenology has been developed  in many works
\cite{Nomura:2019yft,Okada:2019lzv,Nomura:2019lnr,Criado:2019tzk,Gui-JunDing:2019wap,deMedeirosVarzielas:2020kji,Zhang:2019ngf,Nomura:2019xsb,Kobayashi:2019gtp,Lu:2019vgm,Wang:2019xbo,King:2020qaj,Abbas:2020qzc,Okada:2020oxh,Okada:2020dmb,Ding:2020yen,Nomura:2020opk,Nomura:2020cog,Asaka:2020tmo}
while theoretical investigations are also proceeded \cite{Kobayashi:2019xvz,Nilles:2020kgo,Nilles:2020nnc,Kikuchi:2020nxn,Kikuchi:2020frp}.

%%%%%%%%%%%%%%%%%%%%%%%%%%%%%%%%%%%%%%%%%%% 
%%%%%%%%%%%%%%%%%%%%%%%%%%%%%%%%%%%%%%%%%%%
As well known, in non-Abelian discrete  symmetries of flavors,
residual symmetries   provide interesting phenomenology of flavors.
They arise whenever the modulus $\tau$ breaks the modular group  only partially. 
In this work, we study the hierarchical flavor structure of leptons and quarks  in context with the residual symmetry,
in which the modulus $\tau$ is at fixed points.
We examine the flavor structure of  mass matrices 
of leptons and quarks at nearby fixed points of the modulus $\tau$ in the framework of  the  modular invariant flavor model of $\rm A_4$. 
It is challenging to reproduce
the Pontecorvo-Maki-Nakagawa-Sakata (PMNS) mixing angles \cite{Maki:1962mu,Pontecorvo:1967fh} 
and the CP violating Dirac phase  of  leptons
which is expected to be observed at T2K and NO$\nu$A experiments \cite{T2K:2020,Adamson:2017gxd}, as well as
observed  Cabibbo-Kobayashi-Maskawa (CKM) matrix elements  at nearby  fixed points.

We have already discussed numerically  both mass matrices of leptons and quarks 
in the $\rm A_4$ modular symmetry \cite{Okada:2019uoy,Okada:2020rjb},
where modular forms of weights $2$, $4$ and $6$ are used.
In the same framework, we discuss the flavor structure of the  lepton and quark mass matrices focusing on  nearby fixed points.
For this purpose,
we give   linear forms  of  $Y_1(\tau)$, $Y_2(\tau)$ and $Y_3(\tau)$ 
approximately by performing Taylor expansion of modular forms around fixed points of the modulus $\tau$ in the $\rm A_4$ modular symmetry.

%with reference to the sum of neutrino masses.
%It is found that the sum of neutrino masses is
%crucial  to realize the common $\tau$ for quarks and leptons.

%%%%%%%%%%%%%%%%%%%%%%%%%%%%%%%%%%%%%%%%%%%%%%%%%%%%%%%%%%%%%%%%%%%%%%%%%

The paper is organized as follows.
In section 2,  we give a brief review on the modular symmetry and 
modular forms of weights $2$, $4$ and $6$. 
In section 3, we discuss the residual symmetry of $\rm A_4$ and
 modular forms at fixed points.
In section 4, we present modular forms at nearby fixed points.
In section 5 and 6, we discuss flavor mixing angles  at nearby fixed points
 in lepton mass matrices and quark mass matrices, respectively.
 In section 7, the numerical results and predictions are presented.
Section 8 is devoted to a summary.
In Appendix A, the tensor product  of the $\rm A_4$ group is presented.
In Appendix B, the transformation of  mass matrices are discussed in the arbitrary bases of $S$ and $T$.
In Appendix C, the modular forms are given at nearby fixed points.
In Appendix D,  we present how to obtain  Dirac CP phase, Majorana phases and  the effective mass of the $0\nu\beta\beta$ decay.

%%%%%%%%%%%%%%%%%%%%%%%%%%%%%%%%%%%%%%%%%%%%%%%%%%%%%%%%%%%%%%%%%%%%%%
%%%%%%%%%%%%%%%%%%%%%%%%%%%%%%%%%%%%%%%%%%%%%%%%%%%%%%%%%%%%%%%%%%%%%%
\section{Modular group and modular forms of weights $2$, $4$, $6$}

The modular group $\bar\Gamma$ is the group of linear fractional transformation
$\gamma$ acting on the modulus  $\tau$, 
belonging to the upper-half complex plane as:
\begin{equation}\label{eq:tau-SL2Z}
\tau \longrightarrow \gamma\tau= \frac{a\tau + b}{c \tau + d}\ ,~~
{\rm where}~~ a,b,c,d \in \mathbb{Z}~~ {\rm and }~~ ad-bc=1, 
~~ {\rm Im} [\tau]>0 ~ ,
\end{equation}
which is isomorphic to  $\rm PSL(2,\mathbb{Z})=SL(2,\mathbb{Z})/\{I,-I\}$ transformation.
This modular transformation is generated by $S$ and $T$, 
\begin{eqnarray}
S:\tau \longrightarrow -\frac{1}{\tau}\ , \qquad\qquad
T:\tau \longrightarrow \tau + 1\ ,
\label{symmetry}
\end{eqnarray}
which satisfy the following algebraic relations, 
\begin{equation}
 S^2 =\mathbb{I}\ , \qquad (ST)^3 =\mathbb{I}\ .
\end{equation}

We introduce the series of groups $\Gamma(N)~ (N=1,2,3,\dots)$,
called principal congruence subgroups of ${\rm SL(2,\mathbb{Z})}$, defined by
\begin{align}
\begin{aligned}
\Gamma(N)= \left \{ 
\begin{pmatrix}
a & b  \\
c & d  
\end{pmatrix} \in {\rm SL(2,\mathbb{Z})}~ ,
~~
\begin{pmatrix}
a & b  \\
c & d  
\end{pmatrix} =
\begin{pmatrix}
1 & 0  \\
0 & 1  
\end{pmatrix} ~~({\rm mod} N) \right \}
\end{aligned} .
\end{align}
For $N=2$, we define $\rm \bar\Gamma(2)\equiv \Gamma(2)/\{I,-I\}$.
Since the element $\rm -I$ does not belong to $\Gamma(N)$
% while, since the element $-I$\UTF{0081} does not belong to $\Gamma(N)$,
for $N>2$, we have $\bar\Gamma(N)= \Gamma(N)$.
The quotient groups defined as
$\Gamma_N\equiv \bar \Gamma/\bar \Gamma(N)$
are  finite modular groups.
In these finite groups $\Gamma_N$, $ T^N=\mathbb{I}$  is imposed.
The  groups $\Gamma_N$ with $N=2,3,4,5$ are isomorphic to
$\rm S_3$, $\rm A_4$, $\rm S_4$ and $\rm A_5$, respectively \cite{deAdelhartToorop:2011re}.

Modular forms of  level $N$ are 
holomorphic functions $f(\tau)$  transforming under the action of $\Gamma(N)$ as:
\begin{equation}
f(\gamma\tau)= (c\tau+d)^kf(\tau)~, ~~ \gamma \in \Gamma(N)~ ,
\end{equation}
where $k$ is the so-called as the  modular weight.

Superstring theory on the torus $\rm T^2$ or orbifold $\rm T^2/Z_N$ has the modular symmetry \cite{Lauer:1989ax,Lerche:1989cs,Ferrara:1989qb,Cremades:2004wa,Kobayashi:2017dyu,Kobayashi:2018rad}.
Its low energy effective field theory is described in terms of  supergravity theory,
and  string-derived supergravity theory has also the modular symmetry.
Under the modular transformation of Eq.\,(\ref{eq:tau-SL2Z}), chiral superfields $\phi^{(I)}$ 
transform as \cite{Ferrara:1989bc},
\begin{equation}
\phi^{(I)}\to(c\tau+d)^{-k_I}\rho^{(I)}(\gamma)\phi^{(I)},
\end{equation}
where  $-k_I$ is the modular weight and $\rho^{(I)}(\gamma)$ denotes an unitary representation matrix of $\gamma\in\Gamma_N$.

%%%%%%%%%%%%%%%%%%%%%%%%%%%%%%%%%%%%%%%%
In this paper, we study global supersymmetric models, e.g., 
minimal supersymmetric extensions of the Standard Model (MSSM).
The superpotential which is built from matter fields and modular forms
is assumed to be modular invariant, i.e., to have 
a vanishing modular weight. For given modular forms 
this can be achieved by assigning appropriate
weights to the matter superfields.
%%%%%%%%%%%%%%%%%%%%%%%%%%
%%%%%%%%%%%%%%%%%%%%%%%%%%

The kinetic terms  are  derived from a K\"ahler potential.
The K\"ahler potential of chiral matter fields $\phi^{(I)}$ with the modular weight $-k_I$ is given simply  by 
%%%%%%%%%%%%%%%%%%%%%%%%%%%%%%
\begin{equation}
K^{\rm matter} = \frac{1}{[i(\bar\tau - \tau)]^{k_I}} |\phi^{(I)}|^2,
\end{equation}
%%%%%%%%%%%%%%%%%%%%%%%%%%%
where the superfield and its scalar component are denoted by the same letter, and  $\bar\tau =\tau^*$ after taking the vacuum expectation value (VEV).
%%%%%%%%%%%%%%%%%%%%%%%%%%%
Therefore, 
the canonical form of the kinetic terms  is obtained by 
changing the normalization of parameters \cite{Kobayashi:2018scp}.
The general K\"ahler potential consistent with the modular symmetry possibly contains additional terms \cite{Chen:2019ewa}. However, we consider only the simplest form of
	the K\"ahler potential.
%%%%%%%%%%%%%%%%%%%%%%%%%%%%%%%
%%%%%%%%%%%%%%%%%%%%%%%%%%%%%%%%%%%%%%%%

%%%%%%%%%%%%%%%%%%%%%%%%%%%%%%%%%%%%%%%%
For $\rm \Gamma_3\simeq A_4$, the dimension of the linear space 
${\cal M}_k(\Gamma_3)$ 
of modular forms of weight $k$ is $k+1$ \cite{Gunning:1962,Schoeneberg:1974,Koblitz:1984}, i.e., there are three linearly 
independent modular forms of the lowest non-trivial weight $2$.
These forms have been explicitly obtained \cite{Feruglio:2017spp} in terms of
the Dedekind eta-function $\eta(\tau)$: 
%which is written by 
%%%%%%%%%%%%%%%%%%%%%%%%%%%%%
\begin{equation}
\eta(\tau) = q^{1/24} \prod_{n =1}^\infty (1-q^n)~, 
\quad\qquad  q= \exp \ (i 2 \pi  \tau )~,
\end{equation}
%%%%%%%%%%%%%%%%%%%%%%%%%%
%
where $\eta(\tau)$ is a  so called  modular form of weight~$1/2$. 
% and $\eta(\tau)$ is a modular form of weight~$1/2$.
In what follows we will use the following base of the 
$\rm A_4$ generators  $S$ and $T$ in the triplet representation:
%%%%%%%%%%%%%%%%%%%%%%%%%%%
\begin{align}
\begin{aligned}
\rm S=\frac{1}{3}
\begin{pmatrix}
-1 & 2 & 2 \\
2 &-1 & 2 \\
2 & 2 &-1
\end{pmatrix},
\end{aligned}
\qquad \qquad
\begin{aligned}
\rm T=
\begin{pmatrix}
1 & 0& 0 \\
0 &\omega& 0 \\
0 & 0 & \omega^2
\end{pmatrix}, 
\end{aligned}
\label{STbase}
\end{align}
%%%%%%%%%%%%%%%%%%%%%%%%%%%%%%%%%
%
where $\omega=\exp (i\frac{2}{3}\pi)$ .
The  modular forms of weight 2,  
${\bf Y^{(2)}_3}=(Y_1(\tau),Y_2(\tau),Y_3(\tau))^T$ transforming
as a triplet of $\rm A_4$ can be written in terms of 
$\eta(\tau)$ and its derivative \cite{Feruglio:2017spp}:
%%%%%%%%%%%%%%%%%%%%%%%
\begin{eqnarray} 
\label{eq:Y-A4}
Y_1(\tau) &=& \frac{i}{2\pi}\left( \frac{\eta'(\tau/3)}{\eta(\tau/3)}  +\frac{\eta'((\tau +1)/3)}{\eta((\tau+1)/3)}  
+\frac{\eta'((\tau +2)/3)}{\eta((\tau+2)/3)} - \frac{27\eta'(3\tau)}{\eta(3\tau)}  \right), \nonumber \\
Y_2(\tau) &=& \frac{-i}{\pi}\left( \frac{\eta'(\tau/3)}{\eta(\tau/3)}  +\omega^2\frac{\eta'((\tau +1)/3)}{\eta((\tau+1)/3)}  
+\omega \frac{\eta'((\tau +2)/3)}{\eta((\tau+2)/3)}  \right) , \label{Yi} \\ 
Y_3(\tau) &=& \frac{-i}{\pi}\left( \frac{\eta'(\tau/3)}{\eta(\tau/3)}  +\omega\frac{\eta'((\tau +1)/3)}{\eta((\tau+1)/3)}  
+\omega^2 \frac{\eta'((\tau +2)/3)}{\eta((\tau+2)/3)}  \right)\,.
\nonumber
\end{eqnarray}
%%%%%%%%%%%%%%%%%%%%%
%
% where 
%The overall coefficient in Eq.\,(\ref{Yi}) is one possible choice.
% and cannot be determined essentially.
%It cannot be uniquely determined.
The triplet modular forms of weight 2
% are also  expressed in the $q$ expansions
have the following  $q$-expansions:
%%%%%%%%%%%%%%%%%%%%%%%%%%
\begin{align}
{\bf Y^{(\rm 2)}_3}
=\begin{pmatrix}Y_1(\tau)\\Y_2(\tau)\\Y_3(\tau)\end{pmatrix}=
\begin{pmatrix}
1+12q+36q^2+12q^3+\dots \\
-6q^{1/3}(1+7q+8q^2+\dots) \\
-18q^{2/3}(1+2q+5q^2+\dots)\end{pmatrix}.
\label{Y(2)}
\end{align}
%%%%%%%%%%%%%%%%%%%%%%%
%
% where $Y_i^{(2)}(\tau)$ 
They satisfy also the constraint \cite{Feruglio:2017spp}:
%%%%%%%%%%%%%%%%%%%%%%%%
\begin{align}
(Y_2(\tau))^2+2Y_1(\tau) Y_3(\tau)=0~.
\label{condition}
\end{align}
%%%%%%%%%%%%%%%%%%%%%
%%%%%%%%%%%%%%%%%%%%%
%%%%%%%%%%%%%%%%%%%%

The  modular forms of the  higher weight, $k$, can be obtained
by the $\rm A_4$ tensor products of  the modular forms  with weight 2,
${\bf Y^{(\rm 2)}_3}$, 
as given in Appendix A.
For $k=4$, there are  five modular forms
by the tensor product of  $\bf 3\otimes 3$ as:
\begin{align}
&\begin{aligned}
{\bf Y^{(\rm 4)}_1}=Y_1^2+2 Y_2 Y_3 \ , \quad
{\bf Y^{(\rm 4)}_{1'}}=Y_3^2+2 Y_1 Y_2 \ , \quad
{\bf Y^{(\rm 4)}_{1''}}=Y_2^2+2 Y_1 Y_3=0 \ , \quad
\end{aligned}\nonumber \\
\nonumber \\
&\begin{aligned} {\bf Y^{(\rm 4)}_{3}}=
\begin{pmatrix}
Y_1^{(4)}  \\
Y_2^{(4)} \\
Y_3^{(4)}
\end{pmatrix}
=
\begin{pmatrix}
Y_1^2-Y_2 Y_3  \\
Y_3^2 -Y_1 Y_2 \\
Y_2^2-Y_1 Y_3
\end{pmatrix}\ , 
\end{aligned}
\label{weight4}
\end{align}
where ${\bf Y^{(\rm 4)}_{1''}}$ vanishes due to the constraint of
 Eq.\,(\ref{condition}).
For $k=6$, there are  seven modular forms
by the tensor products of  $\rm A_4$ as:
\begin{align}
&\begin{aligned}
{\bf Y^{(\rm 6)}_1}=Y_1^3+ Y_2^3+Y_3^3 -3Y_1 Y_2 Y_3  \ , 
\end{aligned} \nonumber \\
\nonumber \\
&\begin{aligned} {\bf Y^{(\rm 6)}_3}\equiv 
\begin{pmatrix}
Y_1^{(6)}  \\
Y_2^{(6)} \\
Y_3^{(6)}
\end{pmatrix}
=
\begin{pmatrix}
Y_1^3+2 Y_1 Y_2 Y_3   \\
Y_1^2 Y_2+2 Y_2^2 Y_3 \\
Y_1^2Y_3+2Y_3^2Y_2
\end{pmatrix}\ , \qquad
\end{aligned}
\begin{aligned} {\bf Y^{(\rm 6)}_{3'}}\equiv
\begin{pmatrix}
Y_1^{'(6)}  \\
Y_2^{'(6)} \\
Y_3^{'(6)}
\end{pmatrix}
=
\begin{pmatrix}
Y_3^3+2 Y_1 Y_2 Y_3   \\
Y_3^2 Y_1+2 Y_1^2 Y_2 \\
Y_3^2Y_2+2Y_2^2Y_1
\end{pmatrix}\ . 
\end{aligned}
\label{weight6}
\end{align}
By using these modular forms of weights $2, 4, 6$,
we discuss   lepton and quark mass matrices.
%%%%%%%%%%%%%%%%%%%%%%%%%%%%%%%%%%%
%%%%%%%%%%%%%%%%%%%%%%%%%%%%%%%%%%%
%%%%%%%%%%%%%%%%%%%%%%%%%%%%%%%%%%%

%%%%%%%%%%%%%%%%%%%%%%%%%%%%%%%%%%%%%%%%%%%%%%%%%%%%%%%%%%%%%
%%%%%%%%%%%%%%%%%%%%%%%%%%%%%%%%%%%%%%%%%%%%%%%%%%%%%%%%%%%%%
 \section{Residual symmetry of $\rm A_4$ at fixed points}
 \subsection{Modular forms at fixed points}

Residual symmetries arise whenever the VEV of the modulus $\tau$ breaks
the modular group $\overline{\Gamma}$ only partially. 
Fixed points of modulus are the case.
%Those have been investigated in 
% the  modular $\rm S_4$ invariance  \cite{Novichkov:2018ovf},
%and the two modular $\rm S_4$ groups \cite{King:2019vhv}, 
%where viable models of have been composed. 
%Also  fixed points  have been investigated in $\rm A_4$ and  $\rm A_5$ %invariance \cite{Novichkov:2018yse,deAnda:2018ecu,Novichkov:2018nkm}.
% The systematic study of the fixed points
%  has been presented in the modular finite groups \cite{Gui-JunDing:2019wap,deMedeirosVarzielas:2020kji}.
There are only 2 inequivalent finite points in the fundamental domain
 of $\overline{\Gamma}$,
 namely,  $\tau = i$ and 
  $ \tau =\omega=-1/2+ i \sqrt{3}/2$.
  The first point is  invariant under the $S$ transformation
 $\tau=-1/\tau$.  In the case of $\rm A_4$ symmetry, the subgroup $\mathbb{Z}_2^{S}=\{ I, S \}$ is preserved at $ \tau = i$.
The second point is the left cusp in the fundamental domain of the modular group,
which is invariant under the  $ST$ transformation $\tau=-1/(\tau+1)$.
Indeed, $\mathbb{Z}_3^{ST}=\{  I, ST,(ST)^2 \}$ is  one of  subgroups of 
$\rm A_4$ group. 
The right cusp at  
$ \tau =-\omega^2=1/2+ i \sqrt{3}/2$
is related to $\tau=\omega$ by the $T$ transformation.
There is also infinite point $ \tau = i \infty$,
in which  the subgroup  $\mathbb{Z}^T_3=\{ I,T,T^2 \}$ of $\rm A_4$
is preserved.

It is possible to calculate the values of the  $\rm A_4$ triplet modular 
forms of weight 2, 4 and 6 at  $\tau=i$,  $\tau=\omega$ and  $\tau=i\infty$.
The results are summarized in Table~\ref{tb:modularforms}.

If  a residual symmetry of $\rm A_4$ 
is preserved in  mass matrices of leptons and quarks,
we have commutation relations 
between the mass matrices and the generator $ G \equiv S,\, T,\, ST$ as:
\begin{align}
[M_{RL}^\dagger M_{RL},\, { G}]=0 \, , 
\qquad\qquad  [M_{LL},\, { G}]=0 \,  ,
\label{commutator}
\end{align}
where $M_{RL}$ denotes the  mass matrix of charged leptons and quarks, $M_E$ and  $M_q$, 
on the other hand, $M_{LL}$ denotes the left-handed Majorana neutrino mass matrix  $M_\nu$.

%%%%%%%%%%%
%$\mathbb{Z}_2^{S}=\{ \rm I, S \}$  ($\tau = i$)
%%%%%%%%%%%

Therefore,  the mass matrices
$M_E^\dagger M_E$, $M_q^\dagger M_q$ and  $ M_\nu$
could be diagonal in  the diagonal base of G at the fixed points.
The hierarchical structures of flavor mixing are easily realized
near those fixed points.
However, we should  be careful with 
the generator $S$, in  which  two eigenvalues are degenerate. At $\tau = i$, 
one $(2\times 2)$ submatrix of the mass matrix respecting $S$ are not diagonal in general since two eigenvalues of $S$ are degenerate
such as $(-1,1,-1)$.
Therefore, the $S$ symmetry  provides us an advantage to reproduce  the large mixing angle of neutrinos  as discussed in section 5.
%%%%%%%%%%%%%%%%%%%%%%%%%%%%%%%%%%%%%%%%%%%%%%%%%%%%%   
\begin{table}[t!]
	\centering
	\begin{tabular}{|c|c|c|c|c|} \hline 
		\rule[14pt]{0pt}{1pt}
	$k$ & $\bf r$	& $\tau=i$ &$\tau=\omega$ &	$\tau=i\infty$\\ \hline 
		\rule[14pt]{0pt}{2pt}
		%%%%%%%%%%%%%%%%%%%%%%%%%%%%%%%%%%%%%%%%%%%%%%%%%%%%%%%%%%%%%%%%%%
		%%%%%%%%%%%%%%%%%%%%%%%%%%%%%%%%%%%%%%%%%%%%%%%%%%%%%%%%%%%%%%%%%
		 $2$& $\bf 3$	& $Y_0\,(1,1-\sqrt{3}, -2+\sqrt{3})$ &$Y_0\,(1,\omega, -\frac{1}{2}\omega^2)$
		  & $Y_0\,(1,0,0)$\\ \hline
		\rule[14pt]{0pt}{2pt}
		%%%%%%%%%%%%%%%%%%%%%%%%%%%%%%%%%%%%%%%%%%%%%%%%%%%%%%%%%%%%%%%%%%
		 $4$ &$\bf 3$	&$( 6-3\sqrt{3})Y_0^2\,(1,1,1)$  &
		 	$\frac{3}{2}Y_0^2\,(1, -\frac{1}{2}\omega, \omega^2)$
		 &$Y_0^2\,(1,0,0)$\\
		\rule[14pt]{0pt}{2pt}
		 &$\{\bf 1, 1'\}$	&   $Y_0^2\,\{6\sqrt{3}-9,
		 \ \  9-6\sqrt{3} \}$ &$\{0,\quad \frac{9}{4}Y_0^2\,\omega\}$
		 & $\{ Y_0^2, \quad 0\}$\\ \hline
		\rule[14pt]{0pt}{2pt}
		%%%%%%%%%%%%%%%%%%%%%%%%%%%%%%%%%%%%%%%%%%%%%%%%%%%%%%%%%%%%%%%%%%
	 	 $6$& $\bf 3$	& 	$3Y_0^3\, (-3+2\sqrt{3}, -9+5\sqrt{3},12-7\sqrt{3})$
	 	  & 0
	   &$Y_0^3\, (1,0,0)$\\
		\rule[14pt]{0pt}{2pt}
		& $\bf 3'$	& $3Y_0^3 \,(-12+7\sqrt{3}, 3-2\sqrt{3},9-5\sqrt{3})$  &$\frac{9}{8} Y_0^3\,(-1,2\omega, 2\omega^2)$ 	 & 0\\
		\rule[14pt]{0pt}{2pt}
		& $\bf 1$	& 0 & $\frac{27}{8}Y_0^3$ & $Y_0^3$\\ \hline
		\rule[14pt]{0pt}{2pt}
		& $Y_0$	& $Y_1(i)=1.0225...$ & $Y_1(\omega)=0.9486...$
		  &$Y_1(i\infty)=1$\\ \hline
	\end{tabular}
	\caption{Modular forms of weight  $k=2$, $k=4$ and $k=6$  at  fixed points of $\tau$.
	}
	\label{tb:modularforms}
\end{table}

%%%%%%%%%%%%%%%%%%%%%%%%%%%%%%%%%%%%%%%%%%%%%%%%%%%%%%%%%%%%%
%%%%%%%%%%%%%%%%%%%%%%%%%%%%%%%%%%%%%%%%%%%%%%%%%%%%%%%%%%%%%
\subsection{Diagonal base of $S$ and $ST$}
\subsubsection{Diagonal base of $S$}

 The modular forms of Eq.\,(\ref{eq:Y-A4}) is obtained 
 in the base of Eq.\,(\ref{STbase}) for $S$ and $T$.
 In order to present the mass matrices in the diagonal base of $S$,
we move to the diagonal base of $S$ as follows:
\begin{align}
\begin{aligned}
V_{S1}\, { S}\, V_{S1}^\dagger=
\begin{pmatrix} 
-1 & 0 & 0 \\
0 & 1 & 0 \\
0 &0 &-1
\end{pmatrix},  \quad 
V_{S2}\, { S}\, V_{S2}^\dagger=
\begin{pmatrix} 
1 & 0 & 0 \\
0 & -1 & 0 \\
0 &0 &-1
\end{pmatrix},  \quad
V_{S3}\, { S} \, V_{S3}^\dagger=
\begin{pmatrix} 
-1 & 0 & 0 \\
0 & -1 & 0 \\
0 &0 &1
\end{pmatrix}, 
\end{aligned}
\end{align}
where
\begin{align}
\begin{aligned}
V_{Si}\equiv P_i\begin{pmatrix} 
\ \frac{2}{\sqrt{6}} & -\frac{1}{\sqrt{6}} & -\frac{1}{\sqrt{6}} \\
\frac{1}{\sqrt{3}} &\frac{1}{\sqrt{3}} & \frac{1}{\sqrt{3}} \\
0 &-\frac{1}{\sqrt{2}} &\ \frac{1}{\sqrt{2}}
\end{pmatrix},
\quad P_1=
\begin{pmatrix} 
1 & 0 & 0 \\
0 & 1 & 0 \\
0 &0 &1
\end{pmatrix}, \ \ 
P_2=
\begin{pmatrix} 
0 & 1 & 0 \\
1 & 0 & 0 \\
0 &0 &1
\end{pmatrix}, \ \
P_3=
\begin{pmatrix} 
1 & 0 & 0 \\
0 & 0 & 1 \\
0 &1 &0
\end{pmatrix} \,.
\end{aligned}
\label{Sdiagonal}
\end{align}
Then, the  generator $T$ is not anymore diagonal.
%{\footnotesize
%	\begin{align}
%	V_{S1} T V_{S1}^\dagger=\begin{pmatrix} 
%	\frac{1}{2} & \frac{1}{\sqrt{2}} & \frac{1+2\omega}{2\sqrt{3}} \\
%	\frac{1}{\sqrt{2}} &0 & -\frac{1+2\omega}{\sqrt{6}}  \\
%	\frac{1+2\omega}{2\sqrt{3}}& -\frac{1+2\omega}{\sqrt{6}} &\ -\frac{1}{2}
%	\end{pmatrix}, \
%	V_{S2} T V_{S2}^\dagger=\begin{pmatrix} 
%	0 & \frac{1}{\sqrt{2}} & -\frac{1+2\omega}{\sqrt{6}} \\
%	\frac{1}{\sqrt{2}} &\frac{1}{2} & \frac{1+2\omega}{2\sqrt{3}}  \\
%	-\frac{1+2\omega}{\sqrt{6}}& \frac{1+2\omega}{2\sqrt{3}} & -\frac{1}{2}
%	\end{pmatrix},\
%	V_{S3} T V_{S3}^\dagger=\begin{pmatrix} 
%	\frac{1}{2} &  \frac{1+2\omega}{2\sqrt{3}}&\frac{1}{\sqrt{2}}  \\
%	\frac{1+2\omega}{2\sqrt{3}}&-\frac{1}{2} & -\frac{1+2\omega}{\sqrt{6}}\\
%	\frac{1}{\sqrt{2}}& -\frac{1+2\omega}{\sqrt{6}} &0
%	\end{pmatrix}.
%	\label{Tnew}
%	\end{align}
%}
%%%%%%%%%%%%%%%%%%%%%%%%%%%%%%%%%%  

If there is a residual symmetry of $\rm A_4$ in the Dirac mass matrix $M_{RL}$, 
for example,  $\mathbb{Z}_2^{S}=\{  I, S \}$,
the generator $S$ commutes with $M_{RL}^\dagger M_{RL}$,
\begin{align}
\begin{aligned}
\left [M_{RL}^\dagger M_{RL}\, ,\,  S\right ]=0 \, .
\end{aligned}
\end{align}
Therefore, the mass matrix is expected to be diagonal
in the diagonal base of $S$.
However, the eigenvalue $-1$ of $S$ is degenerated, and so
one pair among  off diagonal terms of  $M_{RL}^\dagger M_{RL}$ is not necessarily to vanish depending on $V_i$ of Eq.\,(\ref{Sdiagonal}).
For diagonal matrices $S=(-1,1,-1)$, $(1,-1,-1)$ and  $(-1,-1,1)$, those are:
\begin{align} 
M_{RL}^\dagger M_{RL}\,=\ 
\begin{pmatrix}
\times  & 0&  \times \\
0& \times  &  0\\
\times & 0 & \times 
\end{pmatrix}\, , \quad
\begin{pmatrix}
\times  & 0 & 0\\
0  & \times & \times\\
0 & \times & \times 
\end{pmatrix}\, , \quad
\begin{pmatrix}
\times  & \times & 0\\
\times  & \times & 0\\
0 & 0 & \times 
\end{pmatrix}\, , 
\label{zeros}
\end{align}
respectively,
where $"\times"$ denotes non-vanishing entry.
Thus, one flavor mixing angle appears  even if 
there exists the $\mathbb{Z}_2^{S}=\{  I, S \}$ symmetry..

%%%%%%%%%%%%%%%%%%%%%%%%%%%%%%%%%%%
\subsubsection{Diagonal base of $ST$ and $T$}
%%%%%%%%%%%%%%%%%%%%%%%%%%%%%%%%%%%
If there exists the residual symmetries of the $\rm A_4$
group $\mathbb{Z}_3^{ST}=\{  I, ST, (ST)^2 \}$ or
$\mathbb{Z}_3^{T}=\{  I, T, T^2 \}$,
 we have
\begin{align}
\begin{aligned}
\left [M_{RL}^\dagger M_{RL}\, ,\,  ST\right ]=0 \, ,
\qquad\qquad 
\left [M_{RL}^\dagger M_{RL}\, ,\,  T\right ]=0 \, ,
\end{aligned}
\end{align} 
respectively, 
which lead to the diagonal  $M_{RL}^\dagger M_{RL}$
because $ST$ and $T$ have  three different eigenvalues.

The generator $T$ is already diagonal in the original base of Eq.\,(\ref{STbase}).
On the other hand,
we can move to the diagonal base of $ST$ by the unitary transformation $V_{ST}$
 as follows:
%%%%%%%%%%%%%%%%%%%%%%%%%%%%%%%%%%
\begin{align}
\begin{aligned}
V_{STi}\, {ST} \, V_{STi}^\dagger=P_i
\begin{pmatrix} 
\omega^2 & 0 & 0 \\
0 & \omega & 0 \\
0 &0 &1
\end{pmatrix}P_i^T\,, 
\end{aligned} 
\end{align}
where
\begin{align}
\begin{aligned}
V_{STi}=\frac{1}{3}P_i
\begin{pmatrix} 
-2 \omega^2  & -2 \omega & 1 \\
- \omega^2  & \ 2 \omega & 2 \\
\ 2 \omega^2 &- \omega &2
\end{pmatrix}\,, \qquad 
P_4=
\begin{pmatrix} 
0  & 0 & 1 \\
0  &1 & 0 \\
1  & 0 & 0
\end{pmatrix}\,, \qquad
P_5=
\begin{pmatrix} 
0  & 0 & 1 \\
1  &0 & 0 \\
0  & 1 & 0
\end{pmatrix}\,,
\end{aligned}
\label{STdiagonal}
\end{align}
and $P_i(i=1,2,3)$ are given in Eq.\,(\ref{Sdiagonal}).
The order of eigenvalues of $ST$ depends on $P_i$. We have 
eigenvalues $(\omega,\, \omega^2,\, 1)$ for $P_2$,
 $(\omega^2,\, 1, \, \omega)$ for $P_3$,  
  $(1,\, \omega,\, \omega^2)$ for $P_4$
  and  $(1,\, \omega^2,\, \omega)$ for $P_5$, respectively.

%%%%%%%%%%%%%%%%%%%%%%%%%%%%%%%%%%%%%
In the diagonal bases of $S$ and $ST$, the Dirac mass matrix $\hat M_{RL}$ is given by the unitary transformation  as (See Appendix B):
\begin{align}
\hat M_{RL}=  M_{RL} V_{Si}^\dagger\,, \qquad
 \hat M_{RL}=  M_{RL} V_{STi}^\dagger \ ,
\label{diagonalbase1}
\end{align}
respectively.
%%%%%%%%%%%%%%%%%%%%%%%%%%%%%%%%%%
On the other hand, the Majorana mass matrix $M_{LL}$
is given as:
\begin{align}
\hat M_{LL}= V_{Si} M_{LL} V_{Si}^\dagger \,, \qquad
\hat M_{LL}= V_{STi} M_{LL} V_{STi}^\dagger \,,
\label{diagonalbase2}
\end{align}
respectively.
%%%%%%%%%%%%%%%%%%%%%%%%%%%%%%%%%%%%%%%%%%%%%%%%%%%%%%%%%%%%%%%%%%%%%
We will discuss the  lepton and quark   mass matrices 
in the diagonal bases of the generators by using 
these transformations.

%%%%%%%%%%%%%%%%%%%%%%%%%%%%%%%%%%%%%%%%%%%%%%%%%%%%%%%%%%%%%%%%%%%%%
%%%%%%%%%%%%%%%%%%%%%%%%%%%%%%%%%%%%%%%%%%%%%%%%%%%%%%%%%%%%%%%%%%%%%
\section{Modular forms at nearby fixed points}

The mass matrices of leptons and quarks have simple flavor structures
due to simple  modular forms at  fixed points.
At $\tau=i$, those mass matrices have one flavor mixing angle
%which can be  $0$--$45^\circ$ by choosing other parameters
because the representation of
$S$ for the $\rm A_4$ triplet has two degenerate eigenvalues.
On the other hand, at  $\tau=\omega$ and $\tau=i\infty$,
the square of the mass matrix is diagonal one because 
$ST$ and $T$ of the $\rm A_4$ triplet have three different eigenvalues.
Therefore, the modulus $\tau$ should deviate from the fixed point to reproduce
the observed  PMNS and CKM matrix elements.
We present the explicit modular forms 
by performing Taylor expansion around fixed points.

%%%%%%%%%%%%%%%%%%%%%%%%%%%%%%%%%%%%%%%%%%%%%%%%%%%%%%%%%%%%%%%%% 
\subsection{Modular forms at nearby   $\tau=i$}    

Let us discuss the behavior of modular forms at nearby $\tau=i$.
We consider linear approximation of the modular forms $Y_1(\tau)$, $Y_2(\tau)$ and $Y_3(\tau)$ by performing Taylor expansion around $\tau=i$.
We parametrize $\tau$ as:
\begin{align}
\begin{aligned}
\tau=i +\epsilon \ , 
\end{aligned}
\label{epsilonS}
\end{align}
where $|\epsilon|$ is supposed  as $|\epsilon|\ll 1$.
We obtain the ratios of the modular forms approximately as:
\begin{align}
\begin{aligned}
\frac{Y_2(\tau)}{Y_1(\tau)}\simeq (1+\epsilon_1)\, (1-\sqrt{3}) \, , \quad 
\frac{Y_3(\tau)}{Y_1(\tau)}\simeq (1+\epsilon_2)\, (-2+\sqrt{3}) \, ,
\quad \epsilon_1=\frac{1}{2} \epsilon_2=2.05\,i\,\epsilon\,.
\end{aligned}
\label{epS120}
\end{align}
These approximate  forms are  agreement with exact numerical values within  $0.1\,\%$
for $|\epsilon|\leq 0.05$.
Details are given in  Appendix C.1.
%%%%%%%%%%%%%%%%%%%%%%%%%%%%%%%%%%%%%%%%%%%%%%%
%%%%%%%%%%%%  Higher weights  %%%%%%%%%%%%%%%%%
%%%%%%%%%%%%%%%%%%%%%%%%%%%%%%%%%%%%%%%%%%%%%%%
The  higher weight modular forms $ Y_i^{(k)}$
in Eqs.\,(\ref{weight4}) and (\ref{weight6}) are also given
in terms of $\epsilon_1$ and  $\epsilon_2$ in Appendix C.1.

%%%%%%%%%%%%%%%%%%%%%%%%%%%%%%%%%%%%%%%%%%%%%%%%%%%%%%%%%%%%%%%%%%%%%%
%%%%%%%%%%%%%%%%%%%%%%%%%%%%%%%%%%%%%%%%%%%%%%%%%%%%%%%%%%%%%%%%%%%%%%
%%%%%%%%%%%%%%%%%%%%%%%%%%%%%%%%%%%%%%%%%%%%%%%%%%%%%%%%%%%%%%%%%%%%%%
 \subsection{Modular forms at nearby  $\tau=\omega$}
%Let us present the behavior of modular forms at nearby $\tau=\omega$.
We perform linear approximation of the modular forms $Y_1(\tau)$, $Y_2(\tau)$ and $Y_3(\tau)$ by performing Taylor expansion around $\tau=\omega$.
We parametrize $\tau$ as:
\begin{align}
\begin{aligned}
\tau= \omega+\epsilon=-\frac{1}{2}+\frac{\sqrt{3}}{2}\,i+\epsilon \, , 
\end{aligned}
\label{epST0}
\end{align}
where we suppose $|\epsilon|\ll 1$.
We obtain the ratios of modular forms approximately as:
\begin{align}
\begin{aligned}
\frac{Y_2(\tau)}{Y_1(\tau)}\simeq \omega\,(1+\,\epsilon_1) \, , \quad 
\frac{Y_3(\tau)}{Y_1(\tau)}\simeq -\frac{1}{2}\omega^2 \, 
(1+\, \epsilon_2)\,  ,
\quad \epsilon_1=\frac{1}{2} \epsilon_2=2.1\,i\,\epsilon\,.
\end{aligned}
\label{epST120}
\end{align}
These approximate  forms are  agreement with exact numerical values within  $1\,\%$
for $|\epsilon|\leq 0.05$.
Details are given in  Appendix C.2.

%%%%%%%%%%%%%%%%%%%%%%%%%%%%%%%%%%%%%%%%%%%%%%%
%%%%%%%%%%%%  Higher weights  %%%%%%%%%%%%%%%%%
%%%%%%%%%%%%%%%%%%%%%%%%%%%%%%%%%%%%%%%%%%%%%%%
The  higher weight modular forms $ Y_i^{(k)}$
in Eqs.\,(\ref{weight4}) and (\ref{weight6}) are also given
in terms of $\epsilon_1$ and  $\epsilon_2$ in Appendix C.2.

%%%%%%%%%%%%%%%%%%%%%%%%%%%%%%%%%%%%%%%%%%%%%%%%%%%%%%%%%%%%%%%%%%%%%%
  \subsection{Modular forms  towards  $\tau=i\infty$}    
  
We show the behavior of modular forms at large ${\rm Im}\tau$, where
the magnitude of $q=\exp{(2\pi i\tau)}$ is suppressed.
Taking leading terms of  Eq.\,(\ref{Y(2)}), we can express modular forms 
approximately  as:
\begin{align}
Y_1(\tau)\simeq 1+ 12 p\,\epsilon\,,\quad  
Y_2(\tau)\simeq -6 p^{\frac{1}{3}}\,\epsilon^{\frac{1}{3}}\,,
\quad  Y_3(\tau)\simeq -18 p^{\frac{2}{3}}\,\epsilon^{\frac{2}{3}}\,, \quad
p=e^{2\pi i\, {\rm Re}\, \tau}\,,\quad 
\epsilon=e^{-2\pi\, {\rm Im}\, \tau}\,.
\label{epsilonT0}
\end{align}
Indeed, we obtain  $\epsilon=3.487\times 10^{-6}$ for ${\rm Im}\,\tau=2$.
The leading correction is $\epsilon^{\frac{1}{3}}=0.0152$ in $Y_{2}(\tau)$
while other corrections of $\epsilon^{\frac{2}{3}}$ and $\epsilon$
is negligibly small.
Then,
\begin{align}
|Y_1(2i)|\simeq 1.00004\,,\qquad |Y_2(2i)|\simeq 0.09098\,,\qquad |Y_3(2i)|\simeq 0.00413\,,
\end{align}
which agree with exact values within $0.1\%$.
%%%%%%%%%%%%%%%%%%%%%%%%%%%%%%%%%%%%%%%%%%%%%%%%%%%%%
%%%%%%%%%%%%%%%%%%%%%%%%%%%%%%%%%%%%%%%%%%%%%%%%%%%%%%%%
%%%%%%%%%%%%%%%%%%%  weight 4 and 6  %%%%%%%%%%%%%%%%%%%
%%%%%%%%%%%%%%%%%%%%%%%%%%%%%%%%%%%%%%%%%%%%%%%%%%%%%%%% 
Higher weight modular forms $ Y_i^{(k)}$
in Eqs.\,(\ref{weight4}) and (\ref{weight6})
are also given 
in terms of $p$ and $\epsilon$ approximately in Appendix C.3.
%%%%%%%%%%%%%%%%%%%%%%%%%%%%%%%%%%%%%%%%%%%%%%%%%%%%%%%%%%%%%%%%%%%%%%
%%%%%%%%%%%%%%%%%%%%%%%%%%%%%%%%%%%%%%%%%%%%%%%%%%%%%%%%%%%%%%%%%%%%%%
%%%%%%%%%%%%%%%%%%%%%%%%%%%%%%%%%%%%%%%%%%%%%%%%%%%%%%%%%%%%%%%%%%%%%%
%%%%%%%%%%%%%%%%%%%%%%%%%%%%
%%%%%%%%%%%%%%%%%%%%%%%%%%%%
%%%%%%%%%%%%%%%%%%%%%%%%%%%%
%%%%%%%%% Lepton  %%%%%%%%%%
%%%%%%%%%%%%%%%%%%%%%%%%%%%%
%%%%%%%%%%%%%%%%%%%%%%%%%%%%
%%%%%%%%%%%%%%%%%%%%%%%%%%%%
\section{Lepton mass matrices in the $\rm A_4$ modular invariance}

%%%%%%%%%%%%%%%%%%%%%%%%%%%
%%%%%%%%%%%%%%%%%%%%%%%%%%%
%%%%%%%%%%%%%%%%%%%%%%%%%%%
\subsection{Model of lepton mass matrices}
%Let us apply  modular forms of section 4
%at nearby fixed points  to the lepton mass matrix 
%in Refs.\cite{Okada:2018yrn,Okada:2019uoy}.
Let us discuss models of the lepton mass matrices.
There are freedoms for the assignments of irreducible representations 
of $\rm A_4$ and modular weights to charged leptons and Higgs doublets.
The simplest assignment has been given in the conventional $\rm A_4$ model 	\cite{Altarelli:2005yp,Altarelli:2005yx}, in which	
three left-handed leptons are components of the triplet of the $\rm A_4$ group, 
but   three right-handed charged leptons,
($e^c, \mu^c, \tau^c$) are three different singlets 
$\bf (1,1'',1')$ of  $\rm A_4$, respectively.

Supposing neutrinos to be Majorana particles,
we present the neutrino mass matrix  through the Weinberg operator.
The simple one is given by assigning the $\rm A_4$ triplet and weight $-2$ to the lepton doublets
%%%%%%%%%%%%%%%%%%%%%%%%%%%%%%%%%%
\footnote{There is a possible  assignment of  weight $-1$ to the lepton doublets 
	of the $\rm A_4$ triplet. The neutrino mass matrix is given in terms of
	weight 2 modular forms through the Weinberg operator.
	However, this case is too simple to reproduce the lepton mixing angles
	 as discussed in Ref.\cite{Kobayashi:2018scp}.
},
%%%%%%%%%%%%%%%%%%%%%%%%%%%%%%%%%%
where the Higgs fields are supposed to be $\rm A_4$ singlets with weight $0$.
On the other hand, 
the charged lepton mass matrix depends on the assignment of weight for 
the right-handed charged leptons.
If those weights are $0$ for all right-handed charged leptons,
the charged lepton mass matrix are given in terms of only 
the weight $2$ modular forms of Eq.\,(\ref{eq:Y-A4}).
That is the simplest one.

Alternatively, we also consider
weight $4$ and $6$ modular forms of Eqs.\,(\ref{weight4}) and (\ref{weight6})
in addition to weight $2$ modular forms by taking non-vanishing weights.
The assignment is summarized in Table 2.

%%%%%%%%%%%%%%%%%%%%%%%%%%%%%%%%%%%%%%%%%%%%%%%%%%%%%%%
\begin{table}[H]
	\centering
	\begin{tabular}{|c||c|c|c|c|c|c|c|} \hline
		&$L$&$(e^c,\mu^c,\tau^c)$&$H_u$&$H_d$&$\bf Y_3^{(\rm 6)}, 
		Y_{3'}^{(\rm 6)}$ & $\bf Y_3^{(\rm 4)}, Y_1^{(\rm 4)},  Y_{\rm 1'}^{(4)}$
		&$\bf Y_3^{(\rm 2)}$ \\  \hline\hline 
		\rule[14pt]{0pt}{0pt}
		$SU(2)$&$\bf 2$&$\bf 1$&$\bf 2$&$\bf 2$&$\bf 1$& $\bf 1$&$\bf 1$\\
		$\rm A_4$&$\bf 3$& \bf (1,\ 1$''$,\ 1$'$)&$\bf 1$&$\bf 1$
		& $\bf 3$& $\bf 3\,, \quad  1\, \quad  1'$ &$\bf 3$\\
		$-k_I$&$ -2$&I: $(\ \ 0, \ \  \ 0, \ \ \ 0)$&0&0& 
		$k=6$ &$k=4$ &$k=2$ \\
		& &I\hspace{-.01em}I: $(-4, \ \  -2, \, \  0)$ & & & & &  \\
		\hline
	\end{tabular}	
	\caption{ Assignments of representations and  weights
		$-k_I$ for MSSM fields and  modular forms.
	}
	\label{tb:lepton}
\end{table}
%%%%%%%%%%%%%%%%%%%%%%%%%%%%%%%%%%%%%
%%%%%%%%%%%%%%%%%%%%%%%%%%%%%%%%%%%%%%%
\subsubsection{Neutrino mass matrix}
Let us begin with discussing the  neutrino mass matrix.
The superpotential of the neutrino mass term, $w_\nu$ is  given as:
\begin{align}
w_\nu&=-\frac{1}{\Lambda}(H_u H_u LL{\bf Y_r^{(\rm k)}})_{\bf 1}~,
\label{Weinberg}
\end{align}
where $L$ is the left-handed $\rm A_4$ triplet leptons,
 $H_u$ is the Higgs doublet, and $\Lambda$ is a relevant cut off scale.
Since the  left-handed lepton doublet has weight $-2$, the superpotential
is given in terms of  modular forms of weight $4$, ${\bf Y_3^{(\rm 4)}}$,
${\bf Y_1^{(\rm 4)}}$ and  ${\bf Y_{1'}^{(\rm 4)}}$.
%%%
By putting the VEV of the neutral component of $H_u$, $v_u$ and taking $(\nu_e, \nu_\mu,\nu_\tau)$ for left-handed neutrinos of $L$,
we have

\begin{align}
%\begin{aligned}
w_\nu &=\frac{v_u^2}{\Lambda}
\left [ 
\begin{pmatrix}
2\nu_e\nu_e-\nu_\mu\nu_\tau-\nu_\tau\nu_\mu\\
2\nu_\tau\nu_\tau-\nu_e\nu_\mu-\nu_\mu\nu_\tau\\
2\nu_\mu\nu_\mu-\nu_\tau\nu_e-\nu_e\nu_\tau
\end{pmatrix} \otimes
{\bf Y_3^{(\rm 4)}}  \right . \nonumber \\
& \left .  + \ 
(\nu_e\nu_e+\nu_\mu\nu_\tau+\nu_\tau\nu_\mu)
\otimes g_{\nu 1}{\bf Y_1^{(\rm 4)}}
+
(\nu_e\nu_\tau+\nu_\mu\nu_\mu+\nu_\tau\nu_e)
\otimes g_{\nu 2}{\bf Y_{1'}^{(\rm 4)}}
\right ]  \nonumber \\
=&\frac{v_u^2}{\Lambda}
\left[(2\nu_e\nu_e-\nu_\mu\nu_\tau-\nu_\tau\nu_\mu)Y_1^{(\rm 4)}+
(2\nu_\tau\nu_\tau-\nu_e\nu_\mu-\nu_\mu\nu_e)Y_3^{(\rm 4)}
+(2\nu_\mu\nu_\mu-\nu_\tau\nu_e-\nu_e\nu_\tau)Y_2^{(\rm 4)}\right .
\nonumber \\
& \left .  + \ 
(\nu_e\nu_e+\nu_\mu\nu_\tau+\nu_\tau\nu_\mu)
g_{\nu 1}{\bf Y_1^{(\rm 4)}}
+
(\nu_e\nu_\tau+\nu_\mu\nu_\mu+\nu_\tau\nu_e)
g_{\nu 2}{\bf Y_{1'}^{(\rm 4)}}
\right ]   \ , 
% \end{aligned}
\end{align}
where ${\bf Y_3^{(\rm 4)}}$, ${\bf Y_1^{(\rm 4)}}$ and ${\bf Y_{1'}^{(\rm 4)}}$
are given in Eq.\,(\ref{weight4}), and  $g_{\nu 1}$, $g_{\nu 2}$ are complex parameters.
The neutrino mass matrix is written as follows:
\begin{align}
M_\nu=\frac{v_u^2}{\Lambda} \left [
\begin{pmatrix}
2Y_1^{(4)} & -Y_3^{(4)} & -Y_2^{(4)}\\
-Y_3^{(4)} & 2Y_2^{(4)} & -Y_1^{(4)} \\
-Y_2^{(4)} & -Y_1^{(4)} & 2Y_3^{(4)}
\end{pmatrix}
+g_{\nu 1} {\bf Y_{1}^{(\rm 4)}  }
\begin{pmatrix}
1 & 0 &0\\ 0 & 0 & 1 \\ 0 & 1 & 0
\end{pmatrix}
+g_{\nu 2} {\bf Y_{1'}^{(\rm 4)} }
\begin{pmatrix}
0 & 0 &1\\ 0 & 1 & 0 \\ 1 & 0 & 0
\end{pmatrix}
\right ]_{LL} \, .
\label{neutrinomassmatrix}
\end{align}
%%%%%%%%%%%%%%%%%%%%%%%%%%%%%%%%%%%%%%%%%%%%%%%

\subsubsection{Charged lepton mass matrix}

The relevant superpotentials of the charged lepton masses are given
for two cases as follows:
%%%%%%%%%%%%%%%%%%%%%%%%%%%%%%%%%%%%%%%%%%%%%%%%%%%%%  %%%%%%%%%%%%%%%%%%%%%%%%%%%%%%%%%%%%%%%%%%%%%%%%%%%%%
\begin{align}
{\rm I}
\,:\quad
w_E&=\alpha_e e^c H_d {\bf Y^{(\rm 2)}_3}L+
\beta_e \mu^c H_d {\bf Y^{(\rm 2)}_3}L+
\gamma_e \tau^c H_d {\bf Y^{(\rm 2)}_3}L~,\\
{\rm I\hspace{-.01em}I}\,:\quad
w_E&=\alpha_e e^c H_d {\bf Y^{(\rm 6)}_3} L+
\alpha'_e e^c H_d {\bf Y_{3'}^{(\rm 6)}} L+
\beta_e \mu^c H_d {\bf Y^{(\rm 4)}_3}L+
\gamma_e \tau^c H_d {\bf Y^{(\rm 2)}_3}L~, 
\end{align}
%%%%%%%%%%%%%%%%%%
where $L$ is the left-handed $\rm A_4$ triplet leptons,
$H_d$ is the Higgs doublet.

The  charged lepton mass matrices $M_E$  are given as: 
%%%%%%%%%%%%%%%%%%%%%%%%%%%%%%%%%%%%%%%%%%%%%%% 
\begin{align}
%%%%%%%%%%%%%%%%%%%%%%%%%%%%%%%%%%%%%%%%%%%%%%% 
& \begin{aligned}
{\rm  I}: \quad M_E=v_d
\begin{pmatrix}
\alpha_e & 0 & 0 \\
0 &\beta_e & 0\\
0 & 0 &\gamma_e
\end{pmatrix} \left [
\begin{pmatrix}
Y_1^{(2)} & Y_3^{(2)} & Y_2^{(2)} \\
Y_2^{(2)} & Y_1^{(2)} &  Y_3^{(2)} \\
Y_3^{(2)} &  Y_2^{(2)}&  Y_1^{(2)}
\end{pmatrix}
\right ]_{RL}, 
\end{aligned}
\label{ME222}\\
\\
& \begin{aligned}
{\rm I\hspace{-.01em}I}: \quad M_E=v_d
\begin{pmatrix}
\alpha_e & 0 & 0 \\
0 &\beta_e & 0\\
0 & 0 &\gamma_e
\end{pmatrix} \left [
\begin{pmatrix}
Y_1^{(6)}+g_e Y_1^{'(6)} & Y_3^{(6)} +g_e Y_3^{'(6)} 
& Y_2^{(6)}+g_e Y_2^{'(6)} \\
Y_2^{(4)} & Y_1^{(4)} &  Y_3^{(4)} \\
Y_3^{(2)} &  Y_2^{(2)}&  Y_1^{(2)}
\end{pmatrix}
\right ]_{RL},
\end{aligned}
\label{ME642}
%%%%%%%%%%%%%%%%%%%%%%%%%%%%%%%%%%%%%%%%%%%%%%%%
\end{align}
respectively,
where coefficients $\alpha_e$, $\beta_e$ and $\gamma_e$ are real parameters
while $g_e$ is complex one, and $v_d$ is  VEV of the neutral component of $H_d$.

Model parameters of leptons are
$\alpha_e$, $\beta_e$, $\gamma_e$, $(g_e)$,  $g_{\nu 1}$ and $g_{\nu 2}$
in addition to the modulus $\tau$.
We examine these mass matrices around the fixed points.

%%%%%%%%%%%%%%%%%%%%%%%%%%%%%%%%%%%%%%%%%%
%\footnote{If $\tau$ is scanned as a free parameter in the fundamental region,
%	there are many patterns of the predicted PMNS.}.
%%%%%%%%%%%%%%%%%%%%%%%%   
%%%%%%%%%%%%%%%%%%%%%%%%%%%%%%%%%%%%%%%%%%%%%%%%%
%%%%%%%%%%%%%%%%%%%%%%%%%%%%%%%%%%%%%%%%%%%%%%%%%
\subsection{Lepton mass matrix at $\tau=i$}
%%%%%%%%%%%%%%%%%%%%%%%%%%%%%%%%%%%%%%%%%%%%%%%%%%%
%%%%%%%%%%%%%%%  Neutrino mass matrix %%%%%%%%%%%%%
%%%%%%%%%%%%%%%%%%%%%%%%%%%%%%%%%%%%%%%%%%%%%%%%%%%
%%%%%%%%%%%%%%%%%%%%%%%%%%%%%%%%%%%%%%%%%%%%%%%%%%%
\subsubsection{Neutrino mass matrix at $\tau=i$}

We get the neutrino mass matrix at $\tau=i$  by putting modular forms 
in Table 1 into Eq.\,(\ref{neutrinomassmatrix}) as:
\begin{align}
M_\nu=\frac{v_u^2}{\Lambda} (6-3\sqrt{3})Y_0^2\left [
\begin{pmatrix}
2 & -1 & 1\\
-1 & 2 & -1 \\
-1 & -1 & 2
\end{pmatrix}
+g_{1} 
\begin{pmatrix}
1 & 0 &0\\ 0 & 0 & 1 \\ 0 & 1 & 0
\end{pmatrix}
+g_{2} 
\begin{pmatrix}
0 & 0 &1\\ 0 & 1 & 0 \\ 1 & 0 & 0
\end{pmatrix}
\right ] \ ,
\label{Neutrino}
\end{align}
where 
\begin{align}
g_1=\frac{6\sqrt{3}-9}{6-3\sqrt{3}} g_{\nu 1}=\sqrt{3} g_{\nu 1}\ , \qquad 
g_2=\frac{9-6\sqrt{3}}{6-3\sqrt{3}} g_{\nu 2}=-\sqrt{3} g_{\nu 2}\ .
\end{align}
%%%%%%%%%%%%%%%%%%%%%%%%%%%%
We move to the  disgonal basis of $S$.
% which is denoted as  $\hat S$.
By using the unitary transformation of Eq.\,(\ref{STdiagonal}), $V_{S2}$,
the mass matrix is transformed  as:
%we have  present  $V^*M_\nu V_2^\dagger $ as:
\begin{align}
\hat M_\nu\equiv V_{S2}^* M_\nu V_{S2}^\dagger=\frac{v_u^2}{\Lambda}Y_0^2
\begin{pmatrix}
\ \ g_1 + g_2  & 0 & 0\\
0 & 3+g_1-\frac{1}{2}g_2 & \frac{\sqrt{3}}{2}g_2 \\
0 & \frac{\sqrt{3}}{2}g_2 & 3-g_1+\frac{1}{2}g_2 \ \ 
\end{pmatrix} .
\label{Neutrinonew}
\end{align}
Off diagonal entries of (2,3) and (3,2) are non-zero as discussed in Eq.\,(\ref{zeros}). 
At the limit of vanishing $g_1$ and $g_2$,
the lightest neutrino mass is zero and other ones are degenerated.

%%%%%%%%%%%%%%%%%%%%%%%%%%%%%%%%%%%
In order to discuss the flavor mixing angle, 
we show $\hat M_\nu^\dagger  \hat M_\nu$ as
\begin{align}
\begin{aligned}
{\cal M}_\nu^{2(0)}\equiv \hat M_\nu^\dagger \hat M_\nu= \left (\frac{v_u^2}{\Lambda} Y_0^2 \right )^2
\end{aligned}
\begin{aligned}
\begin{pmatrix}
|g_1 + g_2|^2  & 0 & 0\\
0 & G_\nu+6  {\rm Re}[g_1]-3{\rm Re}[g_2] & \frac{\sqrt{3}}{2}\left ( 6 \ {\rm Re}[g_2]+2i \ {\rm Im}[g_1^*g_2]\right )\\
0& \frac{\sqrt{3}}{2}\left ( 6 \ {\rm Re}[g_2]- 
2i\ {\rm Im}[g_1^*g_2]\right )& G_\nu-6 {\rm Re}[g_1]+3{\rm Re}[g_2] 
\end{pmatrix} ,
\end{aligned}
\label{Neutrinonew2}
\end{align}
where
\begin{align}
G_\nu= 9+|g_1|^2+|g_2|^2-{\rm Re}[g_1^*g_2] \ .
\end{align}
The imaginary part of this matrix is factored out by using a phase matrix $P_\nu$ as:
\begin{align}
\begin{aligned}
\left (\frac{v_u^2}{\Lambda} Y_0^2 \right )^2 
\end{aligned}
\begin{aligned} P_\nu
\begin{pmatrix}
|g_1 + g_2|^2  & 0 & 0\\
0 & G_\nu+6  {\rm Re}[g_1]-3{\rm Re}[g_2] & \sqrt{3}\sqrt{ 9  ({\rm Re}[g_2])^2+ ({\rm Im}[g_1^*g_2] )^2}\\
0&\sqrt{3}\sqrt{ 9  ({\rm Re}[g_2])^2+ ({\rm Im}[g_1^*g_2] )^2} 
&G_\nu-6 {\rm Re}[g_1]+3{\rm Re}[g_2] \ \ 
\end{pmatrix} P_\nu^* \ ,
\end{aligned}
\label{Neutrino-real}
\end{align}
where 
\begin{align}
P_\nu=
\begin{pmatrix}
1  & 0 & 0\\
0 & 1 & 0 \\
0 & 0 &\ \ \  e^{-i\phi^\nu} \ \ 
\end{pmatrix} ,
\label{P}
\end{align}
with 
\begin{align}
\tan \phi^\nu=\frac{{\rm Im}[g_1^*g_2]}{3\ {\rm Re}[g_2]} \ .
\label{phinu}
\end{align}
On the other hand, mass eigenvalues $m_{01}^{2}$, $m_{02}^{2}$ and $m_{03}^{2}$
of ${\cal M}_\nu^{2(0)}$ satisfy:
\begin{align}
m_{01}^2=|g_1+g_2|^2, \
m_{02}^2+ m_{03}^2=18+2(|g_1|^2+|g_2|^2)-2{\rm Re}(g_1^*g_2), \  
m_{02}^2 m_{03}^2=|9-g_1^2-g_2^2+g_1g_2|^2 ,
\label{mass20}
\end{align}
in the unit of $({v_u^2}/{\Lambda})^2 Y_0^4$.
The mixing angle between the 2nd- and 3rd-family, $\theta_{23}^\nu$ is given as:
\begin{align}
\tan 2\theta_{23}^\nu=
\frac{1}{\sqrt{3}}
\frac{\sqrt{9  ({\rm Re}[g_2])^2+ 
		({\rm Im}[g_1^*g_2] )^2}}{ {\rm Re}[g_2]-2{\rm Re}[g_1]} \ . 
\label{theta23}
\end{align}
If we put ${\rm Re}[g_2]=2{\rm Re}[g_1]$,  
we obtain the maximal mixing angle $\theta_{23}^\nu=45^\circ$.
Thus, the large mixing angle is easily obtained by choosing relevant
parameters $g_1$ and $g_2$.
It is also noticed that  $\theta_{23}^\nu$ vanishes for $g_2=0$.
Thus,  $\theta_{23}^\nu$ could be $0$--$45^\circ$ depending on  $g_1$ and $g_2$.

%%%%%%%%%%%%%%%%%%%%%%%%%%%%%%%%%%%%%%%%%%%%%%%%%%%%%%%%%
\subsubsection{Neutrino mass matrix at nearby $\tau=i$}

As discussed in the previous subsubsection,  the large  $\theta_{23}^\nu$ is
easily reproduced at $\tau=i$.
The large flavor mixing angle between the 1st- and 2nd-family,
$\theta_{12}^\nu$ is also  realized at nearby $\tau=i$.
Mass matrix of  neutrinos  in Eq.\,(\ref{neutrinomassmatrix}),
$M_\nu$ 
are corrected due to the deviation from the fixed point of $\tau=i$.
Putting modular forms of Eq.\,(\ref{epS120}) (see also Appendix C.1)
into $M_\nu$,
 the corrections to Eq.\,(\ref{Neutrinonew2}) are given by only a small variable $\epsilon$ in Eq.\,(\ref{epS120})
 in the diagonal base of $S$.
In the 1st order approximation of $\epsilon$,
the correction  ${\cal M}_\nu^{2(1)}$ is given as:
\begin{align}
\begin{aligned} 
{\cal M}_\nu^{2(1)}=\left (\frac{v_u^2}{\Lambda} Y_0^2\right )^2
\begin{pmatrix}
0 & \delta_{\nu 2} & \delta_{\nu 3} \\ 
\delta_{\nu 2}^* & \delta_{\nu 4} & \delta_{\nu 5} \\ 
\delta_{\nu 3}^* & \delta_{\nu 5}^* & \delta_{\nu 6} 
\end{pmatrix} , \qquad
\end{aligned}
\label{mass2-1st}
\end{align}
where   $\delta_{\nu i}$\,$(i=2$--$6)$ are given
in terms of $\epsilon$, $g_1$ and  $g_2$.
Due to the 1st order perturbation of $\epsilon$, we can obtain the mixing angle  $\theta_{12}^\nu$, which vanishes in the 0th order of perturbation.
In order to estimate the flavor mixing angles, we present relevant    $\delta_{\nu i}$ explicitly as:

%%%%%%%%%%%%%%%%%%%%%%%%%%%%
%%%%%%%%%%%%%%%%%%%%%%%%%%%%
\begin{align}
&  \begin{aligned}
\delta_{\nu 2}&= \frac{-1}{\sqrt{2}}\{(g_{1}^*+g_{2}^*)
[(1+\sqrt{3})\epsilon_1+\epsilon_2]+
\epsilon_1^* [(3+g_{1})(1+\sqrt{3})-2g_{2}]+
\epsilon_2^*[(3+g_{1})+(1-\sqrt{3})g_{2}]
\} \nonumber\\
&\simeq -3.34 (g_{1}^*+g_{2}^*)\epsilon_1-(10.04+3.35 g_1-2.45 g_2)\epsilon_1^* \,,
\end{aligned}\nonumber\\
&   \begin{aligned}
\delta_{\nu 3}&=  \frac{1}{\sqrt{6}}\{(g_{1}^*+g_{2}^*)
[(3-\sqrt{3})\epsilon_1+(2\sqrt{3}-3)\epsilon_2]+
\epsilon_1^* [(3-\sqrt{3})(3-g_{1})-2\sqrt{3}g_{2}] \\
&+\epsilon_2^*[(2\sqrt{3}-3)(3-g_{1})-(3-\sqrt{3})g_{2}]
\}\simeq 0.90(g_{1}^*+g_{2}^*)\epsilon_1+(2.69-0.90 g_1-2.45 g_2)\epsilon_1^*\,  ,
\end{aligned}
%   &\begin{aligned}
%   \delta_{\nu 5}=\sqrt{3}(\epsilon_1 g_{\nu 2}^*+\epsilon_1^*g_{\nu 2})
%   +\frac{1}{2}(2\sqrt{3}-3)(\epsilon_2 g_{\nu 2}^*+\epsilon_2^*g_{\nu 2})\,.
%   \end{aligned}
\label{deltanuS}
\end{align} 
where $\epsilon_1=2.05i \epsilon$, and $\epsilon_2=2\epsilon_1$ in Eq.\,(\ref{epS120})  is used in last approximate equalities. 
%%%%%%%%%%%%%%%%%%%%%%%%%%%%%%%%%%%%%%%%%%%%%%%%%%
% for eigenvalues $m_{1}^{2(0)}$, $m_{2}^{2(0)}$ and $m_{3}^{2(0)}$, %respectively, which are
%  \begin{align}
%  m_{1}^{2(0)}=|g_1+g_2|^2, \
%  m_{2}^{2(0)}+ m_{3}^{2(0)}=18+2(|g_1|2+|g_2|^2)-2{\rm Re}(g_1^*g_2), \  
%  m_{2}^{2(0)} m_{3}^{2(0)}=|9-g_1^2-g_2^2+g_1g_2|^2 ,
%  \end{align}
%  in the unit of $({v_u^2}/{\Lambda})^2 Y_0^4$.
%%%%%%%%%%%%%%%%%%%%%%%%%%%%%%%%%%%%%%%%%%%%%%%%%%%%%%

Let us estimate  the mixing angles, $\theta_{12}^\nu$
and $\theta_{13}^\nu$  in terms of $\delta_{\nu 2}$ and $\delta_{\nu 3}$.
The eigenvectors of the lowest order in   ${\cal M}_\nu^{2(0)}$ is given,
\begin{align}
\begin{aligned} u_{\nu 1}^{(0)}=
\begin{pmatrix}
1\\  0\\ 0
\end{pmatrix} , \qquad
u_{\nu 2}^{(0)}= \begin{pmatrix}
0\\  \cos\theta^\nu_{23}\\ -\sin\theta^\nu_{23} e^{-i\phi^\nu}
\end{pmatrix} , \qquad
u_{\nu 3}^{(0)}=\begin{pmatrix}
0\\ \sin\theta^\nu_{23}\\ \cos\theta^\nu_{23} e^{-i\phi^\nu}
\end{pmatrix} ,
\end{aligned}
\label{vectorneutrino}
\end{align}
for eigenvalues $m_{01}^2$, $m_{02}^2$ and $m_{03}^2$
of Eq.\,(\ref{mass20}), respectively.

We can calculate corrections of eigenvectors  in the 1st order of $\epsilon$.
In order to estimate the non-vanishing mixing angle between the 1st- and 2nd-family,
we calculate the eigenvector of 1st order, $u_{\nu 2}^{(1)}$, which is given
\begin{align}
u_{\nu 2}^{(1)}= C^{\nu}_{21} u_{\nu 1}^{(0)} + C^{\nu}_{23} u_{\nu 3}^{(0)} \ ,
\end{align} 
where 
\begin{align}
C^{\nu}_{ji}=\frac{\langle u_{\nu j}^{(0)}| {\cal M}_\nu^{2(1)} |u_{\nu i}^{(0)}\rangle} {m^2_{0j}-m^2_{0i}} \ .
\end{align} 
Therefore, the non-vanishing (1-2) mixing appears
at the first component of $u_{\nu 2}^{(1)}$ as:
%, which is given by Eqs.\,(\ref{mass2-1st}) and  (\ref{vectorneutrino}), as:
\begin{align}
u_{\nu 2}^{(1)}[1,1]=C^{\nu}_{21}=\frac{ \delta_{\nu 2}^*\cos\theta^\nu_{23}  -\delta_{\nu 3}^*\sin\theta^\nu_{23} e^{i\phi^\nu}}{m^2_{02}-m^2_{01}} \ .
\end{align} 
%In order to estimate the magnitude of the (1-2) mixing angle,
%which is the first component of $u_{\nu 2}^{(1)}$,
Here, we take $2 g_1= g_2$, which leads to the maximal mixing
$\theta_{23}^\nu=45^\circ$ as seen in Eq.\,(\ref{theta23}).
Then, the  mass squares   are given  from Eq.\,(\ref{mass20}) as: 
\begin{align}
m_{01}^2=9 |g_1^2|, \qquad 
m_{02}^2=3 \left (3+ |g_1^2|-2\sqrt{3} |{\rm Re}g_1| \right ), \qquad 
m_{03}^2=3 \left (3+ |g_1^2|+2\sqrt{3} |{\rm Re}g_1| \right )\, ,
\end{align} 
in the unit of $({v_u^2}/{\Lambda})^2 Y_0^4$.
Supposing  NH of  neutrino masses, we take the observed ratio of
$\Delta m^2_{\rm atm}/\Delta m^2_{\rm sol}=34.2$, which leads to 
 $g_1=0.61$ by neglecting the imaginary part of $g_1$.
Then,
$\delta_{\nu 2}^*$ and  $\delta_{\nu 3}^*$ are given
in terms of $\epsilon$
 by using $\epsilon_1=2.05\, i\,\epsilon$ in Eq.\,(\ref{epS120})
 as follows:
\begin{align}
\delta_{\nu 2}^*=-18.6 \,i\, \epsilon -12.6 \,i\,\epsilon^* \ , \qquad
\delta_{\nu 3}^*=-1.76 \,i\, \epsilon -0.52 \,i\, \epsilon^* \ .
\end{align} 
Neglecting $\delta_{\nu 3}$ because of $|\delta_{\nu 2}^*|\gg
 |\delta_{\nu 3}^*|$, we have
\begin{align}
u_{\nu 2}^{(1)}[1,1]\simeq \frac{ \delta_{\nu 2}^*\cos\theta^\nu_{23}  }{m^2_{02}-m^2_{01}}= -i\,\frac{ 18.6 \, \epsilon +12.6 \,\epsilon^*  }{0.383 \sqrt{2}}\ ,
\end{align} 
where  $\theta_{23}^\nu=45^\circ$ is put.
We obtain $u_{\nu 2}^{(1)}[1,1]\simeq 0.55$
($\theta_{12}^\nu \simeq 35^\circ$)
by putting $\epsilon=0.05\,i$.
Thus, the large (1-2) mixing angle could be reproduced
by the correction terms  in the neutrino mass matrix
due to the small deviation from  $\tau=i$.
It is remarked that the sum of three neutrino masses is around $110$\,meV
 taking  $2g_1= g_2=1.22$.

On the other hand, the non-vanishing  (1-3) mixing is derived as:
\begin{align}
u_{\nu 3}^{(1)}[1,1]=C^{\nu}_{31}=\frac{ \delta_{\nu 2}^*\sin\theta^\nu_{23}  + \delta_{\nu 3}^*\cos\theta^\nu_{23} e^{i\phi^\nu}}{m^2_{03}-m^2_{01}} \ .
\end{align} 
Since $({m^2_{03}-m^2_{01}})$ is $30$ times larger than
$({m^2_{02}-m^2_{01}})$, $u_{\nu 3}^{(1)}[1,1]$ is suppressed
compared with $u_{\nu 2}^{(1)}[1,1]$.
Indeed, the  (1-3) mixing angle is ${\cal O}(0.01)$.
Therefore, the observed $\theta_{13}\sim 0.15$ of  the PMNS matrix
should  be derived from the charged lepton sector.
It is noted that the correction to the (2-3) mixing is also ${\cal O}(0.01)$
because  $u_{\nu 3}^{(1)}[2,1]$ is suppressed due to the large $({m^2_{03}-m^2_{01}})$.

%%%%%%%%%%%%%%%%%%%%%%%%%%%%%%%%%%%%%%%%%%%%%%%%%%%
%%%%%%%%%%%%%%%%%  IH  %%%%%%%%%%%%%%%%%%%%%%%%%%%%
%%%%%%%%%%%%%%%%%%%%%%%%%%%%%%%%%%%%%%%%%%%%%%%%%%%
We can also discuss the case of  IH of the neutrino masses
by taking $\Delta m^2_{\rm atm}/\Delta m^2_{\rm sol}=-34.2$.
The large mixing angles $\theta_{23}^\nu$ and
$\theta_{12}^\nu$ are obtained  if we take 
$g_1=g_2/2=-2.45$.
The sum of three neutrino masses is around $90$\,meV.
%%%%%%%%%%%%%%%%%%%%%%%%%%%%%%%%%%%%%%%%%%%%

Thus,  our neutrino mass matrix is attractive one at nearby $\tau=i$.
Therefore, we should examine the contribution from the charged lepton sector
carefully for both NH and IH of neutrinos.

%%%%%%%%%%%%%%%%%%%%%%%%%%%%%%%%%%%%%%%%%%
%%%%%%%%%%%%%%%%%%%%%%%%%%%%%%%%%%%%%%%%%%
%%%%%%%%%%%%%%%%%%%%%%%%%%%%%%%%%%%%%%%%%%
%%%%%%%%%%%%%%%%%%%%%%%%%%%%%%%%%%%%%%%%%%
%Since  the $U_{e3}$ element of the PMS matrix  is estimated by
%\begin{align}
%U_{e3}=E^*_{11} N_{13}+E^*_{21} N_{23}+E^*_{31} N_{33} \, ,
%\end{align}
%we can calculate it as:
%%%%%%%%%%%%%%%%%%%%%%%%%%%%%
%\begin{align}
%U_{e3}\simeq u_{e1}^{(1)*}[2,1]\sin\theta^\nu_{23}+u_{e1}^{(1)*}[3,1]
%\cos\theta^\nu_{23}e^{-i\phi^\nu}
%\simeq \frac{1}{\sqrt{2}} %(u_{e1}^{(1)*}[2,1]+u_{e1}^{(1)*}[3,1]e^{-i\phi^\nu}),
%\end{align}
%where $N_{13}$ is expected tiny.
%The observed value $|U_{e3}|=0.15$ is possibly reproduced by choosing the %relevant phase
%$\phi^\nu$.

%%%%%%%%%%%%%%%%%%%%%%%%%%%%%%%%%%%%%%%%%%%%%%%%%%%%%%%%%%%%%
%%%%%%%%%%%%%%%%%%%%%%%%%%%%%%%%%%%%%%%%%%%%%%%%%%%%%%%%%%%%%
%%%%%%%%%%%%%%%%%%% Charged lepton I %%%%%%%%%%%%%%%%%%%%%%%%
%%%%%%%%%%%%%%%%%%%%%%%%%%%%%%%%%%%%%%%%%%%%%%%%%%%%%%%%%%%%%

\subsubsection{Charged lepton mass matrix I at  $\tau=i$}

The charged lepton mass matrix I is the  simplest one,
which is given by using only weight 2 modular forms.  It is given
at fixed points of $\tau=i$ in the base of $S$ of Eq.\,(\ref{STbase}) as follows:
%%%%%%%%%%%%%%%%%%%%%%%%%%%%%%%
\begin{align}
&\begin{aligned}
M_E=v_d
\begin{pmatrix}
\alpha_e & 0 & 0 \\
0 &\beta_e & 0\\
0 & 0 &\gamma_e
\end{pmatrix} 
\begin{pmatrix}
Y_1 & Y_3& Y_2 \\ Y_2 & Y_1&  Y_3\\ Y_3 &  Y_2&  Y_1
\end{pmatrix}
\end{aligned}
\begin{aligned}
=&
\begin{pmatrix}
\tilde \alpha_e & 0 & 0 \\
0 &\tilde \beta_e & 0\\
0 & 0 &\tilde\gamma_e
\end{pmatrix} 
\begin{pmatrix}
1&-2+\sqrt{3} & 1-\sqrt{3}\\
1-\sqrt{3}& 1 & -2+\sqrt{3} \\
-2+\sqrt{3} & 1-\sqrt{3}& 1
\end{pmatrix}\, ,
\end{aligned}
\end{align}
where $\tilde \alpha_e= v_dY_0 \alpha_e$,
$\tilde \beta_e=  v_dY_0  \beta_e$ and 
$\tilde \gamma_e = v_dY_0  \gamma_e$.
%%%%%%%%%%%%%%%%%%%%%%%%%%%%%%%%%%%%%%%%%%
We move to  the diagonal base of $S$.
By using the unitary transformation of Eq.\,(\ref{Sdiagonal}),
the mass matrix is transformed  as presented in Eq.\,(\ref{diagonalbase1}).
Then, we have:
\begin{align}
&{\cal M}_E^{2(0)}\equiv  V_{S2} M_E^\dagger M_E V_{S2}^\dagger = \frac{3}{2}
\begin{aligned}
\begin{pmatrix}
0& 0 & 0 \\
0 &  \tilde\alpha_e^2 + 2(2- \sqrt{3})\tilde \beta_e^2+(7-4\sqrt{3})\tilde \gamma_e^2
& -(2- \sqrt{3})\tilde(\alpha_e^2 -2\tilde \beta_e^2+\tilde \gamma_e^2 )\\
0 & -(2- \sqrt{3})\tilde(\alpha_e^2 -2\tilde \beta_e^2+\tilde \gamma_e^2 ) &  (7-4\sqrt{3})\tilde\alpha_e^2 + 2(2- \sqrt{3})\tilde \beta_e^2+\tilde \gamma_e^2 
\end{pmatrix}
\end{aligned},
\label{chargedlepton-III}
\end{align}
which is a real matrix with rank 2.

%%%%%%%%%%%%%%%%%%%%%%%%%%%%%%%%%%%%%%%%%%%
Since the lightest charged lepton is massless at $\tau=i$, the small deviation from $\tau=i$ is required to obtain the electron mass.
It is remarked that
the flavor mixing  between 2nd- and 3rd-family appears
 at the fixed point $\tau=i$  as seen in  Eq.\,(\ref{chargedlepton-III}).
It is given as:
\begin{align}
\tan 2\theta_{23}^e =
-2 \,\frac{(2- \sqrt{3})\,(\tilde \alpha_e^2-2 \tilde \beta_e^2+\tilde \gamma_e^2)}
{2(2\sqrt{3}-3)\,(\tilde \gamma_e^2-\tilde \alpha_e^2)} =-\frac{1}{\sqrt{3}} 
 \frac{\tilde \alpha_e^2-2 \tilde \beta_e^2+\tilde \gamma_e^2}
{\tilde \gamma_e^2-\tilde \alpha_e^2}  \,,
\label{theta12e-III}
\end{align}
which leads to
$\theta_{23}^e\simeq 15^\circ$ for $\tilde \alpha_e\gg\tilde\beta_e,\tilde\gamma_e$,
$\theta_{23}^e\simeq -15^\circ$ for $\tilde \gamma_e\gg\tilde\beta_e,\tilde\alpha_e$,
$\theta_{23}^e\simeq 
45^\circ$ for $\tilde \beta _e\gg\tilde\alpha_e\gg \tilde\gamma_e$
and $\theta_{23}^e\simeq 
-45^\circ$ for $\tilde \beta _e\gg\tilde\gamma_e\gg \tilde\alpha_e$,
respectively.
This mixing angle leads to $\theta_{23}$ of the PMNS matrix
 by cooperating with the neutrino mixing angle  $\theta_{23}^\nu$ 
 in Eq.\,(\ref{theta23}).
%%%%%%%%%%%%%%%%%%%%%%%%%%%%%%%%%%%%%%%%%%%%%%%%%%%
%%%%%%%%%%%%%%%%%%%%%%%%%%%%%%%%%%%%%%%%%%%%%%%%%%%
\subsubsection{Charged lepton mass matrix I at nearby $\tau=i$}

In order to obtain the electron mass,  $\tau$ should be deviated a little bit from the fixed point $\tau=i$.
%%%%%%%%%%%%%%%%%%%%%%%%%%%%%%%%%%%%%%%%%% 
By using modular forms at nearby  $\tau=i$ in  Eq.\,(\ref{epS120}),
 we obtain the additional contribution  ${\cal M}_E^{2(1)}$ to  ${\cal M}_E^{2(0)}$ in Eq.\,(\ref{chargedlepton-III}) of order $\epsilon$ as:     
%%%%%%%%%%%%%%%%%%%%%%%%%%%%%%%%%%
%%%%%%%%%%%%%%%%%%%%%%%%%%%%%%%%%%
\begin{align}
\begin{aligned}
{\cal M}_E^{2(1)}\simeq 
\begin{pmatrix}
0 &  \delta_{e2}& \delta_{e3}\\ 
\delta_{e2}^* &  \delta_{e4}& \delta_{e5}\\ 
\delta_{e3}^* &  \delta_{e5}^*& \delta_{e6}
\end{pmatrix} \,,
\end{aligned}
\label{M2ES-III}
\end{align}
where $\delta_{ei}$ are given in terms of
$\epsilon$,  $\tilde \alpha_e^2$, $\tilde \beta_e^2$ and 
$\tilde \gamma_e^2$.
In order to estimate the flavor mixing angles, we present relevant  $\delta_{ei}$ as:
\begin{align}
& \begin{aligned}
\delta_{e 2}&=\frac{1}{\sqrt{2}} \{
[(\sqrt{3}-1)\epsilon_1^* +(\sqrt{3}-2)\epsilon_2^* ] \tilde\alpha_e^2
+ [(4-2\sqrt{3})\epsilon_1^* +(3\sqrt{3}-5)\epsilon_2^* ] \tilde\beta_e^2 \\
&+ [(3\sqrt{3}-5)\epsilon_1^* +(7-4\sqrt{3})\epsilon_2^* ] \tilde\gamma_e^2
\} \simeq \frac{1}{\sqrt{2}} \epsilon_1^* 
[(3\sqrt{3}-5)  \tilde\alpha_e^2 +2(2\sqrt{3}-3)  \tilde\beta_e^2  +(9-5\sqrt{3})  \tilde\gamma_e^2]\, ,
\end{aligned}\\
%%%%%%%%%%%%
&  \begin{aligned}
\delta_{e 3}&= \frac{1}{\sqrt{6}} \{ 
[(9-5\sqrt{3})\epsilon_1^* +(7\sqrt{3}-12)\epsilon_2^* ] \tilde\alpha_e^2 +
[(4\sqrt{3}-6)\epsilon_1^* +(9-5\sqrt{3})\epsilon_2^* ] \tilde\beta_e^2 \\
&+ [(\sqrt{3}-3)\epsilon_1^* +(3-2\sqrt{3})\epsilon_2^* ] \tilde\gamma_e^2 
\}\simeq
\frac{\sqrt{6}} {2}\epsilon_1^* 
[(3\sqrt{3}-5)  \tilde\alpha_e^2 +2(2-\sqrt{3})  \tilde\beta_e^2  +(1-\sqrt{3})  \tilde\gamma_e^2] \, ,
\end{aligned}
\label{deltaES-III}
\end{align}    
where  $\epsilon_2=2\epsilon_1$ in Eq.\,(\ref{epS120}) 
is used in the last approximate equalities.
%%%%%%%%%%%%%%%%%%%%%%%%%%%%%%%%%%%%%%%%%
The mixing angle of 1st- and 2nd-family as:
\begin{align}
\tan 2\theta_{12}^e =
\ \frac{2|\delta_{e2}|}{\frac{3}{2}
	[\tilde\alpha_e^2 + 2(2-\sqrt{3})\tilde \beta_e^2+(7-4\sqrt{3})\tilde \gamma_e^2]}\simeq \frac{4}{3\sqrt{2}} \,\frac{9-5\sqrt{3}}{7-4\sqrt{3}}
|\epsilon_1^*| \simeq \frac{4}{3\sqrt{2}} (3+\sqrt{3})|\epsilon_1^*|\simeq 4.46\, |\epsilon_1^*|\,, 
\label{limit-theta12EIII}
\end{align}
where the denominator comes from the $(2,2)$ element of Eq.\,(\ref{chargedlepton-III}).
In the last approximate equality, 
we take $ \tilde\gamma_e\gg \tilde\alpha_e, \tilde\beta_e$,
which is the case  in the numerical fits of section 7.
We estimate $\theta_{12}^e$ to be $0.22$ at $|\epsilon_1|=|2.05\,i\,\epsilon|=0.1$.
This magnitude of $\theta_{12}^e$ leads to $\theta_{13}\simeq 0.15$ of the PMNS matrix by cooperating with the neutrino mixing angle  $\theta_{23}^\nu$ 
in Eq.\,(\ref{theta23}).
The mixing angle between 1st- and 3rd-family $\theta_{13}^e$ is found to be much smaller than  $\theta_{12}^e$ in the similar calculation.

In conclusion, 
the charged lepton mass matrix I combined with the neutrino mass matrix 
of Eq.\,(\ref{neutrinomassmatrix}) is expected to be consistent with the observed  three PMNS mixing angles at nearby   $\tau=i$.
Indeed, this case works well for both NH and IH as seen in
numerical results of section 7.
The output of the Dirac CP violating phase and the sum of neutrino masses
will tested in the future experiments.

%%%%%%%%%%%%%%%%%%%%%%%%%%%%%%%%%%%%%%%%%%%%%%%%%%%%%%%%%%%%%%%
%%%%%%%%%%%%%%%%%%%   Charged lepton II  %%%%%%%%%%%%%%%%%%%%%%
%%%%%%%%%%%%%%%%%%%%%%%%%%%%%%%%%%%%%%%%%%%%%%%%%%%%%%%%%%%%%%%

\subsubsection{Charged lepton mass matrix I\hspace{-.01em}I at $\tau=i$}

We  discuss  another  charged lepton mass matrix I\hspace{-.01em}I at $\tau=i$,
which is\,:
\begin{align}
&\begin{aligned}
& M_E=v_d
\begin{pmatrix}
\alpha_e & 0 & 0 \\
0 &\beta_e & 0\\
0 & 0 &\gamma_e
\end{pmatrix} 
\begin{pmatrix}
Y_1^{(6)}+g_e Y_1^{'(6)} & Y_3^{(6)}+g_e Y_3^{'(6)}& Y_2^{(6)}+g_e Y_2^{'(6)} \\
Y_2^{(4)} & Y_1^{(4)} &  Y_3^{(4)} \\
Y_3^{(2)} &  Y_2^{(2)}&  Y_1^{(2)}
\end{pmatrix}
\end{aligned}
\nonumber\\
\nonumber\\
= v_q
&\begin{aligned}
\begin{pmatrix}
\tilde \alpha_e & 0 & 0 \\
0 &\tilde \beta_e & 0\\
0 & 0 &\tilde\gamma_e
\end{pmatrix} 
\begin{pmatrix}
2\sqrt{3}-3+g_e (7\sqrt{3}-12) & 12-7\sqrt{3}+g_e (9-5\sqrt{3})& 5\sqrt{3}-9+g_e (3-2\sqrt{3}) \\
1 &1 &  1\\
-2+\sqrt{3} & 1-\sqrt{3}& 1
\end{pmatrix},
\end{aligned}
\label{charged-leptonIIi}
\end{align}
where $\tilde \alpha_e=3 v_d^2Y_0^3 \alpha_e$,
$\tilde \beta_e=(6-3\sqrt{3}) v_d^2Y_0^2 \beta_e$ and 
$\tilde \gamma_e =v_d^2Y_0 \gamma_e$.

We move to  the diagonal base of $S$.
The mass matrix $M_E^\dagger M_E$ is transformed by the unitary transformation
$V_{S2}$ as:
\begin{align}
{\cal M}_E^{2(0)}\equiv V_{S2} M_E^\dagger M_E V_{S2}^\dagger = \frac{3}{2}
&\begin{aligned}
\begin{pmatrix}
2\tilde\beta^2_e & 0& 0 \\
0 & A\tilde\gamma_e^2+3(A+ B_{1e}+ |g_e|^2 C) \tilde\alpha_e^2&  
-D\tilde\gamma_e^2 -3 (B_{2e}+Ag_e+ Cg_e^*) \tilde\alpha_e^2) \\
0 &  - D\tilde\gamma_e^2 -3 (B_{2e}+Ag_e^*+ Cg_e) \tilde\alpha_e^2)  &
\tilde\gamma_e^2 +3(C+ B_{1e}+ |g_e|^2 A) \tilde\alpha_e^2
\end{pmatrix}
\end{aligned},
\label{Chargedlepton-I}
\end{align}
where
\begin{align}
&\begin{aligned}
A=7-4\sqrt{3}\, , \quad B=26-15 \sqrt{3} \, , \quad
C=97-56\sqrt{3} \, , \quad
D=2- \sqrt{3} \, , 
\end{aligned}\nonumber\\
&\begin{aligned}
B_{1e}=B (g_e+g_e^*)=2 B \,  {\rm Re}[g_e] , \qquad  B_{2e}=B \,(1+|g_e|^2) \, ,
\quad A^2=C\,, \quad D^2=A\,, \quad A+C=4B\,.
\end{aligned}
\label{ABCD}
\end{align}
The flavor mixing between the 2nd- and 3rd-family
appears at the $\tau=i$ as well as the  charged lepton mass matrix I.

The mass eigenvalues satisfy
\begin{align}
&  m_{e1}^2=3 \tilde\beta_e^2 \, , \qquad\qquad    m_{e2}^2 m_{e3}^2=81(97-56\sqrt{3}) 
\tilde\alpha_e^2 \tilde\gamma_e^2 \,, \nonumber\\
&m_{e2}^2+ m_{e3}^2=6(2-\sqrt{3})  \tilde\gamma_e^2+
3(78-45\sqrt{3}) (2+2{\rm Re}[g_e]+|g_e|^2) \tilde\alpha_e^2  \,.
\label{mass2E}
\end{align} 
The imaginary part of the matrix in
Eq.\,(\ref{Chargedlepton-I}) is factored out by using a phase matrix $P_e$ as:
\begin{align}
\frac{3}{2}\  P_e
\begin{pmatrix}
2\tilde\beta^2_e & 0& 0 \\
0 & A\tilde\gamma_e^2+3(A+ B_{1e}+ |g_e|^2 C) \tilde\alpha_e^2&  
-\sqrt{[D\tilde\gamma_e^2 +3 (B_{2e}+E_e)\tilde\alpha_e^2)]^2 + F_e^2 \tilde\alpha_e^4}\\
0 &  -\sqrt{[D\tilde\gamma_e^2 +3 (B_{2e}+E_e)\tilde\alpha_e^2)]^2 + F_e^2 \tilde\alpha_e^4} &
\tilde\gamma_e^2 +3(C+ B_{1e}+ |g_e|^2 A) \tilde\alpha_e^2
\end{pmatrix} P_e^* \ ,
\end{align}
where 
\begin{align}
E_e=(A+C){\rm Re}[g_e] \ , \qquad\qquad F_e=(A-C){\rm Im}[g_e] \ ,
\end{align}
and 
\begin{align}
P_e=
\begin{pmatrix}
1  & 0 & 0\\
0 & 1 & 0 \\
0 & 0 &\ \ \  e^{-i\phi^e} \ \ 
\end{pmatrix} ,
\label{Pe}
\end{align}
with 
\begin{align}
\tan \phi^e=\frac{F_e \tilde\alpha_e^2}{D\tilde\gamma_e^2 +3 (B_4+E_e)\tilde\alpha_e^2} \, .
\label{phie}
\end{align}
The mixing angle $\theta_{23}^e$ is given as:
\begin{align}
\tan 2\theta_{23}^e =
\frac{-\sqrt{[D\tilde\gamma_e^2 +3 (B_{2e}+E_e)\tilde\alpha_e^2)]^2 + F_e^2 \tilde\alpha_e^4}}{(2\sqrt{3}-3)\tilde\gamma_e^2
	+3(45-26\sqrt{3})(1-|g_e|^2)\tilde\alpha_e^2} \ . 
\label{theta23e}
\end{align}
Neglecting the imaginary part of $g_e$ ($g_e={\rm Re}[g_e]$), it is given simply
as:
\begin{align}
\tan 2\theta_{23}^e =
-\frac{1}{\sqrt{3}}\ \frac{\tilde\gamma_e^2+3(7-4\sqrt{3})(1+4g_e+g_e^2)\tilde\alpha_e^2}{\tilde\gamma_e^2-3(7-4\sqrt{3})(1-g_e^2)\tilde\alpha_e^2} \ .
\label{appro-theta23e}
\end{align}
%where $|g_e|$ is supposed to be ${\cal O}(1)$.
We take $\tilde\beta_e^2 \ll \tilde\alpha_e^2, \tilde\gamma_e^2$ due to the
mass hierarchy of the charged lepton masses.
There are two possible choices of  $\tilde\alpha_e^2 \ll \tilde\gamma_e^2$
and  $\tilde\gamma_e^2 \ll \tilde\alpha_e^2$.

In the case of  $\tilde\alpha_e^2 \ll \tilde\gamma_e^2$,
\begin{align}
\tan 2\theta_{23}^e \simeq 
-\frac{1}{\sqrt{3}}\ [1+6(7-4\sqrt{3})(1+2g_e)\frac{\tilde\alpha_e^2}{\tilde\gamma_e^2}] \ . 
\label{approtheta23e}
\end{align}
At the limit of $\tilde\alpha_e^2/\tilde\gamma_e^2=0$,
we obtain $\theta_{23}^e=-15^\circ$.
%%%%%%%%%%%%%%%%%%%%%%%%%%%%%%%%%%%%%%%%%%%%%%%%%%%%%%%%%%%%%
%The charged lepton masses are:  
%\begin{align}  
%m_{e2}^2\simeq \frac{81(97-56\sqrt{3})}{6(2-\sqrt{3})} 
%\tilde\alpha_e^2=\frac{27}{2} (26-15\sqrt{3}) \simeq 0.26 \,
%  \tilde\alpha_e^2 \, , \qquad 
%m_{e3}^2\simeq 6(2-\sqrt{3})  \tilde\gamma_e^2 \simeq 
% 1.61 \tilde\gamma_e^2 \, .
%\end{align} 
%\begin{align}  
%\frac{m_{e2}^2}{m_{e3}^2}\simeq 0.161\, 
%\frac{ \tilde\alpha_q^2}{\tilde\gamma_q^2} \, .
%\end{align} 
%Putting observed Yukawa couplings at GUT scale,
%\begin{align}  
%& y_e=2.07526\times 10^{-6} \, , \qquad  y_\mu=4.381\times 10^{-4}\,  , \qquad 
%y_\tau=7.480\times 10^{-3}\,  ,
%\end{align} 
%we get 
%\begin{align}
%\frac{\tilde\alpha_e^2}{\tilde\gamma_e^2}\simeq 0.021  \, .
%\label{algaratio}
%\end{align}
%Putting  this value of $\tilde\alpha_e^2/\tilde\gamma_e^2$
%with  $g_e=3 \, (1)$,  we get $-15.8 \, (-15.3)^\circ$.
%%%%%%%%%%%%%%%%%%%%%%%%%%%%%%%%%%%%%%%%%%%%%%%%%

% we   obtain $\tan 2\theta_{23}^e=-1/\sqrt{3}$ in Eq.(\ref{theta23e}), that is $\theta_{23}^e=-15^\circ$.
On the other hand, in the case of   $\alpha_e^2\gg \gamma_e^2  $,
Eq.\,(\ref{appro-theta23e}) turns to
\begin{align}
\tan 2\theta_{23}^e \simeq 
\frac{1}{\sqrt{3}}\ \frac{1+4g_e+g_e^2}{1-g_e^2}
%=\frac{1}{\sqrt{3}}\
% \left [1-2g_e \left (\frac{1}{g_e-1}+\frac{1}{g_e^2-1}\right )\right ] 
\, ,
\label{Chargedleptonmixing-I}
\end{align}
which gives $|\theta_{23}^e|=0$--$45^\circ$ by choosing
relevant $g_e$.
Thus, the large  $\theta_{23}^e$ is obtained easily.

%%%%%%%%%%%%%%%%%%%%%%%%%%%%%%%%%%%%%%%%%%%%%%%%%%%%%%%%%%%%%%%
%%%%%%%%%%%%%%%%%%%%%%%%%%%%%%%%%%%%%%%%%%%%%%%%%%%%%%%%%%
\subsubsection{Charged lepton mass matrix I\hspace{-.01em}I at nearby  $\tau=i$}  

The mass matrix of the charged lepton 
in Eq.\,(\ref{charged-leptonIIi}), $M_E$ 
is corrected due to the deviation from the fixed point of $\tau=i$.
In the 1st order approximation of $\epsilon$,
the correction  ${\cal M}_E^{2(1)}$ to ${\cal M}_E^{2(0)}$ of Eq.\,(\ref{Chargedlepton-I}) is  given by the following matrix:
\begin{align}
\begin{aligned} {\cal M}_E^{2(1)}=
\begin{pmatrix}
\delta_{e1} & \delta_{e2} & \delta_{e3} \\ 
\delta_{e2}^* & \delta_{e4} & \delta_{e5} \\ 
\delta_{e3}^* & \delta_{e5}^* & \delta_{e6} 
\end{pmatrix}
\end{aligned}\,,
\label{masse2-III1st}
\end{align}
where $\delta_{ei}$ are given in terms of
$\epsilon$, $g_e$, $\tilde \alpha_e^2$, $\tilde \beta_e^2$ and 
$\tilde \gamma_e^2$.
By the 1st order perturbation of $\epsilon$, we can obtain the mixing angle $\theta_{12}^e$, which vanishes in the 0th order of perturbation.
In order to estimate the flavor mixing angles, we present relevant  $\delta_{ei}$ as:

%%%%%%%%%%%%%%%%%%%%%%%%%%%%%%%
\begin{align}
&\begin{aligned}
\delta_{e 2}&=\frac{3}{\sqrt{2}}  \tilde\alpha_e^2 (g_e^*-1)
\{\,[\,(11\sqrt{3}-19)+(41\sqrt{3}-71)g_e\,]\epsilon_1^*
-[\,(15\sqrt{3}-26)+(56\sqrt{3}-97)g_e\,]\epsilon_2^* \,\} \nonumber\\
&+ \frac{1}{\sqrt{2}}  \tilde\gamma_e^2 [\,(3\sqrt{3}-5)\epsilon_1^*
+(7-4\sqrt{3})\epsilon_2^*\,] \simeq  (0.193+0.052 g_e)\tilde\alpha_e^2 (g_e^*-1)\epsilon_1^*+0.240 \tilde\gamma_e^2 \epsilon_1^*\,,
\end{aligned}\\
&  \begin{aligned}
\delta_{e 3}&=\frac{1}{\sqrt{6}}  \tilde\alpha_e^2 (g_e^*-1)
\{\,[\,3(71\sqrt{3}-123)+3(19\sqrt{3}-33)g_e\,]\epsilon_1^*
-[\,3(97\sqrt{3}-168)+(26\sqrt{3}-45)g_e\,]\epsilon_2^* \,\} \\
&+ \frac{1}{\sqrt{2}}  \tilde\gamma_e^2 [\,(1-\sqrt{3})\epsilon_1^*
+(\sqrt{3}-2)\epsilon_2^*\,]\simeq 
-(0.052+0.138 g_e)\tilde\alpha_e^2 (g_e^*-1)\epsilon_1^*-0.897 \tilde\gamma_e^2 \epsilon_1^* \,,
\end{aligned}
\label{deltaeS}
\end{align}    
where  ${\cal O} (\tilde\beta_e^2) $ is neglected  and 
 $\epsilon_2=2\epsilon_1$  of Eq.\,(\ref{epS120}) is taken in last approximate equalities. 
%%%%%%%%%%%%%%%%%%%%%%%%%%%%%%%%%%%%%%%%%%%%%%%%%%

Let us  discuss the mixing angles of $\theta_{12}^e$ and  $\theta_{13}^e$ of the charged lepton flavors, which vanish in the leading terms of the mass matrix.
As seen in Eq.\,(\ref{deltaeS}),
both $\delta_{e2}$ and  $\delta_{e3}$ are  of  
${\cal O}(\tilde\alpha_e^2,\tilde\gamma_e^2)\times\epsilon_1$ for
$g_e={\cal O}(1)$. 
Suppose $\tilde\gamma_e^2\ll \tilde\alpha_e^2$
to realize the hierarchy of charged lepton masses in Eq.\,(\ref{mass2E})
%%%%%%%%%%%%%%%%%%%%%%%%%%%%%%%%%%%%%
\footnote{Indeed,  a successful numerical result  is obtained for 
	$\tilde\gamma_e^2\ll \tilde\alpha_e^2$ in section 7.} .
%%%%%%%%%%%%%%%%%%%%%%%%%%%%%%%%%%%%%
Then, we have  mass eigenvalues from Eq.\,(\ref{mass2E}) as:
%   m_{e2}^2 m_{e3}^2=81(97-56\sqrt{3}) 
%\tilde\alpha_e^2 \tilde\gamma_e^2 , \ \ 
%m_{e2}^2+ m_{e3}^2=6(2-\sqrt{3})  \tilde\gamma_e^2+
%3(78-45\sqrt{3}) (2+2{\rm Re}[g_e]+|g_e|^2) \tilde\alpha_e^2  . \nonumber
\begin{align}
&m_{e1}^2=3 \tilde\beta_e^2 \, , \quad   
m_{e2}^2\simeq \frac{9(2-\sqrt{3})}{2+2{\rm Re}[g_e]+|g_e|^2} 
\tilde\gamma_e^2\, , \quad
&m_{e3}^2\simeq 3(78-45\sqrt{3})(2+2{\rm Re}[g_e]+|g_e|^2)\tilde\alpha_e^2 \, , 
\label{emass2}
\end{align} 
which lead to 
\begin{align}
\frac{m_{e2}^2}{m_{e3}^2}\simeq \frac{7+4\sqrt{3}}{(2+2{\rm Re}[g_e]+|g_e|^2)^2} 
\frac{\tilde\gamma_e^2}{\tilde\alpha_e^2}\, .
\label{emassratio}
\end{align} 
The mixing angles between 1st- and 2nd-family $\theta_{12}^e$
and  between 1st- and 3rd-family $\theta_{13}^e$
are given approximately as:
\begin{align}
\theta_{12}^e \simeq \left |\frac{\delta_{e 2}}{m_{e2}^2}\right | \,,
\qquad\qquad 
\theta_{13}^e  \simeq \left |\frac{\delta_{e 3}}{m_{e3}^2} \right | \,,
\label{mixing12-13}
\end{align} 
where
\begin{align}
\delta_{e 2}\simeq  (0.193+0.052 g_e) \tilde\alpha_e^2 (g_e^*-1)\epsilon_1^*\,,
\qquad
\delta_{e 3}\simeq -(0.052+0.138 g_e)\tilde\alpha_e^2 (g_e^*-1)\epsilon_1^* \,,
\end{align} 
respectively.
Substituting  mass eigenvalues of Eq.\,(\ref{emass2}) into mixing angles in Eq.\,(\ref{mixing12-13}), we can estimate  magnitudes of $\theta_{12}^e$
and $\theta_{13}^e$.
The mixing angle of $\theta_{12}^e$ is given as:
\begin{align}
\theta_{12}^e &\simeq 
\left | \frac{(0.193+0.052 g_e)  (g_e^*-1)}
{9(2-\sqrt{3})} (2+2{\rm Re}[g_e]+|g_e|^2)
\frac{\tilde\alpha_e^2}{\tilde\gamma_e^2} \epsilon_1^*\right | \nonumber\\
&= \left | \frac{(0.193+0.052 g_e)  (g_e^*-1) (26+15\sqrt{3})}
{9\,(2+2{\rm Re}[g_e]+|g_e|^2)}
\frac{m_{e3}^2}{m_{e2}^2} \epsilon_1^*\right |
\simeq 1.7\,
\frac{\left | (0.193+0.052 g_e)  (g_e^*-1)\right |}{2+2{\rm Re}[g_e]+|g_e|^2} 
\left |\epsilon_1^*\right |\times 10^3\,,
\label{12E}
\end{align} 
where the mass ration of  Eq.\,(\ref{emassratio}) is used to 
remove the ratio $\tilde\gamma_e^2/\tilde\alpha_e^2$.
In the last equality,  observed masses of the tauon and the muon are input.
Suppose the magnitude of  $|\epsilon_1^*|$ to be $0.02$ as a typical value.
As seen in Eq.\,(\ref{12E}),  $\theta_{12}^e$ depends on  $g_e$.
Indeed, $\theta_{12}^e$ vanishes at $g_e=1$ or $-3.62$ while
it is  of order one if  $|g_e|\ll 1$ or $|g_e|\gg 1$.
%Indeed, ${\rm Re}[g_e]\simeq 1$ is required to reproduce the observed mixing %angles  in the numerical discussion in section 7.
On the other hand,  $\theta_{13}^e$ is suppressed
due to the factor of $1/m_{e3}^2$ as seen Eq.\,(\ref{mixing12-13}).
%%%%%%%%%%%%%%%%%%%%%%%%%%%%%%%%%%%%%%%%%%%%%%%%%%%%%%%%%%%%%%%%%%%%%%

In conclusion, 
the charged lepton mass matrix I\hspace{-.01em}I combined with the neutrino mass matrix 
of Eq.\,(\ref{neutrinomassmatrix}) is expected to be consistent with the observed  three PMNS mixing angles at nearby   $\tau=i$ 
as well as   charged lepton mass matrix I.
Indeed, this case works well for  NH,
but  it leads to the sum of neutrino masses larger than $120$\,meV for IH
 as seen in numerical results of section 7.

\subsection{Lepton mass matrix at  $\tau=\omega$}

\subsubsection{Neutrino mass matrix at $\tau=\omega$}

Let us consider the neutrino mass matrix at $\tau=\omega$,
where  there exists the residual symmetry of the $\rm A_4$
group $\mathbb{Z}_3^{ST}=\{ I, ST,(ST)^2 \}$.
By putting the modular forms in Table 1 into Eq.\,(\ref{neutrinomassmatrix}),
the neutrino mass matrix is written as:
\begin{align}
M_\nu=\frac{v_u^2}{\Lambda}Y_0^2\left [
\frac{3}{2}\begin{pmatrix}
2 & -\omega^2 & \frac{1}{2}\omega\\
-\omega^2 & -\omega & -1 \\
\frac{1}{2}\omega & -1 & 2 \omega^2
\end{pmatrix}
+\frac{9}{4}\omega g_{\nu 2} 
\begin{pmatrix}
0 & 0 &1\\ 0 & 1 & 0 \\ 1 & 0 & 0
\end{pmatrix}
\right ] \ ,
\label{NeutrinotauST}
\end{align}
%%%%%%%%
where the $g_{\nu 1}$ term of Eq.\,(\ref{neutrinomassmatrix}) disappears 
because of ${\bf Y^{(\rm 4)}_1}=0$ at $\tau=\omega$.
We move to the  diagonal base of $ST$.
By using the unitary transformation of Eq.\,(\ref{STdiagonal}), $V_{ST4}$
or  $V_{ST5}$,
the neutrino mass matrix is transformed  as:
\begin{align}
\begin{aligned}
{\cal M}_\nu^{2(0)}\equiv V_{ST4(5)}\hat M_\nu^\dagger \hat M_\nu V_{ST4(5)}^\dagger= \left (\frac{9}{4}\frac{v_u^2}{\Lambda} Y_0^2 \right )^2
\end{aligned}
\begin{aligned}
\begin{pmatrix}
|2 +  g_{\nu 2}|^2  & 0 & 0\\
0 & |1-  g_{\nu 2}|^2  & 0\\
0& 0& |1-  g_{\nu 2}|^2
\end{pmatrix} .
\end{aligned}
\label{Neutrinonew2ST}
\end{align}
The neutrino mass matrix is diagonal and two neutrinos are degenerated at $\tau=\omega$.
Three neutrino masses are  degenerate if  $g_{\nu 2}=-0.5$.
Then, large flavor mixing angles are possibly reproduced  if small off diagonal elements are generated by the deviation from $\tau=\omega$.

\subsubsection{Neutrino mass matrix at nearby  $\tau=\omega$}

%%%%%%%%%%%%%%%%%%%%%%%%%%%%%%%%%%%%%%%%%%%%%%%%%%%%%%%
%%%%%%%%%%%%%%%%%%%% Neutrino tau=\omega%%%%%%%%%%%%%%%
%%%%%%%%%%%%%%%%%%%%%%%%%%%%%%%%%%%%%%%%%%%%%%%%%%%%%%%
Neutrino mass matrix
in Eq.\,(\ref{neutrinomassmatrix}),  $M_\nu$ 
is corrected due to the deviation from the fixed point of $\tau=\omega$.
After putting modular forms of  Eq.\,(\ref{epST120})  and moving to the  diagonal base of $ST$ by $V_{ST4}$, the corrections to Eq.\,(\ref{Neutrinonew2ST}) are given by only a small variable $\epsilon$ of in Eq.\,(\ref{epST120}).
In the 1st order approximation of $\epsilon$,
the correction  ${\cal M}_\nu^{2(1)}$ to ${\cal M}_\nu^{2(0)}$ of Eq.\,(\ref{Neutrinonew2ST}) is given by the following matrix:
\begin{align}
\begin{aligned}\quad  
{\cal M}_\nu^{2(1)}=\left (\frac{9}{4}\frac{v_u^2}{\Lambda} Y_0^2\right )^2
\begin{pmatrix}
\delta_{\nu 1} & \delta_{\nu 2} & \delta_{\nu 3} \\ 
\delta_{\nu 2}^* & \delta_{\nu 4} & \delta_{\nu 5} \\ 
\delta_{\nu 3}^* & \delta_{\nu 5}^* & \delta_{\nu 6} 
\end{pmatrix} , \qquad
\end{aligned}
\label{correctionneutrino-tauL}
\end{align}
where  $\delta_{\nu i}$ are given
in terms of $\epsilon$, $g_{\nu 1}$ and  $g_{\nu 2}$.
By the 1st order perturbation of $\epsilon$, we can obtain the mixing angle  $\theta_{12}^\nu$, which vanishes in the 0th order of perturbation.
In order to estimate the flavor mixing angles, we present off diagonal elements,   $\delta_{\nu 2}$, $\delta_{\nu 3}$
and $\delta_{\nu 5}$ as:
%%%%%%%%%%%%%%%%%%%%%%%%%%%%
%%%%%%%%%%%%%%%%%%%%%%%%%%%%
\begin{align}
&  \begin{aligned}
\delta_{\nu 2}&= 
\frac{3}{4} (2+g_{\nu 2}^*)\epsilon_1+
\frac{3}{8}  (1+6g_{\nu 1}^*)(1-g_{\nu 2})\epsilon_1^*
- \frac{21}{8} (2+g_{\nu 2}^*)\epsilon_2
-\frac{3}{4} (4-3g_{\nu 1}^*) (1-g_{\nu 2})\epsilon_2 ^* \\
& \simeq 
-\frac{9}{2}(2+g_{\nu 2}^*)\epsilon_1
-\frac{9}{8}  (5-6g_{\nu 1}^*)(1-g_{\nu 2})\epsilon_1^* \, ,
\end{aligned} \nonumber\\
&   \begin{aligned}
\delta_{\nu 3}&= \frac{3}{8} (1+6g_{\nu 1}) (2+g_{\nu 2}^*)\epsilon_1+
\frac{3}{4}(1-g_{\nu 2})\epsilon_1^*
- \frac{3}{4}(4-3g_{\nu 1})(2+g_{\nu 2}^*)\epsilon_2
-\frac{21}{8}(1-g_{\nu 2})\epsilon_2 ^* \\
&\simeq -\frac{9}{8}(5-6 g_{\nu 1}) (2+g_{\nu 2}^*)\epsilon_1-
\frac{9}{2} (1-g_{\nu 2})\epsilon_1^* \, ,
\end{aligned} \nonumber\\
&\begin{aligned}
\delta_{\nu 5}&=\frac{3}{4} (1-3g^*_{\nu 1}) (1-g_{\nu 2})\epsilon_1^*
-\frac{3}{2}(1-g^*_{\nu 2})\epsilon_1
- \frac{3}{4}(8+3g^*_{\nu 1})(1-g_{\nu 2})\epsilon_2^*
+\frac{21}{4}(1-g^*_{\nu 2})\epsilon_2  \\
&\simeq -\frac{9}{4}(5+3 g^*_{\nu 1}) (1-g_{\nu 2})\epsilon_1^*+
9 (1-g^*_{\nu 2})\epsilon_1 \, ,
\end{aligned}
\label{deltanuST}
\end{align} 
where $\epsilon_1=2.1i\,\epsilon$, and $\epsilon_2=2\epsilon_1$ of Eq.\,(\ref{epST120})
is used for last approximate equalities. 
If we move to the  diagonal base of $ST$ by using $V_{ST5}$ instead of  $V_{ST4}$, 
we obtain the corrections by exchanging the above results as:
\begin{align} 
\delta_{\nu 2} \leftrightarrow  \delta_{\nu 3}\,, \qquad 
\delta_{\nu 5} \leftrightarrow  \delta_{\nu 5}^*\,.
\end{align} 
Indeed, we  move to the  diagonal base of $ST$ by using $V_{ST5}$
for the charged lepton mass matrix I\hspace{-.01em}I in 
section 5.3.5.

It is noticed  that the off-diagonal elements are enhanced 
by large coefficients in front of $\epsilon_1$ and $\epsilon_1^*$. For example,  $|\delta_{\nu 5}|$ 
could be comparable to diagonal element if $|\epsilon_1|=0.1$ is taken.
Since the 2nd and 3rd eigenvalues are  degenerated as seen in Eq.\,(\ref{Neutrinonew2ST}), the large (2--3) mixing angle is easily
obtained due to those corrections. The large (1--2) mixing angle is also possible by choosing
relevant $g_{\nu 1}$ and $g_{\nu 2}$.
The  (1--3) mixing angle is   relatively small due to the fixed mass square difference $\Delta m_{31}^2$.
%which should be  canceled out by the mixing  the charged lepton sector.
On the other hand, the sum of neutrino masses may increase if mass eigenvalues become quasi-degenerate.
 Then, its cosmological  upper-bound provides a crucial test
for the lepton mass matrices. 
Therefore, we  should examine the contribution from the charged lepton sector
carefully for both NH and IH of neutrinos to judge
it  working  well or not.
Indeed,  we will see in section 7 that 
the model of the charged lepton mass matrix I is excluded by the sum of neutrino masses while the model with the charged lepton mass matrix I\hspace{-.01em}I
is consistent with it for both NH and IH of neutrino masses.

%%%%%%%%%%%%%%%%%%%%%%%%%%%%%%%%%%%%%%%%%%%%%%%%%%%%%%%%%%%%%%%%%%%
%%%%%%%%%%%%%%%%%%%%%%%%%%%%%  I %%%%%%%%%%%%%%%%%%%%%%%%%%%%%%%%%%
%%%%%%%%%%%%%%%%%%%%%%%%%%%%%%%%%%%%%%%%%%%%%%%%%%%%%%%%%%%%%%%%%%%
\subsubsection{Charged lepton mass matrix I  at  $\tau=\omega$}

We discusses the charged lepton mass matrix I at the fixed point  $\tau=\omega$
by using modular forms in Table 1.
In the base of $S$ and $T$ of Eq.\,(\ref{STbase}),
the charged lepton mass matrix I in Eq.\,(\ref{ME222}) 
is given as:
%%%%%%%%%%%%%%%%%%%%%%%%%%%%%%%
\begin{align}
M_E=
\begin{pmatrix}
\tilde \alpha_e & 0 & 0 \\
0 &\tilde \beta_e & 0\\
0 & 0 &\tilde\gamma_e
\end{pmatrix} 
\begin{pmatrix}
1& -\frac{1}{2}\omega^2 & \omega \\
\omega& 1& -\frac{1}{2}\omega^2  \\
-\frac{1}{2}\omega^2 &\omega& 1 \\
\end{pmatrix}\, ,
\label{chargedleptonST}
\end{align}
where
$\tilde \alpha_e=  v_d Y_0 \alpha_e$,
$\tilde \beta_e=   v_d Y_0  \beta_e$ and
$\tilde \gamma_e = v_d Y_0 \gamma_e$.
%%%%%%%%%%%%%%%%%%%%%%%%%%%%%%%%%%%%%%%%%%%%%%%%%%%%%%%%%%%%%%%%%%%%%%%%%
By using the unitary transformation of Eq.\,(\ref{STdiagonal}), $V_{ST4}$, like the case of the neutrino mass matrix,
 $M_E^\dagger M_E$ is transformed  as:
%%%%%%%%%%%%%%%%%%%%%%%%%%%%
%%%%%%%%%%%%%%%%%%%%%%%%%%%%
\begin{align}
{\cal M}_E^{2(0)}\equiv  V_{ST4} M_E^\dagger M_E V_{ST4}^\dagger 
= \frac{9}{4}\,
\begin{aligned}
\begin{pmatrix}
\tilde\alpha_e^2&0 & 0 \\
0&\tilde\gamma_e^2  &   0  \\
0 & 0 & \tilde\beta_e^2 \\
\end{pmatrix}
\end{aligned}\,.
\label{chargedleptonST-III}
\end{align}
It is remarked that it is diagonal one as well as the neutrino mass matrix
 in Eq.\,(\ref{Neutrinonew2ST}).

%%%%%%%%%%%%%%%%%%%%%%%%%%%%%%%%%%%%%%%%%%%%%%%%%%%%%%%%%%%%%%%%%%%
\subsubsection{Charged lepton mass matrix I at nearby $\tau=\omega$}

%%%%%%%%%%%%%%%%%%%%%%%%%%%%
%%%%%%%%%%%%%%%%%%%%%%%%%%%%%%%%%%%%%%%%%%%%%%%%%%%%%%%
The charged lepton  mass matrix
in Eq.\,(\ref{chargedleptonST}),  $M_E$ 
is corrected due to the deviation from the fixed point of $\tau=\omega$.
After putting modular forms of  Eq.\,(\ref{epST120})  and moving to the  diagonal base of $ST$ by $V_{ST4}$,
the correction ${\cal M}_\nu^{2(1)}$ to ${\cal M}_\nu^{2(0)}$ of Eq.\,(\ref{chargedleptonST-III}) is given
in the 1st order approximation of $\epsilon$ as:
\begin{align}
\begin{aligned} 
{\cal M}_E^{2(1)}=
\begin{pmatrix}
\delta_{e1} & \delta_{e2} & \delta_{e3} \\  
\delta_{e2}^* & \delta_{e4} & \delta_{e5} \\ 
\delta_{e3}^* & \delta_{e5}^* & \delta_{e6} 
\end{pmatrix}\, , 
\end{aligned}
\end{align}
where 
\begin{align}
&\begin{aligned}
\delta_{e2}=   i\,\tilde\alpha_e^2 (\epsilon_1-\frac{1}{2}\epsilon_2)
+\frac{1}{2} i\, \tilde\gamma_e^2(\epsilon_1^*+\epsilon_2^*)
=  \frac{3}{2}\, i\,\tilde\gamma_e^2 \, \epsilon_1^* \, ,
\end{aligned}\\
&\begin{aligned}
\delta_{e3}=  \frac{1}{2}\, i\,\tilde\alpha_e^2 (\epsilon_1+\epsilon_2)
+ i\, \tilde\beta_e^2(\epsilon_1^*-\frac{1}{2}\epsilon_2^*)
=  \frac{3}{2}\, i\,\tilde\alpha_e^2 \, \epsilon_1 \, ,
\end{aligned}\\
&\begin{aligned}
\delta_{e5}=  -i\, \tilde\gamma_e^2(\epsilon_1-\frac{1}{2}\epsilon_2)
-\frac{1}{2}\, i\,\tilde\beta_e^2 (\epsilon_1^*+\epsilon_2^*)
=  -\frac{3}{2}\, i\,\tilde\beta_e^2 \, \epsilon_1^* \, ,
\end{aligned}
\end{align}
where $\epsilon_2=2\epsilon_1$ of Eq.\,(\ref{epST120})
is used for last  equalities. 
Due to $\tilde\beta_e^2\gg \tilde\gamma_e^2\gg\tilde\alpha_e^2$,
 mixing angles $\theta_{ij}^e$ are easily obtained 
 by using $\epsilon_1=2.1 i\epsilon$ as follows:
\begin{align}
\theta_{12}^e\simeq \theta_{23}^e\simeq  \frac{2}{3}\, |\epsilon_1|\simeq \frac{4.2}{3}|\epsilon|\, ,
\end{align}
which are smaller than $0.1$,
moreover, $\theta_{13}^e$ is highly suppressed
due to the factor $\tilde\alpha_e^2/\tilde\beta_e^2$.
Thus, the flavor mixing angles of the charged lepton are very small
 at nearby the fixed point $\tau=\omega$.
 The PMNS mixing angles come from mainly the neutrino sector
 in this case.
 Therefore,  the increase of the sum of neutrino masses is unavoidable
 since  mass eigenvalues become quasi-degenerate in order to reproduce
 large mixing angles.
%%%%%%%%%%%%%%%%%%%%%%%%%%%%%%%%%%%%%%%%%%%%%%%
%%%%%%%%%%%%%%%%%%%%%%%%% II %%%%%%%%%%%%%%%%%%
%%%%%%%%%%%%%%%%%%%%%%%%%%%%%%%%%%%%%%%%%%%%%%%
\subsubsection{Charged lepton mass matrix I\hspace{-.01em}I at  $\tau=\omega$}

We discusses the charged lepton mass matrix I\hspace{-.01em}I at the fixed point  $\tau=\omega$
by using modular forms in Table 1.
The charged lepton mass matrix  I\hspace{-.01em}I
in Eq.\,(\ref{ME642})  is given as:
%%%%%%%%%%%%%%%%%%%%%%%%%%%%%%%
\begin{align}
\begin{aligned}
M_E=
\begin{pmatrix}
\tilde \alpha_e & 0 & 0 \\
0 &\tilde \beta_e & 0\\
0 & 0 &\tilde\gamma_e
\end{pmatrix} 
\begin{pmatrix}
g_e& -2\omega^2 g_e & -2\omega g_e \\
-\frac{1}{2}\omega &1 &  \omega^2\\
-\frac{1}{2}\omega^2 & \omega& 1
\end{pmatrix}\, ,
\end{aligned}
\label{chargedleptonST2}
\end{align}
where $\tilde \alpha_e= (9/8) v_d  Y_0^3 \alpha_d$,
$\tilde \beta_e= (3/2) v_d Y_0^2 \beta_q$ and 
$\tilde \gamma_e =v_dY_0 \gamma_e$.
%%%%%%%%%%%%%%%%%%%%%%%%%%%%%%%%%%%%%%%%%%
By using the unitary transformation of Eq.\,(\ref{STdiagonal}), $V_{ST5}$,
which is different from the case of 
the charged lepton mass matrix I,  $M_E^\dagger M_E$ is transformed  as:
\begin{align}
{\cal M}_E^{2(0)}\equiv  V_{ST5} M_E^\dagger M_E V_{ST5}^\dagger 
= \frac{9}{4}\,
\begin{aligned}
\begin{pmatrix}
0&0 & 0 \\
0&0 &   0  \\
0 & 0 & 4\tilde\alpha_e^2 |g_e|^2+\tilde\beta_e^2+\tilde\gamma_e^2 \\
\end{pmatrix}
\end{aligned}\,,
\label{ChargedleptonST-I}
\end{align}
which gives two massless charged leptons.

\subsubsection{Charged lepton mass matrix  I\hspace{-.01em}I at nearby $\tau=\omega$}  
The charged lepton  mass matrix
in Eq.\,(\ref{chargedleptonST2}),  $M_E$ 
is corrected due to the deviation from the fixed point of $\tau=\omega$.
After putting modular forms of  Eq.\,(\ref{epST120}) and moving to the  diagonal base of $ST$ by $V_{ST5}$,
the correction  ${\cal M}_\nu^{2(1)}$ to ${\cal M}_\nu^{2(0)}$ of Eq.\,(\ref{ChargedleptonST-I}) is given
 as:
\begin{align}
\begin{aligned}   {\cal M}_E^{2(1)}=
\begin{pmatrix}
0 & 0 & \delta_{e3} \\ 
0 & 0 & \delta_{e5} \\ 
\delta_{e3}^* & \delta_{e5}^* & \delta_{e6} 
\end{pmatrix} 
\end{aligned}
\label{chargedST-I}
\end{align}
where $\delta_{ei}$ are given in terms of
$\epsilon$, $g_e$, $\tilde \alpha_e^2$, $\tilde \beta_e^2$ and 
$\tilde \gamma_e^2$.
By the 1st order perturbation of $\epsilon$, we can obtain the mixing angles $\theta_{23}^e$ and $\theta_{13}^e$, which vanish in the 0th order of perturbation.
In order to estimate the flavor mixing angles, we present  $\delta_{e3}$
and   $\delta_{e5}$ as:
%%%%%%%%%%%%%%%%%%%%%%%%%%%%%%%
\begin{align}
&  \begin{aligned}
\delta_{e 3}&= -2\tilde\alpha_e^2g_e(2+g_e^*) (\epsilon_1^*+\epsilon_2^*)
+ \frac{1}{6}\tilde\beta_e^2(\epsilon_1^*-8\epsilon_2^*) +\frac{1}{2} i\,\tilde\gamma_e^2(\epsilon_1^*+\epsilon_2^*) \\
&\simeq   [-6\tilde\alpha_e^2g_e(2+g_e^*) 
-\frac{5}{2}\tilde\beta_e^2 +\frac{3}{2} i\,\tilde\gamma_e^2]\,\epsilon_1^* \,,
\end{aligned}\nonumber \\
&\begin{aligned}
\delta_{e 5}=  \tilde\alpha_e^2|g_e|^2 (-4\epsilon_1^*+2\epsilon_2^*)
+ \tilde\beta_e^2(\frac{1}{3}\epsilon_1^*-\frac{7}{6}\epsilon_2^*) +i\,\tilde\gamma_e^2(\epsilon_1^*-\frac{1}{2}\epsilon_2^*) 
\simeq -2 \tilde\beta_e^2 \epsilon_1^*\, ,
\end{aligned}
\label{deltaeST}
\end{align}    
%where  $\epsilon_2=2\epsilon_1$ of Eq.(\ref{epST12}) is used in last %approximate equalities, 
%%%%%%%%%%%%%%%%%%%%%%%%%%%%
where $\epsilon_2=2\epsilon_1$ of Eq.\,(\ref{epST120})  is used in last approximate equalities. 
If $\tilde\beta_e^2\gg \tilde\alpha_e^2|g_e|^2, \tilde\gamma_e^2$, 
 mixing angles  $\theta_{23}^e$ and 
$\theta_{13}^e$  are  given :
\begin{align}
\theta_{23}^e\simeq  \frac{8}{9}\, |\epsilon_1|\simeq \frac{17}{9}|\epsilon|\, ,\qquad
\theta_{13}^e\simeq  \frac{10}{9}\, |\epsilon_1|\simeq\frac{21}{9}|\epsilon|\, ,\qquad
\end{align}
where $\epsilon_1= 2.1 i\, \epsilon$ in Eq.\,(\ref{epST120}) is taken.
Therefore, these mixing angles are  at most  $0.1$.
It is noticed that $\theta_{12}^e$ vanishes.

On the other hand, 
if $\tilde\alpha_e^2|g_e^2|\gg \tilde\beta_e^2, \tilde\gamma_e^2$, 
the mixing angle  $\theta_{13}^e$ is  given :
\begin{align}
\theta_{13}^e\simeq  \frac{2}{3}\,\left|\frac{2+g_e^*}{g_e}\epsilon_1\right| \simeq  \frac{8.4}{3}\,\left|\frac{1}{g_e}\epsilon\right|\, ,
\end{align}
where $|g_e|$ is supposed to be much smaller than 1 in the last equality.
Therefore, $\theta_{13}^e$ is enhanced by taking $|g_e|\simeq 0.1$.
It could be  of order $1$ if $|\epsilon|=0.05$. 
Thus, the flavor mixing angle $\theta_{13}^e$ contributes significantly  to the PMNS mixing angle  $\theta_{13}$.
%the PMNS mixing angles  $\theta_{12}$ and $\theta_{13}$ significantly
%at nearby the fixied point $\tau=\omega$.
%On the other hand, $\theta_{23}^e$ is highly suppressed.
%The PMNS mixing angle $\theta_{23}$ almost comes from the neutrino sector in %this case.

Indeed, we obtain the allowed region of $|\epsilon|\simeq 0.1$ 
with $|g_e|\simeq 0.2$  for NH of neutrinos by  performing numerical scan
  in section 7. 
However, for IH of neutrinos,  $|\epsilon|\simeq 0.15$ is obtained  with large  $|g_e|=5$--$10$.

%In order to obtain the non-vanishing charged lepton mass and  (1--2) flavor %mixing angle, we need the second-order perturbation.

\subsection{Lepton mass matrix at $\tau= i\infty$}

%%%%%%%%%%%%%%%%%%%%%%%%%%%%%%%%%%%%%%%%%%%%%%%%%%%%%%%
\subsubsection{Neutrino mass matrix at $\tau= i\infty$}

Let us consider the neutrino mass matrix at $\tau=i\infty$,
where  there exists the residual symmetries of the $\rm A_4$
group $\mathbb{Z}_3^{T}=\{  I, T, T^2 \}$.
By putting the modular forms in Table 1 into Eq.\,(\ref{neutrinomassmatrix}),
the neutrino mass matrix is written as:
\begin{align}
M_\nu=\frac{v_u^2}{\Lambda}Y_0^2\left [
\begin{pmatrix}
2 & 0 & 0\\
0 & 0 & -1 \\
0 & -1 & 0
\end{pmatrix}
+g_{\nu 1} 
\begin{pmatrix}
1 & 0 &0\\ 0 & 0 & 1 \\ 0 & 1 & 0
\end{pmatrix}
\right ] \ ,
\label{NeutrinotauT}
\end{align}
where the $g_{\nu 2}$ term of Eq.\,(\ref{neutrinomassmatrix}) disappears 
because of ${\bf Y^{\rm (4)}_{1'}}=0$ at $\tau=i\infty$.
%%%%%%%%
Since $T$ is already in the diagonal base as seen in Eq.\,(\ref{STbase}),
we can write down $M_\nu^\dagger  M_\nu$ straightforward as follows:
\begin{align}
\begin{aligned}
{\cal M}_\nu^{2(0)}\equiv  M_\nu^\dagger  M_\nu= \left (\frac{v_u^2}{\Lambda} Y_0^2 \right )^2
\end{aligned}
\begin{aligned}
\begin{pmatrix}
|2 + g_{\nu 1}|^2  & 0 & 0\\
0 & |1- g_{\nu 1}|^2  & 0\\
0& 0& |1- g_{\nu 1}|^2
\end{pmatrix} ,
\end{aligned}
\label{Neutrinonew2T}
\end{align}
which is a diagonal matrix as well as the neutrino mass matrix at  $\tau=\omega$
in Eq.\,(\ref{Neutrinonew2ST}). 
Three neutrino masses are  degenerate if  $g_{\nu 1}=-0.5$.
Then, large flavor mixing angles are possibly reproduced  if small off diagonal elements are generated due to finite effect of  $\tau$.

%%%%%%%%%%%%%%%%%%%%%%%%%%%%%%%%%%%%%%%%%%%%%%%%%%%%%%%%%
\subsubsection{Neutrino mass matrix towards $\tau=i\infty$}

Neutrino mass matrix
in Eq.\,(\ref{neutrinomassmatrix}),  $M_\nu$ 
is given  from the finite correction of $\tau=i\infty$.
Taking account of  modular forms of  Eq.\,(\ref{epsilonT0}), the corrections to Eq.\,(\ref{Neutrinonew2T}) are given by only a small variable $\epsilon$ of in Eq.\,(\ref{epsilonT0}).
In the 1st order approximation of $\epsilon$,
the correction ${\cal M}_\nu^{2(1)}$ to ${\cal M}_\nu^{2(0)}$ of Eq.\,(\ref{Neutrinonew2T}) is given in terms of 
\begin{align}
\delta=-6\,e^{\frac{2}{3}\pi i\,{\rm Re}\,\tau}e^{-\frac{2}{3}\pi\,{\rm Im}\,\tau}\,.
\label{deltaT}
\end{align}  
It is given by the following matrix:
\begin{align}
& \begin{aligned}
{\cal M}_\nu^{2(1)}\simeq  \left (\frac{v_u^2}{\Lambda} Y_0^2 \right )^2
\begin{pmatrix}
0&  -\delta^*\,(1-g_{\nu 1})(1+2g_{\nu 2}^*)&
\delta\,(2+g_{\nu 1}^*)(1+2g_{\nu 2})
\\ 
-\delta\,(1-g_{\nu 1}^*)(1+2g_{\nu 2})& 0  &  
2\delta^*\,(1-g_{\nu 1})(1-g_{\nu 2}^*)\\ 
\delta^*\,(2+g_{\nu 1})(1+2g_{\nu 2}^*)
& 2\delta\,(1-g_{\nu 1}^*)(1-g_{\nu 2})
&0
\end{pmatrix} .
\end{aligned}
\label{M2nu1T}
\end{align}  
If we take $\rm Im \tau=1.6$,
we get $|\delta|\simeq 0.21$,
which is derived in Eq.\,(\ref{deltaT}). 
Thus,
  the large (2--3) mixing angle is easily obtained
  since 2nd and 3rd eigenvalues are degenerated as seen in Eq.\,(\ref{Neutrinonew2T}).
The large (1--2) mixing angle is also possible by choosing
relevant $g_{\nu 1}$ and $g_{\nu 2}$.
The  (1--3) mixing angle is expected relatively small due to the fixed mass square difference $\Delta m_{31}^2$.
Then, the cosmological  upper-bound of the sum of neutrino masses is a crucial criterion to test neutrino mass matrices.
In section 7, we will see that 
both charged lepton mass matrix I and  I\hspace{-.01em}I 
satisfy the  sum of neutrino masses less than the cosmological upper-bound
$120$\,meV
for NH of neutrinos, but they do not satisfy it for IH.
%%%%%%%%%%%%%%%%%%%%%%%%%%%%%%%%%%%%%%%%%%%%%%%%%%%%%%%%%
%%%%%%%%%%%%%%%%%%%%%%%%%%%%%%%%%%%%%%%%%%%%%%%%%%%%%%%%%

\subsubsection{Charged lepton mass matrix I and I\hspace{-.01em}I 
	at $\tau= i\infty$}

The charged lepton mass matrices of  I and I\hspace{-.01em}I
in Eqs.\,(\ref{ME222}) and (\ref{ME642})  
are simple   at    $\tau=i\infty$
since the  modular forms of weight $2$, $4$ and $6$ are given in the $T$ diagonal base. Putting them of Table 1 into the charged lepton mass matrices  in Eqs.\,(\ref{ME222}) and (\ref{ME642}), we obtain
%%%%%%%%%%%%%%%%%%%%%%%%%%%%%%%%%%%%%%%%%%%%%%%%%%%%%%%%%
\begin{align}
\begin{aligned}
M_E=
\begin{pmatrix}
\tilde \alpha_e & 0 & 0 \\
0 &\tilde \beta_e & 0\\
0 & 0 &\tilde\gamma_e
\end{pmatrix} \, ,
\end{aligned}
\label{chargedleptonT0}
\end{align}
where 
$\tilde \alpha_e=  v_dY_0 \alpha_e$,
$\tilde \beta_e= v_d Y_0 \beta_e$ and 
$\tilde \gamma_e =v_dY_0 \gamma_e$
for the case  I and 
$\tilde \alpha_e=  v_d Y_0^3 \alpha_e$,
$\tilde \beta_e= v_d Y_0^2 \beta_e$ and 
$\tilde \gamma_e =v_d Y_0 \gamma_e$ for the case I\hspace{-.01em}I.
%%%%%%%%%%%%%%%%%%%%%%%%%%%%%%%%%%%%%%%%%%
The mass matrix $M_E^\dagger M_E$ is given as:
\begin{align}
{\cal M}_E^{2(0)}\equiv   M_E^\dagger M_E 
= 
\begin{aligned}
\begin{pmatrix}
\tilde\alpha_e^2&0 & 0 \\
0&\tilde\beta_e^2 &   0  \\
0 & 0 & \tilde\gamma_e^2 \\
\end{pmatrix}
\end{aligned}\,.
\label{chargedleptonT}
\end{align}
The flavor mixing appears through the finite effect of ${\rm Im }\,[\tau]$.

%%%%%%%%%%%%%%%%%%%%%%%%%%%%%%%%%%%%%%%%%%%%%%%%%%%%%%%%%%%%%%%%%
\subsubsection{Charged lepton mass matrix I and 	I\hspace{-.01em}I towards  $\tau=i\infty$} 
%%%%%%%%%%%%%%%%%%%%%%%%%%%%%%%%%%%%%%%%%%%%%%%%%%%%%%%
%%%%%%%%%%%%%%%%%%%%%  Charged Lepton %%%%%%%%%%%%%%%%% 
%%%%%%%%%%%%%%%%%%%%%%%%%%%%%%%%%%%%%%%%%%%%%%%%%%%%%%%
The charged lepton mass matrices of  I and I\hspace{-.01em}I
in Eqs.\,(\ref{ME222}) and (\ref{ME642})  
are given  from the finite correction of $\tau=i\infty$.
By using modular forms of  Eq.\,(\ref{epsilonT0}), the corrections to Eq.\,(\ref{chargedleptonT}) are given by only a small variable $\epsilon$ of  Eq.\,(\ref{epsilonT0}).
In the 1st order approximation of $\epsilon$,
the correction ${\cal M}_E^{2(1)}$ to ${\cal M}_E^{2(0)}$ of Eq.\,(\ref{chargedleptonT}) is given in terms of $\delta$ of Eq.\,(\ref{deltaT})
as:
%%%%%%%%%%%%%%%%%%%%%  ME-I %%%%%%%%%%%%%%%%
\begin{align}
{\cal M}_E^{2(1)}\simeq  
\begin{pmatrix}
0 &  \delta^*\, \tilde\beta_e^2 &
\delta \, \tilde\alpha_e^2 \\ 
\delta\, \tilde\beta_e^2&  0 &  
\delta^*\, \tilde\gamma_e^2 \\ 
\delta^* \, \tilde\alpha_e^2
&   \delta\, \tilde\gamma_e^2
&0
\end{pmatrix} \, ,
\label{Iinfty}
\end{align}  
for  the charged lepton mass matrix I.
On the other hand, for the charged lepton mass matrix I\hspace{-.01em}I, 
it is: 
%%%%%%%%%%%%%%%%%  ME-II  %%%%%%%%%%%
\begin{align}
& \begin{aligned}
{\cal M}_E^{2(1)}\simeq  
\begin{pmatrix}
0&  -\delta^*\,\tilde\beta_e^2&
(1+2\,g_e)\,\delta\,\tilde\alpha_e^2\\ 
-\delta\,\tilde\beta_e^2& 0  &  
\delta^*\,\tilde\gamma_e^2\\ 
(1+2\,g_e^*)\,\delta\,\tilde\alpha_e^2
& \delta\,\tilde\gamma_e^2
&0
\end{pmatrix} .
\end{aligned}
\label{IIinfty}
\end{align}

In both charged lepton mass matrices I and
I\hspace{-.01em}I,
 (1--2) and (2--3) families mixing angles $\theta_{23}^e$,
$\theta_{12}^e$, are given as:
\begin{align}
\theta_{12}^e\simeq 
\frac{|\delta^*|\,\tilde\beta_e^2}{\tilde\beta_e^2 }
\simeq |\delta|\,, \qquad\qquad 
\theta_{23}^e\simeq \frac{|\delta^*|\,\tilde\gamma_e^2}{\tilde\gamma_e^2}
=|\delta| \,,
\qquad 
\end{align}
respectively,
where  $\tilde\gamma_e^2 \gg\tilde\beta_e^2\gg\tilde\alpha_e^2$.
If we take $\rm Im \tau=1.6$,
the magnitude of $\theta_{12}^e\simeq |\delta|\simeq 0.21$.
This magnitude of $\theta_{12}^e$  
contributes significantly to the  PMNS mixing angle $\theta_{13}$.
On the other hand, 
the  mixing angle $\theta_{13}^e$ between 1st- and 3rd-family is highly suppressed
due to the factor $\tilde\alpha_e^2/\tilde\gamma_e^2$.

It is remarked that the mass matrix of Eq.\,(\ref{IIinfty}) is agreement with 
Eq.(\ref{Iinfty})  in the case of $|g_e|\ll 1$ apart from the minus sign in front of (1,2) and (2,1) entries.
However,
this minus sign of the charged lepton mass matrix I\hspace{-.01em}I spoils to
reproduce  large mixing angles of the PMNS matrix, $\theta_{12}$ and $\theta_{23}$
together although the charged lepton mass matrix I is successful to reproduce
the observed  PMNS mixing angles.

Alternatively,  the observed PMNS mixing angles can be reproduced
in  the charged lepton mass matrix I\hspace{-.01em}I if  a large
mixing angle for  $\theta_{13}^e$ is obtained by taking $|g_e|\gg 1$ with  $\tilde\alpha_e^2 \gg\tilde\beta_e^2,\,\tilde\gamma_e^2$. 
This case is shown numerically in section 7.

%%%%%%%%%%%%%%%%%%%%%%%%%%%%%%%%%%%%%%%%%%%%%%%%%%%%%%%%%%%%%%%%%%%%%%
%%%%%%%%%%%%%%%%%%%%%%%%%%%%%%%%%%%%%%%%%%%%%%%%%%%%%%%%%%%%%%%%%%%%%%
%%%%%%%%%%%%%%%%%%%%%%%%%%%%%%%%%%%%%%%%%%%%%%%%%%%%%%%%%%%%%%%%%%%%%%
%%%%%%%%%%%%%%%%%%%%%%%%%%%%%%%%%%%%%%%%%%%%%%%%%%%%%%%%%%%%%%%%%%%%%%
%%%%%%%%%%%%%%%%%%%%%%%%%%%%%%%%%%%%%%%%%%%%%%%%%%%%%%%%%%%%%%%%%%%%%%
%%%%%%%%%%%%%%%  Quark %%%%%%%%%%%%%%%%%%%%%%%%%%%%%%%%%%%%%%%%%%%%%%%
%%%%%%%%%%%%%%%%%%%%%%%%%%%%%%%%%%%%%%%%%%%%%%%%%%%%%%%%%%%%%%%%%%%%%%
%%%%%%%%%%%%%%%%%%%%%%%%%%%%%%%%%%%%%%%%%%%%%%%%%%%%%%%%%%%%%%%%%%%%%%
%%%%%%%%%%%%%%%%%%%%%%%%%%%%%%%%%%%%%%%%%%%%%%%%%%%%%%%%%%%%%%%%%%%%%%
\section{Quark mass matrices in the $\rm A_4$ modular invariance}

If  flavors of quarks and leptons are originated from a same two-dimensional compact space, 
the leptons and quarks have same flavor symmetry and the same value of
the modulus  $\tau$.
Therefore, the modular symmetry provides a new approach 
towards the unification of quark and lepton flavors.
In order to investigate the possibility of the quark/lepton unification, we discuss a $\rm A_4$ modular invariant flavor model for quarks
together with the lepton sector.
%%%%%%%%%%%%%%%%%%%%%%%%%%%%%%%%%%%%%%%%%%%%%%%%%%%%%%%%%%%%
%%%%%%%%%%%%%%%%%%%%%%%%%%%%%%%%%%%%%%%%%%%%%%%%%%%%%%%%%%%%
\subsection{Model of quark mass matrices}

 We take the assignments of $\rm A_4$ irreducible representations and modular weights 
 for  quarks like the charged leptons.
 That is,  three left-handed quarks are components of the triplet of the $\rm A_4$ group,  but   three right-handed quarks,
 ($u^c, c^c, t^c$) and  ($d^c, s^c, b^c$) are three different singlets 
 $\bf (1,1'',1')$ of  $\rm A_4$, respectively.
 Quark mass matrices   depend on   modular weights of the left-handed and the  right-handed quarks since the sum of their weight including modular forms should vanish. Let us fix the weights of left-handed quarks
 to be $-2$ like the left-handed charged leptons.
    If the weight is $0$ for all right-handed quarks 
     like right-handed charged leptons,
   both up-type and down-type mass matrices are given in terms of only 
     the weight $2$ modular forms of Eq.\,(\ref{eq:Y-A4}).
     However, this case is inconsistent with the observed CKM matrix as well known
     \cite{Okada:2019uoy}.
   In order to overcome this failure, we introduce  weight $4$ and $6$ modular forms of Eqs.\,(\ref{weight4}) and (\ref{weight6})
    in addition to weight $2$ modular forms \cite{Okada:2019uoy}.
    We consider one simple model in the case I, where the up-type right-handed quarks have different weights from  the weight $0$ of the right-handed  down-type quarks. 
    The assignment is presented in Table 3, in which
     the weight of right-handed up-type quarks  is $-4$.
     Therefore,  the up-type quark mass matrix is given 
      in terms of the weight  $6$ modular forms, in which  two different 
      triplet modular forms are available.
     This model has already discussed  in Ref.\cite{Okada:2019uoy} numerically.
     We reexamine the flavor structure of these quark mass matrices
      at nearby fixed point explicitly, and then we can understand why this model
      works well.
      
     Alternatively, another quark mass matrix is  also considered as the case I\hspace{-.01em}I.
     In this case,
      weights of  the right-handed  up-type quarks and the down-type ones
       are same ones, which are also discussed numerically 
       in Ref.\cite{Okada:2020rjb}. 
       The modular forms of weight $6$ join  only  in the 1st-family.
    
  %%%%%%%%%%%%%%%%%%%%%%%%%%%%%%%%%%%%%%%%%%%%%%%%
  %\vskip - 0.3 cm  
  \begin{table}[h]
  	\centering
  	\begin{tabular}{|c|c|c|c|c|c|c|} \hline
  		&$Q$&$(u^c,c^c,t^c)$,\  $(d^c,s^c,b^c)$&$H_q$&
  		$\bf Y_3^{(\rm 6)}, \ Y_{3'}^{(\rm 6)}$& $\bf Y_3^{(\rm 4)}$& $\bf Y_3^{(\rm2)}$\\  \hline\hline 
  		\rule[14pt]{0pt}{0pt}
  		$SU(2)$&$\bf 2$&$\bf 1$&$\bf 2$&$\bf 1$&$\bf 1$&$\bf 1$\\
  		\hline
  			\rule[14pt]{0pt}{0pt}
  		$\rm A_4$&$\bf 3$&\bf (1,\ 1$''$,\ 1$'$)&$\bf 1$&$\bf 3 $&$\bf 3$&$\bf 3$\\ \hline
  			\rule[14pt]{0pt}{0pt}
  		$-k_I$	&$ -2$& I\,:  $(-4, -4, -4)$,\quad \ $(0,\, 0,\, 0)$ &0&$k=6$&$k= 4$&$k=2$\\
  		&&I\hspace{-.01em}I\ :$(-4, \,  -2, \, 0)$, 
  		$(-4, \,  -2, \, 0)$& & & &\\
  		 \hline
  	\end{tabular}
  	\caption{Assignments of representations and  weights
  		$-k_I$ for MSSM fields and  modular forms.
  	}
  	\label{tb:weight6}
  \end{table}
  %%%%%%%%%%%%%%%%%%%%%%%%%%%%%%%%%%%%%%%%%%%%%%%%%%%%% 

 The relevant superpotentials of the quark sector are given
 for two cases as follows:
%%%%%%%%%%%%%%%%%%%%%%%%%%%%%%%%%%%%%%%%%%%%%%%%%%%%%  %%%%%%%%%%%%%%%%%%%%%%%%%%%%%%%%%%%%%%%%%%%%%%%%%%%%%
 \begin{align}
 {\rm I}
 \,:\quad
 w_u&=\alpha_u u^c H_u {\bf Y^{(\rm6)}_3} Q+
 \alpha'_u u^c H_u {\bf Y_{3'}^{(\rm 6)}} Q+
 \beta_u c^c H_u {\bf Y^{(\rm 6)}_3}Q +
 \beta'_u c^c H_u {\bf Y_{3'}^{(\rm 6)}} Q\nonumber\\
 &+\gamma_u t^c H_u {\bf Y^{(\rm 6)}_3}Q +
 \gamma'_u t^c H_q {\bf Y_{3'}^{(\rm 6)}} Q\,,\nonumber\\
 w_d&=\alpha_d d^c H_d {\bf Y^{(\rm 2)}_3}Q+
 \beta_d s^c H_d {\bf Y^{(\rm 2)}_3}Q+
 \gamma_d b^c H_d {\bf Y^{(\rm 2)}_3}Q~,\\
{\rm I\hspace{-.01em}I}\,:\quad
  w_q&=\alpha_q q_1^c H_q {\bf Y^{(\rm 6)}_3} Q+
 \alpha'_q q_1^c H_q {\bf Y_{3'}^{(\rm 6)}} Q+
 \beta_q q_2^c H_q {\bf Y^{(\rm 4)}_3}Q+
 \gamma_q q_3^c H_q {\bf Y^{(\rm 2)}_3}Q~,
 \end{align}
%%%%%%%%%%%%%%%%%%%%%%%%%%%%%%%%%%%%%%%%%%%%%%%%%%%%% 
where $q=u,\,d$, and the argument $\tau$ in the modular forms $Y_i(\tau)$ is  omitted.
%%%%%%%%%%%%%%%%%%%%%%%%%%%%%%%%%%%%%%%%%%%%%%%
Couplings  $\alpha_q$,  $\alpha'_q$,$\beta_q$, $\beta'_q$,
 $\gamma_q$ and   $\gamma'_q$ can be adjusted to the  observed quark masses.

%%%%%%%%%%%%%%%%%%%%%%%%%%%%%%%%%%%%%%%%%%%%%%%%
    The quark mass matrices are  written as:
\begin{align}
&\begin{aligned}
{\rm I}\,: \quad 
&M_u=v_u
\begin{pmatrix}
\alpha_u & 0 & 0 \\
0 &\beta_u & 0\\
0 & 0 &\gamma_u
\end{pmatrix} \left [
\begin{pmatrix}
Y_1^{(6)} & Y_3^{(6)}& Y_2^{(6)} \\
Y_2^{(6)} & Y_1^{(6)} &  Y_3^{(6)} \\
Y_3^{(6)} &  Y_2^{(6)}&  Y_1^{(6)}
\end{pmatrix}
+ 
\begin{pmatrix}
g_{u1} & 0 & 0 \\
0 &g_{u2} & 0\\
0 & 0 &g_{u3}
\end{pmatrix}
\begin{pmatrix}
Y_1^{'(6)} & Y_3^{'(6)}& Y_2^{'(6)} \\
Y_2^{'(6)} & Y_1^{'(6)} &  Y_3^{'(6)} \\
Y_3^{'(6)} &  Y_2^{'(6)}&  Y_1^{'(6)}
\end{pmatrix}
\right ]_{RL},
\\
&M_d= v_d
\begin{pmatrix}
\alpha_d & 0 & 0 \\
0 &\beta_d & 0\\
0 & 0 &\gamma_d
\end{pmatrix}
\begin{pmatrix}
Y_1 & Y_3& Y_2\\
Y_2 & Y_1 &  Y_3 \\
Y_3 &  Y_2&  Y_1
\end{pmatrix}_{RL}\,,  
\end{aligned} 
\label{matrixIII}\\
&\begin{aligned}
{\rm I\hspace{-.01em}I}\,:\quad &M_q=v_q
\begin{pmatrix}
\alpha_q & 0 & 0 \\
0 &\beta_q & 0\\
0 & 0 &\gamma_q
\end{pmatrix} 
\begin{pmatrix}
Y_1^{(6)}+g_q Y_1^{'(6)} & Y_3^{(6)} +g_q Y_3^{'(6)} 
& Y_2^{(6)}+g_q Y_2^{'(6)} \\
Y_2^{(4)} & Y_1^{(4)} &  Y_3^{(4)} \\
Y_3^{(2)} &  Y_2^{(2)}&  Y_1^{(2)}
\end{pmatrix}_{RL},
\end{aligned}
\label{matrixI}
%%%%%%%%%%%%%%%%%%%%%%
\end{align}
where 
$g_{u1}=\alpha'_u/\alpha_u$, $g_{u2}=\beta'_u/\beta_u$, $g_{u3}=\gamma'_u/\gamma_u$ and $g_q\equiv \alpha_q'/\alpha_q$.
The VEV of the Higgs field $H_q$ is denoted by  $ v_q$.
Parameters $\alpha_q$,  $\beta_q$,  $\gamma_q$  can be taken to be  real,
on the other hand, $g_{u1}$, $g_{u2}$, $g_{u3}$, $g_{u}$ and $g_{d}$ are  complex parameters.

 These mass matrices turn to the simple ones at the fixed points,
 $\tau=i$, $\tau=\omega$ and $\tau=i\infty$. 
  We discuss them in the diagonal bases of $S$, $ST$ and $T$,
  respectively.

%%%%%%%%%%%%%%%%%%%%%%%%%%%%%%%%%%%%%
%%%%%%%%%%%%%%%%%%%%%%%%%%%%%%%%%%%%%
\subsection{Quark mass matrix  at the fixed point of $\tau=i$}
%%%%%%%%%%%%%%%%%%%%%%%%%%%%%%%
\subsubsection{Quark mass matrix I at $\tau=i$}

The quark matrix {\rm I} is given by using modular forms
in Table 1 at fixed point  $\tau=i$ in the base of $S$ of Eq.\,(\ref{STbase}) as follows:
%%%%%%%%%%%%%%%%%%%%%%%%%%%%%%%
\begin{align}
\begin{aligned}
M_u=&
\begin{pmatrix}
\tilde \alpha_u & 0 & 0 \\
0 &\tilde \beta_u & 0\\
0 & 0 &\tilde\gamma_u
\end{pmatrix} \times \\
& \begin{pmatrix}
2\sqrt{3}-3+g_{u1} (7\sqrt{3}-12) & 12-7\sqrt{3}+g_{u1} (9-5\sqrt{3})&
5\sqrt{3}-9+g_{u1} (3-2\sqrt{3}) \\
5\sqrt{3}-9+g_{u2} (3-2\sqrt{3})&2\sqrt{3}-3+g_{u2} (7\sqrt{3}-12) & 12-7\sqrt{3}+g_{u2} (9-5\sqrt{3}) \\
12-7\sqrt{3}+g_{u3} (9-5\sqrt{3}) &5\sqrt{3}-9+g_{u3} (3-2\sqrt{3})&2\sqrt{3}-3+g_{u3} (7\sqrt{3}-12) 
\end{pmatrix}\, , \\
M_d=&
\begin{pmatrix}
\tilde \alpha_d & 0 & 0 \\
0 &\tilde \beta_d & 0\\
0 & 0 &\tilde\gamma_d
\end{pmatrix} 
\begin{pmatrix}
1&-2+\sqrt{3} & 1-\sqrt{3}\\
1-\sqrt{3}& 1 & -2+\sqrt{3} \\
-2+\sqrt{3} & 1-\sqrt{3}& 1
\end{pmatrix}\, ,
\end{aligned}
\end{align}
where $\tilde \alpha_u=3 v_uY_0^3 \alpha_u$,
$\tilde \beta_u=  3v_uY_0^3  \beta_u$, 
$\tilde \gamma_u =3 v_uY_0^3   \gamma_u$,
$\tilde \alpha_d=(6-3\sqrt{3}) v_d Y_0^2\alpha_d$,
$\tilde \beta_d= (6-3\sqrt{3}) v_d Y_0^2 \beta_d$ and 
$\tilde \gamma_q =(6-3\sqrt{3}) v_d Y_0^2  \gamma_d$.
%%%%%%%%%%%%%%%%%%%%%%%%%%%%%%%%%%%%%%%%%%

We move the quark mass matrix to the diagonal base  of $S$.
By using the unitary transformation of Eq.\,(\ref{Sdiagonal}), $V_{S2}$,
the mass matrix $M_u^\dagger M_u$ is transformed  as:
\begin{align}
&{\cal M}_u^{2(0)}\equiv  V_{S2} M_q^\dagger M_u V_{S2}^\dagger = \frac{9}{2}
\begin{aligned}
\begin{pmatrix}
0& 0 & 0 \\
0 & a_{22} \tilde\alpha_u^2 + b_{22}\tilde \beta_u^2+c_{22}\tilde \gamma_u^2
& a_{23} \tilde\alpha_u^2 + b_{23}\tilde \beta_u^2+c_{23}\tilde \gamma_u^2 \\
0 & a_{23}^* \tilde\alpha_u^2 + b_{23}^*\tilde \beta_u^2+c_{23}^*\tilde \gamma_u^2 & a_{33} \tilde\alpha_u^2 + b_{33}\tilde \beta_u^2+c_{33}\tilde \gamma_u^2 
\end{pmatrix}
\end{aligned}\,.
\label{upquark-III}
\end{align}
Each coefficient is given as:
\begin{align}
& a_{22}=A+2B{\rm Re}[g_{u1}]+C|g_{u1}|^2,\qquad\qquad\qquad\quad\ \
b_{22}= 2B+2(A-B){\rm Re}[g_{u2}]+A|g_{u2}|^2, \nonumber \\
&c_{22}= C+2(C-B){\rm Re}[g_{u3}]+2B|g_{u3}|^2,\qquad\quad\quad  \
a_{23}=-B-A g_{u1}-C g_{u1}^*-B|g_{u1}|^2, \nonumber\\
&b_{23}=2B+(C-B) g_{u2}+(A-B) g_{u2}^*-B|g_{u2}|^2,  \ 
c_{23}=-B+(C-B) g_{u3}+(A-B) g_{u3}^*+2B|g_{u3}|^2,\nonumber\\
&a_{33}=C+2B{\rm Re}[g_{u1}]+A|g_{u1}|^2,\qquad\qquad\qquad\quad\ \
b_{33}= 2B+2(C-B){\rm Re}[g_{u2}]+C|g_{u2}|^2, \nonumber\\
&c_{33}= A+2(A-B){\rm Re}[g_{u3}]+2B|g_{u3}|^2,
\label{cij}
\end{align}
where $A$, $B$ and $C$ are given in Eq.\,(\ref{ABCD}).
 %For example,
%in the case of  $\gamma_u\gg \beta_u,\,\alpha_u$ and $|g_{ui}|\ll 1 $,
On the other hand,
the mass matrix $M_d^\dagger M_d$ is transformed  as:
\begin{align}
&{\cal M}_d^{2(0)}\equiv  V_{S2} M_d^\dagger M_d V_{S2}^\dagger = \frac{3}{2}
\begin{aligned}
\begin{pmatrix}
0& 0 & 0 \\
0 &  \tilde\alpha_d^2 + 2D\tilde \beta_d^2+A\tilde \gamma_d^2
& -D\tilde(\alpha_d^2 -2\tilde \beta_d^2+\tilde \gamma_d^2 )\\
0 & -D\tilde(\alpha_d^2 -2\tilde \beta_d^2+\tilde \gamma_d^2 ) &  A\tilde\alpha_d^2 + 2D\tilde \beta_d^2+\tilde \gamma_d^2 
\end{pmatrix}
\end{aligned}\,.
\label{downquark-III}
\end{align}

%%%%%%%%%%%%%%%%%%%%%%%%%%%%%%%%%%%%%%%%%%%
It is remarked that the lightest quarks are massless for both up-type 
and down-type quarks at $\tau=i$. Therefore,
the small deviation from $\tau=i$ is required to avoid
the massless quark.
There exists a non-vanishing flavor mixing angle $\theta_{23}^u$  at $\tau=i$
as discussed in Eq.\,(\ref{zeros}).
Supposing 
$\tilde \gamma_q\gg \tilde \beta_q, \tilde \alpha_q$,
the mixing angle $\theta_{23}^u$ is given 
from Eq.\,(\ref{upquark-III}) as:
\begin{align}
\tan 2\theta_{23}^u &\simeq 
2 \frac{|-B+(C-B)g_{u3}+(A-B)g_{u3}^*+2B|g_{u3}|^2|}
{(A-C)(1+2{\rm Re}[g_{u3}])} \nonumber \\
&=
2 \frac{\sqrt{[-B+2B{\rm Re}[g_{u3}] +2B|g_{u3}|^2]^2+
		[(C-A) {\rm Im}g_{u3}]^2 }}
{2\sqrt{3}B(1+2{\rm Re}[g_{u3}])} \simeq
\frac{1}{\sqrt{3}}\left |\frac{2g_{u3}^2+2g_{u3}-1}{1+2g_{u3}}\right |
\,,
\label{theta12q-III}
\end{align}
where $A+C=4B$ is used and the imaginary part of $g_q$ is neglected in the last equation
($g_{u3}={\rm Re}[g_{u3}]$).
In this case, $\tan 2\theta_{23}^u$ vanishes at
$g_{u3}=(-1\pm \sqrt{3})/2$, while
$\theta_{23}^u=15^\circ$ at $g_{u3}=0$.

On the other hand,  the mixing angle $\theta_{23}^d$  is simply given
from Eq.\,(\ref{downquark-III})   as:
\begin{align}
\tan 2\theta_{23}^d &\simeq 
2 \frac{D}
{1-A} =\frac{1}{\sqrt{3}} 
\,,
\label{theta12q-III}
\end{align}
which leads to $\theta_{23}^d=15^\circ$.
Since  the observed small CKM mixing angle $\theta_{23}^{\rm CKM}$
(around $2^\circ$) is given by the difference  $(\theta_{23}^d-\theta_{23}^u)$,
the magnitude of $g_{u3}$ should be small in order to realize the enough cancellation
between $\theta_{23}^d$ and $\theta_{23}^u$.
Indeed, $|g_{u3}|$ is in $[0,02,0.07]$ in our numerical result of section 7.

\subsubsection{Quark mass matrix I at nearby   $\tau=i$}

By using the approximate modular forms of weight 2 and 6
 in Eqs.\,(\ref{epS12}) and (\ref{epS666}) of Appendix C.1, we present
the deviations from  ${\cal M}_u^{2(0)}$
and ${\cal M}_d^{2(0)}$ in  Eqs.\,(\ref{upquark-III})
and (\ref{downquark-III}).
%for the quark mass matrix I.
% $\tilde\beta_u^2 \gg \tilde\alpha_u^2,\, \tilde\gamma_u^2$.
Then, 
the additional contribution  ${\cal M}_u^{2(1)}$ to ${\cal M}_u^{2(0)}$  of Eq.\,(\ref{upquark-III})
of order $\epsilon$ is given in terms of $A$, $B$ and $C$
in Eq.\,(\ref{ABCD}) as follows:     
%%%%%%%%%%%%%%%%%%%%%%%%%%%%%%%%%%
%%%%%%%%%%%%%%%%%%%%%%%%%%%%%%%%%%
\begin{align}
\begin{aligned}
{\cal M}_u^{2(1)}\simeq 
\begin{pmatrix}
0 &  \delta_{u2}& \delta_{u3}\\ 
\delta_{u2}^* &  \delta_{u4}& \delta_{u5}\\ 
\delta_{u3}^* &  \delta_{u5}^*& \delta_{u6}
\end{pmatrix} \,,
\end{aligned}
\label{M2uS-III}
\end{align}
where
\begin{align}
&\begin{aligned}
\delta_{u 2}&=\frac{3}{\sqrt{2}}
\{
[\,(A-B+(B-C)g_{u1})\epsilon_1^* +(B+C g_{u1})\epsilon_2^*\, ] (g_{u1}^*-1)\tilde\alpha_u^2\\
&+ (-2B+(B-A)g_{u2})\epsilon_1^* +(C-B-B g_{u2})\epsilon_2^*\, ] (g_{u2}^*-1) \tilde\beta_u^2 \\
&+ (C-B+2Bg_{u3})\epsilon_1^* +(-C+(B-C) g_{u3})\epsilon_2^*\, ] (g_{u3}^*-1) \tilde\gamma_u^2
\} \\
&\simeq
\frac{3}{\sqrt{2}} \epsilon_1^*\{
[\,(A+B)+(B+C) g_{u1}\,](g_{u1}^*-1)\tilde\alpha_u^2
+  [\,2(C-2B)-(A+B) g_{u2}\,](g_{u2}^*-1) \tilde\beta_u^2
\\
&+ [\,-(B+C)+2(2B-C) g_{u3}\,](g_{u3}^*-1) \tilde\gamma_u^2
\}\, , 
\end{aligned}\\
&  \begin{aligned}
\delta_{u 3}&=\frac{3}{\sqrt{2}}
\{
[\,(C-B-(A-B)g_{u1})\epsilon_1^* -(C+B g_{u1})\epsilon_2^*\, ] (g_{u1}^*-1)\tilde\alpha_u^2\\
&+ (-2B+(B-C)g_{u2})\epsilon_1^* +(C-B-C g_{u2})\epsilon_2^*\, ] (g_{u2}^*-1) \tilde\beta_u^2 \\
&+ (A-B+2Bg_{u3})\epsilon_1^* +(B+(B-C) g_{u3})\epsilon_2^*\, ] (g_{u3}^*-1) \tilde\gamma_u^2
\} \\
&\simeq
\frac{3}{\sqrt{2}} \epsilon_1^*\{
- [\,(C+B)+(A+B) g_{u1}\,](g_{u1}^*-1)\tilde\alpha_u^2
+  [\,2(C-2B)+(B+C) g_{u2}\,](g_{u2}^*-1) \tilde\beta_u^2
\\
&+ [\, A+B+2(2B-C) g_{u3}\,](g_{u3}^*-1) \tilde\gamma_u^2
\}\, . 
\end{aligned}
\label{deltauS-III}
\end{align}  
In the  approximate equalities,  
 $\epsilon_2=2\epsilon_1$ in Eq.\,(\ref{epS120}) is put. 
%%%%%%%%%%%%%%%%%%%%%%%%%%%%%%%%%%%%%%%%%
In order to estimate the Cabibbo angle, we calculate the
mixing angle of the 1st- and 2nd-family as:
\begin{align}
\tan 2\theta_{12}^u =
\ \frac{2|\delta_{u2}|}{\frac{9}{2}
	(a_{22}\tilde\alpha_u^2+b_{22}\tilde\beta_u^2+
	c_{22}\tilde\gamma_u^2)}\simeq \frac{4}{3\sqrt{2}} \frac{B+C}{C}
| \epsilon_1^*| \simeq \frac{4}{3\sqrt{2}} (3+\sqrt{3})|\epsilon_1^*|\simeq 4.46\, |\epsilon_1^*|\,, 
\label{limit-theta12uIII}
\end{align}
where the denominator comes from the (2,\,2) element of Eq.\,(\ref{upquark-III}).
In the second approximate equality, 
$ \tilde\gamma_u\gg \tilde\alpha_u, \tilde\beta_u$ and  $|g_{u3}|\ll 1$
are put, while  $c_{22}$ is given in Eq.\,(\ref{cij}).

%%%%%%%%%%%%%%%%%%%%%%%%%%%%%%%%%%%%%%%%%% 
The additional contribution  ${\cal M}_d^{2(1)}$ to  
 ${\cal M}_d^{2(0)}$ of Eq.\,(\ref{downquark-III})
of order $\epsilon$ is:     
%%%%%%%%%%%%%%%%%%%%%%%%%%%%%%%%%%
%%%%%%%%%%%%%%%%%%%%%%%%%%%%%%%%%%
\begin{align}
\begin{aligned}
{\cal M}_d^{2(1)}\simeq 
\begin{pmatrix}
0 &  \delta_{d2}& \delta_{d3}\\ 
\delta_{d2}^* &  \delta_{d4}& \delta_{d5}\\ 
\delta_{d3}^* &  \delta_{d5}^*& \delta_{d6}
\end{pmatrix} \,,
\end{aligned}
\label{M2dS-III}
\end{align}
where
\begin{align}
& \begin{aligned}
\delta_{d 2}&=\frac{1}{\sqrt{2}} \{
[(\sqrt{3}-1)\epsilon_1^* +(\sqrt{3}-2)\epsilon_2^* ] \tilde\alpha_d^2
+ [(4-2\sqrt{3})\epsilon_1^* +(3\sqrt{3}-5)\epsilon_2^* ] \tilde\beta_d^2 \\
&+ [(3\sqrt{3}-5)\epsilon_1^* +(7-4\sqrt{3})\epsilon_2^* ] \tilde\gamma_d^2
\} \simeq \frac{1}{\sqrt{2}} \epsilon_1^* 
[(3\sqrt{3}-5)  \tilde\alpha_d^2 +2(2\sqrt{3}-3)  \tilde\beta_d^2  +(9-5\sqrt{3})  \tilde\gamma_d^2]\, ,
\end{aligned}\\
%%%%%%%%%%%%
&  \begin{aligned}
\delta_{d 3}&= \frac{1}{\sqrt{6}} \{ 
[(9-5\sqrt{3})\epsilon_1^* +(7\sqrt{3}-12)\epsilon_2^* ] \tilde\alpha_d^2 +
[(4\sqrt{3}-6)\epsilon_1^* +(9-5\sqrt{3})\epsilon_2^* ] \tilde\beta_d^2 \\
&+ [(\sqrt{3}-3)\epsilon_1^* +(3-2\sqrt{3})\epsilon_2^* ] \tilde\gamma_d^2 
\}\simeq
\frac{\sqrt{6}} {2}\epsilon_1^* 
[(3\sqrt{3}-5)  \tilde\alpha_d^2 +2(2-\sqrt{3})  \tilde\beta_d^2  +(1-\sqrt{3})  \tilde\gamma_d^2] \, .
\end{aligned}
\label{deltadS-III}
\end{align}    
In the last approximate equalities, $\epsilon_2=2\epsilon_1$ in Eq.\,(\ref{epS120}) is put.
%%%%%%%%%%%%%%%%%%%%%%%%%%%%%%%%%%%%%%%%%
The mixing angle of the 1st- and 2nd-family as:
\begin{align}
\tan 2\theta_{12}^d =
\ \frac{2|\delta_{d2}|}{\frac{3}{2}
	( \tilde\alpha_d^2 + 2D\tilde \beta_d^2+A\tilde \gamma_d^2)}\simeq \frac{4}{3\sqrt{2}} \frac{9-5\sqrt{3}}{A}
|\epsilon_1^*| \simeq \frac{4}{3\sqrt{2}} (3+\sqrt{3})|\epsilon_1^*|\simeq 4.46\, |\epsilon_1^*|\,, 
\label{limit-theta12dIII}
\end{align}
where the denominator comes from the (2,\,2) element of Eq.\,(\ref{downquark-III}).
In the second approximate equality, 
$ \tilde\gamma_d\gg \tilde\alpha_d, \tilde\beta_d$ is taken.
Since the magnitudes  of $\theta_{12}^u$ and $\theta_{12}^d$
in Eqs.\,(\ref{limit-theta12uIII}) and (\ref{limit-theta12dIII}) are almost same, the  phase of $\epsilon_1$ is important to reproduce the Cabibbo angle.
If we take $|\epsilon_1|=0.1$ (see $\tau=i +\epsilon$ and  $\epsilon_1=2.05\,i\,\epsilon$ in Eq.\,(\ref{epS120})), both $\theta_{12}^{u(d)}$ are  approximately $0.22$. Thus, the magnitude of Cabibbo angle is easily reproduced by taking   the relevant phase of $\epsilon$.
Indeed,  the observed CKM elements are  reproduced at
  $\tau\simeq i+(0.05$--$0.09)\,e^{i\phi}$
  % ($\phi\simeq 0$--$2 \pi$)
with relevant $\phi$ as numerically  discussed in section 7.

%%%%%%%%%%%%%%%%%%%%%%%%%%%%
%%%%%%%%%% II %%%%%%%%%%%%%%
%%%%%%%%%%%%%%%%%%%%%%%%%%%%
\subsubsection{Quark mass matrix I\hspace{-.01em}I at $\tau=i$}

Let us discuss the quark mass matrix {\rm I\hspace{-.01em}I}
in Eq.\,(\ref{matrixI}) at fixed points of $\tau$ by using modular forms in Table 1.
At $\tau= i$, both up-type and down-type  quark mass matrices  are given in the base of $S$ of
Eq.\,(\ref{STbase}) as:
%%%%%%%%%%%%%%%%%%%%%%%%%%%%%%%
\begin{align}
\begin{aligned}
M_q=&
\begin{pmatrix}
\tilde \alpha_q & 0 & 0 \\
0 &\tilde \beta_q & 0\\
0 & 0 &\tilde\gamma_q
\end{pmatrix} \times \\
& \begin{pmatrix}
2\sqrt{3}-3+g_q (7\sqrt{3}-12) & 12-7\sqrt{3}+g_q (9-5\sqrt{3})& 5\sqrt{3}-9+g_q (3-2\sqrt{3}) \\
1 &1 &  1\\
-2+\sqrt{3} & 1-\sqrt{3}& 1
\end{pmatrix}\, ,
\end{aligned}
\end{align}
where $\tilde \alpha_q=3 v_qY_0^3 \alpha_q$,
$\tilde \beta_q= (6-3\sqrt{3}) v_qY_0^2 \beta_q$ and 
$\tilde \gamma_q =v_qY_0 \gamma_q \ (q=u,d)$.
%%%%%%%%%%%%%%%%%%%%%%%%%%%%%%%%%%%%%%%%%%

%%%%%%%%%%%%%%%%%%%%%%%%%%%%%%%%%%%%%%%%%%
Let us move them to the diagonal base  of $S$.
%Since $ M_q^\dagger M_q$ commutes with $S$ due to $Z_2$ symmetry,
%$ M_q^\dagger M_q$ is expected to be diagonal in the diagonal basis
%of $S$.  However, two eigenvalues of $S$ is degenerate,
%there is a freedom of the  arbitrary rotaion in the 
%$ M_q^\dagger M_q$.
By using the unitary transformation of Eq.\,(\ref{Sdiagonal}), $V_{S3}$,
the matrix $M_q^\dagger M_q$ is transformed  as  $(M_q V_{S3})^\dagger M_q V_{S3}$.
Then, we have
\begin{align}
{\cal M}_q^{2(0)}&\equiv  V_{S3} M_q^\dagger M_q V_{S3}^\dagger \nonumber\\
&= \frac{3}{2}
\begin{aligned}
\begin{pmatrix}
A\tilde\gamma_q^2+3(A+ B_{1q}+ |g_q|^2 C) \tilde\alpha_q^2&  
-[D\tilde\gamma_q^2 +3 (B_{2q}+Ag_q+ Cg_q^*) \tilde\alpha_q^2) ] & 0\\
-[ D\tilde\gamma_q^2 +3 (B_{2q}+Ag_q^*+ Cg_q) \tilde\alpha_q^2) ] &
\tilde\gamma_q^2 +3(C+ B_{1q}+ |g_q|^2 A) \tilde\alpha_q^2 & 0 \\
0 & 0 &2\tilde\beta^2 \\
\end{pmatrix}
\end{aligned}\,,
\label{quark-I}
\end{align}
with
\begin{align}
&\begin{aligned}
A=7-4\sqrt{3}\, , \quad B=26-15 \sqrt{3} \, , \quad
C=97-56\sqrt{3} \, , \quad
D=2- \sqrt{3} \, , 
\end{aligned}\nonumber\\
&\begin{aligned}
B_{1q}=B (g_q+g_q^*)=2 B \,  {\rm Re}[g_q] , \qquad  B_{2q}=B \,(1+|g_q|^2) \, ,
\quad A^2=C\,, \quad D^2=A\,, \quad A+C=4B\,,
\end{aligned}
\label{ABCDq}
\end{align}
where $A$, $B$, $C$ and $D$ in Eq.\,(\ref{ABCD}) are again presented for convenience.
%%%%%%%%%%%%%%%%%%%%%%%%%%%%%%%
The mass eigenvalues satisfy:
\begin{align} 
   m_{q1}^2 m_{q2}^2=81 C\, \tilde\alpha_q^2 \tilde\gamma_q^2 , \qquad
m_{q1}^2+ m_{q2}^2=6 D\,  \tilde\gamma_q^2+
9B\, (2+2{\rm Re}[g_q]+|g_q|^2) \tilde\alpha_q^2, \qquad
 m_{q3}^2=3 \tilde\beta_q^2 \ .  \label{mass2Q}  
\end{align} 
%%%%%%%%%%%%%%%%%%%%%%%%%%%%%%%

The mixing angle between 1st- and 2nd-family, $\theta_{12}^q$,
is given as:
\begin{align}
\tan 2\theta_{12}^q =-
\frac{\sqrt{[D\tilde\gamma_q^2 +3 (B_2+E_q)\tilde\alpha_q^2)]^2+9 F_q^2 \tilde\alpha_q^4}}{(2\sqrt{3}-3)\tilde\gamma_q^2
	+3(45-26\sqrt{3})(1-|g_q|^2)\tilde\alpha_q^2} \,,
\label{theta12q}
\end{align}
where 
\begin{align}
E_q=(A+C){\rm Re}[g_q]=(104-60\sqrt{3}){\rm Re}[g_q] \, ,
\quad F_q=(A-C)\,{\rm Im}[g_q]
=(52\sqrt{3}-90)\,{\rm Im}[g_q] \, . 
\label{EF}
\end{align}
Neglecting the imaginary part of $g_q$ ($g_q={\rm Re}[g_q]$), it is simply given
as:
\begin{align}
\tan 2\theta_{12}^q =
-\frac{1}{\sqrt{3}}\ \frac{\tilde\gamma_q^2+3(7-4\sqrt{3})(1+4g_q+g_q^2)\tilde\alpha_q^2}{\tilde\gamma_q^2-3(7-4\sqrt{3})(1-g_q^2)\tilde\alpha_q^2} \,. 
\label{limit-theta12q}
\end{align}
where $|g_q|$ is supposed to be ${\cal O}(1)$.
We take $ \tilde\alpha_q^2, \tilde\gamma_q^2 \ll \tilde\beta_q^2 $ due to the
mass hierarchy of quark masses.
There are two possible choices of 
$\tilde\alpha_q^2 \ll \tilde\gamma_q^2$
and  $\tilde\gamma_q^2 \ll \tilde\alpha_q^2$.

In the case of  $\tilde\alpha_q^2 \ll \tilde\gamma_q^2$,
\begin{align}
\tan 2\theta_{12}^q \simeq 
-\frac{1}{\sqrt{3}}\ [1+6(7-4\sqrt{3})(1+2g_q)\frac{\tilde\alpha_q^2}{\tilde\gamma_q^2}]
\simeq  -\frac{1}{\sqrt{3}}\ ,
\label{appro-theta12q}
\end{align}
which gives 
 $\theta_{12}^q=-15^\circ$ at the limit of $\tilde\alpha_q^2/\tilde\gamma_q^2=0$.
This  is common for both up-quark and down-quark mass matrices
because  it is independent of $g_q$.
Then, the flavor mixing (CKM) between 1st- and 2nd-family
vanishes due to the cancellation between up-quarks and down-quarks.

%%%%%%%%%%%%%%%%%%%%%%%%%%%%%%%%%
On the other hand,
in the case of  $\tilde\gamma_q^2 \ll \tilde\alpha_q^2$, we obtain
\begin{align}
\tan 2\theta_{12}^q \simeq 
\frac{1}{\sqrt{3}}\, \frac{1+4 g_q+g_q^2}{1-g_q^2} \, , 
\label{appro-theta12qx}
\end{align} 
where the imaginary part of $g_q$ and
terms of  $\tilde\gamma_q^2$ are neglected.
The Cabibbo angle  could be reproduced by choosing relevant values of
$g_d$ and $g_u$ of order one.
%%%%%%%%%%%%%%%%%%%%%%%%%%%%%%%%%
%%%%%%%%%%%%%%%%%%%%%%%%%%%%%%%%%%%%%%%%%%%%%%%
However, the CKM matrix elements $V_{cb}$ and $V_{ub}$ vanish at $\tau=i$.
In order to obtain desirable CKM matrix, $\tau$
should be deviated from $i$ a little bit.

%%%%%%%%%%%%%%%%%%%%%%%%%%%%
%%%%%%%%%%%%%%%%%%%%%%%%%%%%
\subsubsection{Quark mass matrix I\hspace{-.01em}I  at nearby $\tau=i$}

By using  modular forms  of weight 2, 4 and 6
%in  Eqs.\,(\ref{epS12}), (\ref{epS4A}) and (\ref{epS666})
  in  Appendix C.1, we obtain
the deviation from  ${\cal M}_q^{2(0)}$ in  Eq.\,(\ref{quark-I}).
Then, 
the additional contribution  ${\cal M}_q^{2(1)}$ to ${\cal M}_q^{2(1)}$ of Eq.\,(\ref{quark-I})
of order $\epsilon$ is:
%%%%%%%%%%%%%%%%%%%%%%%%%%%%%%%%%%%%%%%%%%%%%%%%%%%
\begin{align}
\begin{aligned}
{\cal M}_q^{2(1)}\simeq 
\begin{pmatrix}
{\cal O}(\tilde\alpha_q^2,\,\tilde\gamma_q^2,\epsilon_1,\epsilon_2)  & 
{\cal O}(\tilde\alpha_q^2,\,\tilde\gamma_q^2,,\epsilon_1,\epsilon_2)&
\frac{\tilde\beta_q^2}{\sqrt{2}} [(\sqrt{3}-1)\epsilon_1^* +(2-\sqrt{3})\epsilon_2^*]\\ 
{\cal O}(\tilde\alpha_q^2,\,\tilde\gamma_q^2,\epsilon_1,\epsilon_2)& 	{\cal O}(\tilde\alpha_q^2,\,\tilde\gamma_q^2,\epsilon_1,\epsilon_2)  &  
\frac{\tilde\beta_q^2}{\sqrt{6}} [(3+\sqrt{3})\epsilon_1^* +\sqrt{3}\epsilon_2^*] \\ 
\frac{\tilde\beta_q^2}{\sqrt{2}} [(\sqrt{3}-1)\epsilon_1 +(2-\sqrt{3})\epsilon_2]
& \frac{\tilde\beta_q^2}{\sqrt{6}} [(3+\sqrt{3})\epsilon_1 +\sqrt{3}\epsilon_2]
& \tilde\beta_q^2 [4{\rm Re}(\epsilon_1) + 
2(2-\sqrt{3}){\rm Re}(\epsilon_2)]
\end{pmatrix} ,
\end{aligned}
\label{M2Q1st}
\end{align}
where ${\cal O}(\tilde\alpha_q^2,\,\tilde\gamma_q^2,\epsilon_1,\epsilon_2)$ terms are highly suppressed compared with elements (1,3),\,(3,1),\,
(2,3),\,(3,2),\,(3.3) due to $\tilde\beta_q^2 \gg \tilde\alpha_q^2,\, \tilde\gamma_q^2$.
Therefore,
the 2nd- and 3rd-family mixing angle $\theta_{23}^q$ is given as:
\begin{align}
\theta_{23}^q\simeq \frac{ \frac{1}{\sqrt{6}}\tilde\beta_q^2 |(3+\sqrt{3})\epsilon_1^* +\sqrt{3}\epsilon_2^*|}{3\tilde\beta_q^2}
=\frac{3+\sqrt{3}}{\sqrt{6}} \, |\epsilon_1^*|\simeq 2.23 \,|\epsilon^*| \,,
\end{align}
and the 1st- and 3rd-family mixing angle $\theta_{13}^q$ is:
\begin{align}
\theta_{13}^q\simeq \frac{\frac{1}{\sqrt{2}}\tilde\beta_q^2|(\sqrt{3}-1)\epsilon_1^* +(2-\sqrt{3})\epsilon_2^*|}{3\tilde\beta_q^2}
=\frac{3-\sqrt{3}}{3\sqrt{2}} \, |\epsilon_1^*|\simeq 0.613 \,|\epsilon^*|\,,
\end{align}
where $3\tilde\beta_q^2$  in the denominators 
is the (3,\,3) element of Eq.\,(\ref{quark-I}), and 
$\epsilon_2=2\epsilon_1=4.10\,i\,\epsilon$ of  Eq.\,(\ref{epS120}) is used.
The ratio $\theta_{13}^q/\theta_{23}^q\simeq 0.27$ is rather large compared
with observed CKM ratio $|V_{ub}/V_{cb}|\simeq 0.08$.
This rather large  $\theta^q_{13}$ spoils to reproduce
observed CKM elements 
$V_{cb}$ and  $V_{ub}$ at the nearby  fixed point $\tau=i$.

% For the case of I\hspace{-.01em}I of Eq.(\ref{quark-II}) and 
%  I\hspace{-.15em}I\hspace{-.15em}I of Eq.(\ref{quark-III}),
%%%%%%%%%%%%%%%%%%%%%%%%%%%%%%%%%%%%%%%%% 

%%%%%%%%%%%%%%%%%%%%%%%%%%%%%%%%%%%%%%%%%%%%%%%%%%%%%%%%%%%%%%%%%%%%%%%%%
%%%%%%%%%%%%%%%%%%%%%%%%%%%%%%%%%%%%%%%%%%%%%%%%%%%%%%%%%%%%%%%%%%%%%%%%%
\subsection{Quark mass matrix  at the fixed point of $\tau=\omega$}
%%%%%%%%%%%%%%%%%%%%%%%%%%%%%%%%%%%%%%%%%%%%%%%
%%%%%%%%%%%%%%%%%%%%%%%%% I %%%%%%%%%%%%%%%%%%%
%%%%%%%%%%%%%%%%%%%%%%%%%%%%%%%%%%%%%%%%%%%%%%%
\subsubsection{Quark mass matrix I at  $\tau=\omega$}

In the quark mass matrix I
of Eq.\,(\ref{matrixIII}), the up-type and down-type mass matrices  
are given at   $\tau=\omega$ by using modular forms in Table 1:
%%%%%%%%%%%%%%%%%%%%%%%%%%%%%%%
\begin{align}
\begin{aligned}
&M_u=
\begin{pmatrix}
-g_u\tilde \alpha_q & 0 & 0 \\
0 &-g_u\tilde \beta_q & 0\\
0 & 0 &-g_u\tilde\gamma_q
\end{pmatrix} 
\begin{pmatrix}
1& -2\omega^2 & -2\omega \\
-2\omega& 1& -2\omega^2  \\
-2\omega^2& -2\omega& 1\\
\end{pmatrix}\, ,\\
&M_d=
\begin{pmatrix}
\tilde \alpha_d & 0 & 0 \\
0 &\tilde \beta_d & 0\\
0 & 0 &\tilde\gamma_d
\end{pmatrix} 
\begin{pmatrix}
1& -\frac{1}{2}\omega^2 & \omega \\
\omega& 1& -\frac{1}{2}\omega^2  \\
-\frac{1}{2}\omega^2 &\omega& 1 \\
\end{pmatrix}\, ,
\end{aligned}
\label{QuarkI-ST}
\end{align}
where
$\tilde \alpha_u= (9/8) v_u Y_0^3 \alpha_q$,
$\tilde \beta_u=  (9/8) v_uY_0^3  \beta_q$ and
$\tilde \gamma_u =(9/8) v_uY_0^3 \gamma_q$ for up-type quarks, and 
$\tilde \alpha_d=  v_d Y_0 \alpha_d$,
$\tilde \beta_d=   v_d Y_0  \beta_d$ and 
$\tilde \gamma_d = v_d Y_0 \gamma_d$ for down-type quarks, respectively.
%%%%%%%%%%%%%%%%%%%%%%%%%%%%%%%%%%%%%%%%%%%%%%%%%%%%%%%%%%%%%%%%%%%%%%%%%
By using the unitary transformation of Eq.\,(\ref{STdiagonal}), $V_{ST4}$,
the mass matrix $M_u^\dagger M_u$ is transformed  as:
%%%%%%%%%%%%%%%%%%%%%%%%%%%%
%%%%%%%%%%%%%%%%%%%%%%%%%%%%
\begin{align}
{\cal M}_u^{2(0)}\equiv  V_{ST4} M_q^\dagger M_q V_{ST4}^\dagger 
= 9\,
\begin{aligned}
\begin{pmatrix}
|g_{u2}|^2\tilde\beta_u^2&0 & 0 \\
0& |g_{u1}|^2\tilde\alpha_u^2  &   0  \\
0 & 0 & |g_{u3}|^2\tilde\gamma_u^2 \\
\end{pmatrix}
\end{aligned}\,.
\label{upquarkST-III}
\end{align}
The mass matrix $M_d^\dagger M_d$ is transformed  as:
\begin{align}
{\cal M}_d^{2(0)}\equiv  V_{ST4} M_d^\dagger M_d V_{ST4}^\dagger 
= \frac{9}{4}\,
\begin{aligned}
\begin{pmatrix}
\tilde\alpha_d^2&0 & 0 \\
0&\tilde\gamma_d^2  &   0  \\
0 & 0 & \tilde\beta_d^2 \\
\end{pmatrix}
\end{aligned}\,.
\label{downquarkST-III}
\end{align}
It is remarked that both are diagonal ones.

\subsubsection{Quark mass matrix I at nearby $\tau=\omega$}

% I\hspace{-.01em}I, I\hspace{-.15em}I\hspace{-.15em}I
%%%%%%%%%%%%%%%%  Model I %%%%%%%%%%%%%%%%%%%%%%%%%
Quark mass matrix I in Eq.\,(\ref{QuarkI-ST}) 
is corrected due to the deviation from the fixed point of $\tau=\omega$.
By using  modular forms  of weight 2, 4 and 6
in  Appendix C.2, we obtain
the deviations from  ${\cal M}_u^{2(0)}$ and  ${\cal M}_d ^{2(0)}$
in Eqs.\,(\ref{upquarkST-III}) and (\ref{downquarkST-III}).
In the diagonal base of $ST$, the corrections are given by only a small variable $\epsilon$
as seen in Eq.\,(\ref{epST0}).
In the 1st order perturbation of $\epsilon_1$,
the corrections  ${\cal M}_u^{2(1)}$ and  ${\cal M}_d ^{2(1)}$ are given as:
\begin{align}
\begin{aligned} {\cal M}_u^{2(1)}=
\begin{pmatrix}
\delta_{u1}& \delta_{u2} & \delta_{u3} \\ 
\delta_{u2}^*& \delta_{u4} & \delta_{u5}\\ 
\delta_{u3^*}& \delta_{u5}^* & \delta_{u6}
\end{pmatrix}\, ,
\end{aligned} \qquad\qquad  
\begin{aligned} {\cal M}_d^{2(1)}=
\begin{pmatrix}
\delta_{d1}& \delta_{d2} & \delta_{d3} \\ 
\delta_{d2}^*& \delta_{d4} & \delta_{d5}\\ 
\delta_{d3^*}& \delta_{d5}^* & \delta_{d6}
\end{pmatrix}
\end{aligned}\,,
\end{align}
where   off diagonal elements  $\delta_{q2}$,
$\delta_{q3}$ and $\delta_{q5}$  are: 
%%%%%%%%%%%%%%%%%%%%%%%%%%%%%%%
\begin{align}
&\begin{aligned}
\delta_{u2}=   2\tilde\beta_u^2 |g_{u2}|^2(2\epsilon_1-\epsilon_2)    -2\tilde\alpha_u^2(2+g_{u1}^*)g_{u1} (\epsilon_1^*+\epsilon_2^*)
= - 6 (2+g_{u1}^*)g_{u1}\,\epsilon_1^*\,\tilde\alpha_u^2 \, ,
\end{aligned}\\
&\begin{aligned}
\delta_{u3}=  2\tilde\beta_u^2(2+g_{u2})g_{u2}^* (\epsilon_1+\epsilon_2)
+  2\tilde\gamma_u^2 |g_{u3}|^2(-2\epsilon_1^*+\epsilon_2^*)
=  6 (2+g_{u2})g_{u2}^*\,\epsilon_1\,\tilde\beta_u^2 \, ,
\end{aligned}\\
&\begin{aligned}
\delta_{u5}=  2\tilde\gamma_u^3(2+g_{u3}^*)g_{u3} (\epsilon_1^*+\epsilon_2^*)
+  2\tilde\alpha_u^2 |g_{u1}|^2(-2\epsilon_1+\epsilon_2)
=  6 (2+g_{u3}^*)g_{u3}\,\epsilon_1^*\,\tilde\gamma_u^2 \, ,
\end{aligned}\\
&\begin{aligned}
\delta_{d2}=   i\,\tilde\alpha_d^2 (\epsilon_1-\frac{1}{2}\epsilon_2)
+\frac{1}{2} i\, \tilde\gamma_d^2(\epsilon_1^*+\epsilon_2^*)
=  \frac{3}{2}\, i\,  \epsilon_1^* \,\tilde\gamma_d^2 \, ,
\end{aligned}\\
&\begin{aligned}
\delta_{d3}=  \frac{1}{2}\, i\,\tilde\alpha_d^2 (\epsilon_1+\epsilon_2)
+ i\, \tilde\beta_d^2(\epsilon_1^*-\frac{1}{2}\epsilon_2^*)
=  \frac{3}{2}\, i\, \epsilon_1 \,\tilde\alpha_d^2 \,  ,
\end{aligned}\\
&\begin{aligned}
\delta_{d5}=  -\frac{1}{2}\, i\,\tilde\beta_d^2 (\epsilon_1^*+\epsilon_2^*)
- i\, \tilde\gamma_d^2(\epsilon_1-\frac{1}{2}\epsilon_2)
=  -\frac{3}{2}\, i\, \epsilon_1^*\, \tilde\beta_d^2 \, .
\end{aligned}
\end{align}
In last  equalities, $\epsilon_2=2\epsilon_1$ of Eq.\,(\ref{epST120}) is used. 

Taking account of  $\tilde\gamma_u^2\gg\tilde\alpha_u^2\gg \tilde\beta_u^2$
and $\tilde\beta_d^2\gg \tilde\gamma_d^2\gg \tilde\alpha_d^2$
as seen in Eqs.\,(\ref{upquarkST-III}) and (\ref{downquarkST-III}),
 mixing angles $\theta_{12}^q$ and  $\theta_{23}^q$ are given as:
 \begin{align}
 \theta_{12}^u\simeq 
\frac{2}{3}|(2+g_{u1}^*)g_{u1}\,\epsilon_1^*|\,, 
 \qquad  \theta_{23}^u\simeq 
 \frac{2}{3}|(2+g_{u3}^*)g_{u3}\,\epsilon_1^*|\,, 
 \qquad \theta_{12}^d\simeq \theta_{23}^d \simeq
 \frac{2}{3}|\epsilon_1^*|\,,
 \end{align}
respectively,
 while both $\theta_{13}^q\,(q=u,d)$ are
highly suppressed.

 Since up-type quark mixing angles depend on the magnitudes of $g_{u1}$
 and  $g_{u3}$, the magnitudes of CKM matrix elements $V_{us}$ and $V_{cb}$ could be reproduced
 by choosing relevant $g_{u1}$ and $g_{u3}$.
 For example,  we can take $\theta_{12}^u\sim \lambda$ 
 and $\theta_{23}^u\sim\theta_{12}^d\sim \theta_{23}^d\sim  \lambda^2$,
 where $\lambda\simeq 0.2$ is put to reproduce observed $|V_{us}|$, $|V_{cb}|$ and $|V_{ub}|$.
 However, this scheme leads to  $|V_{td}|\sim \lambda^4$, which is much smaller than the observed one.
 Indeed,  the observed   $|V_{td}|$ is not reproduced 
 at nearby $\tau=\omega$  in  section 7.
 
%%%%%%%%%%%%%%%%%%%%%%%%%%%%%%%%%%%%%%%%%%%%%%%
%%%%%%%%%%%%%%%%%%%%%%%%% II %%%%%%%%%%%%%%%%%%
%%%%%%%%%%%%%%%%%%%%%%%%%%%%%%%%%%%%%%%%%%%%%%%
\subsubsection{Quark mass matrix I\hspace{-.01em}I at  $\tau=\omega$}

We discuss the quark mass matrix I\hspace{-.01em}I at the fixed point  $\tau=\omega$
by using modular forms in Table 1.
In the base of $S$ and $T$ of Eq.\,(\ref{STbase}),
it is given at the fixed point  $\tau=\omega$:
%%%%%%%%%%%%%%%%%%%%%%%%%%%%%%%
\begin{align}
\begin{aligned}
M_q=
\begin{pmatrix}
-g_q\tilde \alpha_q  & 0 & 0 \\
0 &\tilde \beta_q & 0\\
0 & 0 &\tilde\gamma_q
\end{pmatrix} 
\begin{pmatrix}
1& -2\omega^2 & -2\omega \\
-\frac{1}{2}\omega &1 &  \omega^2\\
-\frac{1}{2}\omega^2 & \omega& 1
\end{pmatrix}\, ,
\end{aligned}
\label{quarkST-II}
\end{align}
where $\tilde \alpha_q=  (9/8)v_qY_0^3  \alpha_q$,
$\tilde \beta_q= \frac{3}{2} v_q Y_0^2 \beta_q$ and 
$\tilde \gamma_q =v_qY_0 \gamma_q$.
%%%%%%%%%%%%%%%%%%%%%%%%%%%%%%%%%%%%%%%%%%
By using the unitary transformation of Eq.\,(\ref{STdiagonal}), $V_{ST5}$,
the mass matrix $M_q^\dagger M_q$ is transformed  as:
\begin{align}
{\cal M}_q^{2(0)}\equiv  V_{ST5} M_q^\dagger M_q V_{ST5}^\dagger 
= \frac{9}{4}\,
\begin{aligned}
\begin{pmatrix}
0&0 & 0 \\
0&0 &   0  \\
0 & 0 & 4g_q^2\tilde\alpha_q^2+\tilde\beta_q^2+\tilde\gamma_q^2 \\
\end{pmatrix}
\end{aligned}\,,
\label{quarkST-I}
\end{align}
which gives two massless quarks.
Therefore, it seems very difficult to reproduce observed quark masses
and CKM elements even if we shift   $\tau$  from $\tau=\omega$ a little bit
and choose relevant $g_q$.

%%%%%%%%%%%%%%%%%%%%%%%%%%%%%%%%%%%%%%%%%%%%%%%
%%%%%%%%%%%%%%%%%%%%%%%%%%%%%%%%%%%%%%%%%%%%%%%
%%%%%%%%%%%%%%%%%%%%%%%%%%%%%%%%%%%%%%%%%%%%%%%%%%%%%%
\subsubsection{Quark mass matrix I\hspace{-.01em}I at nearby  $\tau=\omega$}    
%%%%%%%%%%%%%%%%%%%%%%%%%%%%%%%%%%%%%%%%%%%%%%%%%%%%%%%
%%%%%%%%%%%%%%  Model III %%%%%%%%%%%%%%%%%%%%%%%%%%%%%%%
%%%%%%%%%%%%%%%%%%%%%%%%%%%%%%%%%%%%%%%%%%%%%%%%%%%%%%%
Quark mass matrix I\hspace{-.01em}I in Eq.\,(\ref{quarkST-II}) 
is corrected due to the deviation from the fixed point of $\tau=\omega$.
By using  modular forms  of weight 2, 4 and 6
in  Appendix C.2, we obtain
the deviation from  ${\cal M}_q^{2(0)}$ 
in Eq.\,(\ref{quarkST-I}).
In the diagonal base of $ST$, the correction is given by only a small variable $\epsilon$ as seen in Eq.\,(\ref{epST0}).
In the 1st order approximation of $\epsilon_i$,
the correction  ${\cal M}_q^{2(1)}$ is given as:
\begin{align}
\begin{aligned} {\cal M}_q^{2(1)}=
\begin{pmatrix}
0& 0 & \delta_{q3} \\ 
0 & 0 & \delta_{q5} \\ 
\delta_{q3}^* & \delta_{q5}^* & \delta_{q6} 
\end{pmatrix}
\end{aligned}\,,
\label{quarkST-I1}
\end{align}
where $\delta_{qi}$ are given in terms of
$\epsilon$, $g_q$, $\tilde \alpha_q^2$, $\tilde \beta_q^2$ and 
$\tilde \gamma_q^2$.
In order to estimate the flavor mixing anles, we present relevant  $\delta_{qi}$ as: 
%%%%%%%%%%%%%%%%%%%%%%%%%%%%%%%
\begin{align}
&  \begin{aligned}
\delta_{q 3}&= -2\tilde\alpha_q^2g_q(2+g_q^*) (\epsilon_1^*+\epsilon_2^*)
+ \frac{1}{6}\tilde\beta_q^2(\epsilon_1^*-8\epsilon_2^*) +\frac{1}{2} i\,\tilde\gamma_q^2(\epsilon_1^*+\epsilon_2^*) \\
&\simeq   -6\tilde\alpha_q^2g_q(2+g_q^*) \epsilon_1^*
-\frac{5}{2}\tilde\beta_q^2\epsilon_1^* +\frac{3}{2} i\,\tilde\gamma_q^2\epsilon_1^* \,,
\end{aligned} \nonumber \\
&\begin{aligned}
\delta_{q 5}=  \tilde\alpha_q^2|g_q|^2 (-4\epsilon_1^*+2\epsilon_2^*)
+ \tilde\beta_q^2(\frac{1}{3}\epsilon_1^*-\frac{7}{6}\epsilon_2^*) +i\,\tilde\gamma_q^2(\epsilon_1^*-\frac{1}{2}\epsilon_2^*)
\simeq -2 \tilde\beta_q^2 \epsilon_1^*\, ,
\end{aligned}
\end{align}
where $\epsilon_2=2\epsilon_1$ of Eq.\,(\ref{epST120})  is used in last approximate equalities. 
By using  Eqs.\,(\ref{quarkST-I}) and (\ref{quarkST-I1}),
we obtain ${\rm Det}[{\cal M}_Q^{2(0)}+{\cal M}_Q^{2(1)}]=0$.
Therefore, it is impossible to reproduce observed quark masses
at nearby $\tau=\omega$  in the 1st order  perturbation  of $\epsilon$.
Indeed, this model cannot reproduce the observed CKM elements at nearby $\tau=\omega$ in section 7.

%%%%%%%%%%%%%%%%%%%%%%%%%%%%%%%%%%%%%%%%%%%%%%%%%%%%%%%%%%%%%%%%%%%
\subsection{Quark mass matrix at  $\tau= i\infty$}
%%%%%%%%%%%%%%%%%%%%%%%%%%%%%%%
\subsubsection{Quark mass matrix I and  I\hspace{-.01em}I at $\tau= i\infty$}
%%%%%%%%%%%%%%%%%%%%%%%%%%%%%%%

The mass matrices of  I and  I\hspace{-.01em}I 
in Eqs.\,(\ref{matrixI}) and (\ref{matrixIII})  
are simply  given by using modular forms in Table 1  at   $\tau=i\infty$
 since the  modular forms of weight $2$, $4$ and $6$  are same.
 Those are both diagonal ones as follows:
%%%%%%%%%%%%%%%%%%%%%%%%%%%%%%%
\begin{align}
\begin{aligned}
M_q=
\begin{pmatrix}
\tilde \alpha_q & 0 & 0 \\
0 &\tilde \beta_q & 0\\
0 & 0 &\tilde\gamma_q
\end{pmatrix} \, ,
\end{aligned}
\label{quarkT-I}
\end{align}
where 
$\tilde \alpha_u=  v_uY_0^3 \alpha_q$,
$\tilde \beta_u= v_u Y_0^3 \beta_u$,
$\tilde \gamma_u =v_uY_0^3 \gamma_u$,
$\tilde \alpha_d=  v_dY_0 \alpha_d$,
$\tilde \beta_d= v_d Y_0 \beta_d$ and
$\tilde \gamma_d =v_dY_0 \gamma_d$
for quark mass matrix  I,
and $\tilde \alpha_q=  v_qY_0^3 \alpha_q$,
$\tilde \beta_q= v_q Y_0^2 \beta_q$ and 
$\tilde \gamma_q =v_qY_0 \gamma_q$ 
for quark mass matrix  I\hspace{-.01em}I.

%%%%%%%%%%%%%%%%%%%%%%%%%%%%%%%%%%%%%%%%%%
In the diagonal base of $T$ of Eq.\,(\ref{STbase}),
the mass matrix $M_q^\dagger M_q$ is given as:
\begin{align}
{\cal M}_q^{2(0)}\equiv   M_q^\dagger M_q 
= 
\begin{aligned}
\begin{pmatrix}
\tilde\alpha_q^2&0 & 0 \\
0&\tilde\beta_q^2 &   0  \\
0 & 0 & \tilde\gamma_q^2 \\
\end{pmatrix}
\end{aligned}\,.
\label{quarknewT}
\end{align}
 Mixing angles appear through the finite effect of ${\rm Im }\,[\tau]$.

%%%%%%%%%%%%%%%%%%%%%%%%%%%%%%%%%%%%%%%%%%%%%%%%%%%%%%%
%%%%%%%%%%%%%%%%%%%%%%%%%%%%%%%%%%%%%%%%%%%%%%%%%%%%%%% 
%%%%%%%%%%%%%%%%%%%%%%%%%%%%%%%%%%%%%%%%%%%%%%%%%%%%%%%%
%%%%%%%%%%%%%%%%%%%%%%%%%%%  I %%%%%%%%%%%%%%%%%%%%%%%%%
%%%%%%%%%%%%%%%%%%%%%%%%%%%%%%%%%%%%%%%%%%%%%%%%%%%%%%%%
\subsubsection{Quark mass matrix I towards  $\tau=i\infty$}   
Quark mass matrix I in Eq.\,(\ref{quarkT-I}) 
is corrected due to the finite effect of  $\tau=i\infty$.
By using modular forms of Eqs.\,(\ref{epT12}), (\ref{epT4})
and (\ref{epT666}) in Appendix C.3, we obtain
the deviation from  ${\cal M}_q^{2(0)}$ in  Eq.\,(\ref{quarknewT})
for the quark mass matrix I.
We present the first order corrections ${\cal M}_q^{2(1)}$ for up-type quarks and down-type quarks  to  ${\cal M}_q^{2(0)}$ of Eq.\,(\ref{quarknewT}), respectively\,:
%%%%%%%%%%%%%%%%%%%%%%%%%%%%%%%%%%%%%%%%%%%%%%%%%%%
\begin{align}
\begin{aligned}
&{\cal M}_u^{2(1)}\simeq  
\begin{pmatrix}
0 & (1+2g_{u2}^*)\tilde\beta_u^2\,  \delta^*&
(1+2\,g_{u1})\,\tilde\alpha_u^2\, \delta\\ 
(1+2g_{u2})\tilde\beta_u^2\,  \delta&  0 &  
(1+2\,g_{u3}^*)\,\tilde\gamma_u^2 \delta^*\\ 
(1+2\,g_{u1}^*)\,\tilde\alpha_u^2\, \delta^*
& (1+2\,g_{u3})\,\tilde\gamma_u^2 \delta
&0
\end{pmatrix} , \\
\\
&{\cal M}_d^{2(1)}\simeq  
\begin{pmatrix}
0 &  \tilde\beta_d^2\, \delta^*&
\tilde\alpha_d^2\, \delta\\ 
\tilde\beta_d^2\, \delta&  0 &  
\tilde\gamma_d^2\,  \delta^*\\ 
\tilde\alpha_d^2\, \delta^*
& \tilde\gamma_d^2\,  \delta
&0
\end{pmatrix} \, ,
\end{aligned}
\label{M2Q1T}
\end{align}
where $\delta$ is given in Eq.(\ref{deltaT}).
%$\delta=-6 p^{\frac{1}{3}}\,\epsilon^{\frac{1}{3}}$.
We obtain mixing angles as:
\begin{align}
\theta_{12}^u\simeq 
|(1+2\,g_{u2}^*)\,\delta^*|\,, 
\qquad \theta_{23}^u
\simeq |(1+2\,g_{u3}^*)\,\delta^*| \,,
\qquad 
\theta_{12}^d\simeq \theta_{23}^d\simeq 
|\delta^*|\,,
\end{align}
respectively.
The   1st- and 3rd-family mixing angle $\theta_{13}^q$ is  suppressed due to the factor $\tilde\alpha_q^2/\tilde\gamma_q^2$ for both up- and down-type quarks.
Since $\theta_{12}^u$ and $\theta_{23}^u$ depend on the magnitudes of $g_{u2}$
and  $g_{u3}$, the CKM matrix elements $V_{us}$ and $V_{cb}$ could be reproduced
by choosing relevant $g_{u2}$ and $g_{u3}$.
For example,  we can take $\theta_{12}^u\sim \lambda$ 
and $\theta_{23}^u\sim\theta_{12}^d\sim \theta_{23}^d\sim  \lambda^2$,
where $\lambda\simeq 0.2$ to reproduce observed $|V_{us}|$, $|V_{cb}|$ and $|V_{ub}|$.
However, this scheme leads to  $|V_{td}|\sim \lambda^4$, which is much smaller than the observed one.
Indeed,  the successful CKM matrix elements are not reproduced 
at large ${\rm Im}\tau$ in the numerical results of section 7.

%%%%%%%%%%%%%%%%%%%%%%%%%%%%%%%%%%%%%%%%%%%%%%%%%%%%%%%%%%%%%%%%%%%%%%
%%%%%%%%%%%%%%%%%%%%%%% Quark mass matrix-II %%%%%%%%%%%%%%%%%%%%%%%%%
%%%%%%%%%%%%%%%%%%%%%%%%%%%%%%%%%%%%%%%%%%%%%%%%%%%%%%%%%%%%%%%%%%%%%%
\subsubsection{Quark mass matrix I\hspace{-.01em}I towards  $\tau=i\infty$}    
Quark mass matrix I\hspace{-.01em}I in Eq.\,(\ref{quarkT-I}) 
is corrected due to the finite effect of  $\tau=i\infty$.
By using modular forms of Eqs\,.(\ref{epT12}), (\ref{epT4})
and (\ref{epT666}) in Appendix C.3, we obtain
the deviation from  ${\cal M}_q^{2(0)}$ in  Eq.\,(\ref{quarknewT})
for the quark mass matrix I\hspace{-.01em}I.
The  first order correction  ${\cal M}_q^{2(1)}$ to ${\cal M}_q^{2(0)}$ of Eq.\,(\ref{quarknewT}) is given as\,:
%%%%%%%%%%%%%%%%%%%%%%%%%%%%%%%%%%%%%%%%%%%%%%%%%%%
\begin{align}
\begin{aligned}
{\cal M}_q^{2(1)}\simeq  
\begin{pmatrix}
0 &  -\delta^*\,\tilde\beta_q^2&
(1+2\,g_q)\,\delta^*\,\tilde\alpha_q^2\\ 
-\delta\,\tilde\beta_q^2&  0 &  
\delta^*\,\tilde\gamma_q^2\\ 
(1+2\,g_q^*)\,\delta\,\tilde\alpha_q^2
& \delta\,\tilde\gamma_q^2
&0
\end{pmatrix} ,
\end{aligned}
\label{M2Q1T}
\end{align}
where  $\tilde\alpha_q^2 \ll \tilde\beta_q^2\ll \tilde\gamma_q^2$.
Therefore,
the   mixing angles $\theta_{12}^q$ and 
$\theta_{23}^q$, are given as:
\begin{align}
\theta_{12}^q\simeq 
\frac{|\delta^*|\,\tilde\beta_q^2}{\tilde\beta_q^2 }
\simeq |\delta^*|\,, \qquad\qquad 
\theta_{23}^q\simeq \frac{|\delta^*|\,\tilde\gamma_q^2}{\tilde\gamma_q^2}
=|\delta^*| \,,
\qquad 
\end{align}
respectively.
On the other hand, 
1st- and 3rd-family mixing angle $\theta_{13}^q$ is highly suppressed
due to the factor $\tilde\alpha_q^2/\tilde\gamma_q^2$.
Since $\theta_{12}^q$ and $\theta_{23}^q$ are the same magnitude
for both up-type and down-type quarks,
it is impossible to reproduce observed CKM mixing angles.

In conclusion of section 6, it is found that the only quark mass matrix I
 works well at nearby $\tau=i$.

%%%%%%%%%%%%%%%%%%%%%%%%%%%%%%%%%%%%%%%%%%%%%%%%%%%%%%%%%
%%%%%%%%%%%%%%%%%%%%%%%%%%%%%%%%%%%%%%%%%%%%%%%%%%%%%%%%%
%%%%%%%%%%%%%%%%%%%%%%%%%%%%%%%%%%%%%%%%%%%%%%%%%%%%%%%%%
%%%%%%%%%%%%%%%%%%%%%%%%%%%%%%%%%%%%%%%%%%%%%%%%%%%%%%%%%
%%%%%%%%%%%%%%%%%%%%%%%%%%%%%%%%%%%%%%%%%%%%%%%%%%%%%%%%%
%%%%%%%%%%%%%%%%%%%%%%%%%%%%%%%%%%%%%%%%%%%%%%%%%%%%%%%%%
%%%%%%%%%%%%%%%%%%%%%%%%%%%%%%%%%%%%%%%%%%%%%%%%%%%%%%%%%
%%%%%%%%%%%%%%%%%%%%%%%%%%%%%%%%%%%%%%%%%%%%%%%%%%%%%%%%%
%%%%%%%%%%%%%%%%%%%%%%%%%%%%%%%%%%%%%%%%%%%%%%%%%%%%%%%%%
%%%%%%%%%%%%%%%%%%%%%%%%%%%%%%%%%%%%%%%%%%%%%%%%%%%%%%%%%
%%%%%%%%%%%%%%%%%%%%%%%%%%%%%%%%%%%%%%%%%%%%%%%%%%%%%%%%%
%%%%%%%%%%%%%%%%%%%%%%%%%%%%%%%%%%%%%%%%%%%%%%%%%%%%%%%%%
%%%%%%%%%%%%%%%%%%%%%%%%%%%%%%%%%%%%%%%%%%%%%%%%%%%%%%%%%
%%%%%%%%%%%%%%%%%%%%%%%%%%%%%%%%%%%%%%%%%%%%%%%%%%%%%%%%%
%%%%%%%%%%%%%%%%%%%%%%%%%%%%%%%%%%%%%%%%%%%%%%%%%%%%%%%%%
%%%%%%%%%%%%%%%%%%%%%%%%%%%%%%%%%%%%%%%%%%%%%%%%%%%%%%%%%

\section{Numerical results at nearby fixed points }

We have presented  analytical  discussions of  lepton and quark mass matrices 
at nearby fixed points of modulus.
In this section, we show  numerical results at the nearby fixed points of $\tau=i$, $\tau=\omega$ and $\tau= i \infty$  to confirm above discussions and give predictions.
 
%%%%%%%%%%%%%%%%%%%%%%%%%%%%%%%%%%%%%%%%%%%%%
 \subsection{Frameworks of numerical calculations}

%%%%%%%%%%%%%%%%%%%%%%%%%%%%%%%%%%%%%%%%%%%%%
%%%%%%%%%%%  Lepton input %%%%%%%%%%%%%%%%%%%
%%%%%%%%%%%%%%%%%%%%%%%%%%%%%%%%%%%%%%%%%%%%%
In order to calculate the left-handed flavor mixing of leptons numerically,
we generate random number for  model parameters.
 The modulus $\tau$ is scanned
around fixed points $\tau= i $ and $\tau=\omega$.
It is also scanned ${\rm Im} \tau\geq 1.2$ towards $\tau= i \infty$. 
We keep the parameter sets, in  which the  neutrino experimental data and charged lepton masses are reproduced
within $3\sigma$ interval of error-bars.
We continue this procedure to obtain enough points for plotting allowed region.

%%%%%%%%%%%%%%%%%%%%%%%%%%%%%%%%%%%%%%%%%%%%%%%%%%%%%%%%%%%%
As input of the neutrino data, we take 
 three mixing angles of the PMNS matrix
 and the observed neutrino mass ratio $\Delta m_{\rm sol}^2/\Delta m_{\rm atm}^2$
 with $3\,\sigma$,
 which   are given by NuFit 4.1 in Table 4 \cite{Esteban:2018azc}.
 Since  there are two possible spectrum of neutrinos masses $m_i$, which are
 the normal  hierarchy (NH), $m_3>m_2>m_1$, and the  inverted  hierarchy (IH),
 $m_2>m_1>m_3$, we investigate both cases.
 We also take account of 
 the sum of three neutrino  masses  $\sum m_i$  since
 it is constrained  by the recent cosmological data
 \cite{Tanabashi:2018oca,Vagnozzi:2017ovm,Aghanim:2018eyx}.
 We impose the constraint of the upper-bound $\sum m_i \leq 120$ meV.
 
 Since the modulus $\tau$ obtains the expectation value
 by the breaking of the modular invariance at the high mass scale,
 the observed masses and lepton mixing angles should be taken at the GUT scale
 by the renormalization group equations (RGEs).
 However,  we have not included  the RGE effects
 in the lepton mixing angles and neutrino mass ratio
 $\Delta m_{\rm sol}^2/\Delta m_{\rm atm}^2$ in our numerical calculations.
 We suppose that those corrections  are very small between 
 the electroweak  and GUT scales.
 This assumption is  confirmed  well in the case of $\tan\beta\leq 5$
 unless neutrino masses are almost degenerate \cite{Criado:2018thu}.
 Since we impose 
 the sum of neutrino masses to be smaller  than $120$meV,
 this criterion is satisfied in our analyses.

On the other hand,
 we also take the charged lepton masses
 at the GUT scale $2\times 10^{16}$ GeV with  $\tan\beta=5$ 
 in the framework of the minimal SUSY breaking scenarios
 \cite{Antusch:2013jca, Bjorkeroth:2015ora}:
 \begin{eqnarray}
 y_e=(1.97\pm 0.024) \times 10^{-6}, \quad 
 y_\mu=(4.16\pm 0.050) \times 10^{-4}, \quad 
 y_\tau=(7.07\pm 0.073) \times 10^{-3},
 \end{eqnarray}
 where lepton masses are  given by $m_\ell=y_\ell v_H$ with $v_H=174$ GeV.
 %%%%%%%%%%%%%%%%%%%%%%%%%%%%%%%%%%%%%%%%%%%%%%%%%%%%%%%%%%%%%%%%%%%%%

%%%%%%%%%%%%%%%%%%%%%%%%%%%%%%%%%%%%%%%%%%%
%%%%%%%%%%%%%%%%%%%%%%%%%%%%%%%%%%%%%%%%%%%
\begin{table}[h]
	\begin{center}
		\begin{tabular}{|c|c|c|}
			\hline 
			\rule[14pt]{0pt}{0pt}
			\  observable \ &  $3\,\sigma$ range for NH  & $3\,\sigma$ range for IH \\
			\hline 
			\rule[14pt]{0pt}{0pt}
			%&& \\
			$\Delta m_{\rm atm}^2$& \ \   \ \ $(2.436$--$ 2.618) \times 10^{-3}\,{\rm eV}^2$ \ \ \ \
			&\ \ $- (2.419$--$2.601) \times 10^{-3}\,{\rm eV}^2$ \ \  \\
			%&& \\
			\hline 
			\rule[14pt]{0pt}{0pt}
			%&& \\
			$\Delta m_{\rm sol }^2$& $(6.79$--$ 8.01) \times 10^{-5}\,{\rm eV}^2$
			& $(6.79$--$ 8.01)  \times 10^{-5}\,{\rm eV}^2$ \\
			%&& \\
			\hline 
			\rule[14pt]{0pt}{0pt}
			%&& \\
			$\sin^2\theta_{23}$&  $0.433$--$ 0.609$ & $0.436$--$ 0.610$ \\
			%&& \\
			\hline 
			\rule[14pt]{0pt}{0pt}
			%&& \\
			$\sin^2\theta_{12}$& $0.275$--$ 0.350$ & $0.275$--$ 0.350$ \\
			%&& \\
			\hline 
			\rule[14pt]{0pt}{0pt}
			%&& \\
			$\sin^2\theta_{13}$&$0.02044$--$ 0.02435$ & $0.02064$--$0.02457$ \\
			%&& \\
			\hline 
		\end{tabular}
		\caption{The $3\,\sigma$ ranges of neutrino  parameters from NuFIT 4.1
			for NH and IH 
			\cite{Esteban:2018azc}. 
		}
		\label{DataNufit}
	\end{center}
\end{table}
%%%%%%%%%%%%%%%%%%%%%%%%%%%%%%%%%%
 %%%%%%%%%%%  Quark Inputs  %%%%%%%
 %%%%%%%%%%%%%%%%%%%%%%%%%%%%%%%%%%
 For the quark sector,
  we also adopt numerical values  of Yukawa couplings of quarks
 at the GUT scale $2\times 10^{16}$ GeV with $\tan\beta=5$ 
 in the framework of the minimal SUSY breaking scenarios
 \cite{Antusch:2013jca, Bjorkeroth:2015ora}:
 \begin{align}
 \begin{aligned}
 &y_d=(4.81\pm 1.06) \times 10^{-6}, \quad y_s=(9.52\pm 1.03) \times 10^{-5}, \quad y_b=(6.95\pm 0.175) \times 10^{-3}, \\
 \rule[15pt]{0pt}{1pt}
 &y_u=(2.92\pm 1.81) \times 10^{-6}, \quad y_c=(1.43{\pm 0.100}) \times 10^{-3}, \quad y_t=0.534\pm 0.0341  ~~,
 \end{aligned}\label{yukawa5}
 \end{align}
 which give quark masses as $m_q=y_q v_H$ with $v_H=174$ GeV.
% In our numerical calculation, we input  $2\sigma$ interval for quark masses.
 
 %%%%%%%%%%%%%%%%%%%%%%%%%%%%%%%%%%%%%%%%%%%%%%%%%%%%%%%%%%%%%%%%%%%%%%%%
 We also use the following CKM mixing angles   at the GUT scale $2\times 10^{16}$ GeV with $\tan\beta=5$ \cite{Antusch:2013jca, Bjorkeroth:2015ora}:
 \begin{align}
 \begin{aligned}
 &\theta_{12}^{\rm CKM}=13.027^\circ\pm 0.0814^\circ ~ , \qquad
 \theta_{23}^{\rm CKM}=2.054^\circ\pm 0.384^\circ ~ ,  \qquad
 \theta_{13}^{\rm CKM}=0.1802^\circ\pm 0.0281^\circ~ .
 \end{aligned}\label{CKM}
 \end{align}
 %%%%%%%%%%%%%%%%%%%%%%%%%%%%%%%%%%%%%%%%%%%%%%
 Here $\theta_{ij}^{\rm CKM}$ is given in the PDG notation
 of the CKM matrix  $V_{\rm CKM}$ \cite{Tanabashi:2018oca}.
 In addition, we impose the recent data of LHCb \cite{Tanabashi:2018oca}:
  \begin{align}
 \left |\frac{V_{ub}}{V_{cb}}\right |=0.079\pm 0.006 \, ,
 \label{Vratio}
 \end{align}
 where  $V_{ij}$'s  are CKM matrix elements.
 This ratio is  stable against radiative corrections.
 The observed CP violating phase is given at the GUT scale as:
 \begin{equation}
 \delta_{\rm CP}^{\rm CKM}=69.21^\circ\pm 6.19^\circ~ ,
 \label{CKMphase}
 \end{equation}
 which is also  in the PDG notation.
 The error intervals in Eqs.\,(\ref{yukawa5}),  (\ref{CKM}),
  (\ref{Vratio}) and (\ref{CKMphase}) represent $1\sigma$ interval.
 %%%%%%%%%%%%%%%%%%%%%%%%%%%%%%%%%%%%%%%%%%%%%%%%%%%%%%%%%
 %%%%%%%%%%%%%%%%%%%%%%%%%%%%%%%%%%%%%%%%%%%%%%%%%%%%%%%%%
 %%%%%%%%%%%%%%%%%%%%%%%%%%%%%%%%%%%%%%%%%%%%%%%%%%%%%%%%%
 %%%%%%%%%%%%%%%%%%%%%%%%%%%%%%%%%%%%%%%%%%%%%%%%%%%%%%%%%
 %%%%%%%%%%%%%%%%%%%%%%%%%%%%%%%%%%%%%%%%%%%%%%%%%%%%%%%%%
 %%%%%%%%%%%%%%%%%%%%%%%%%%%%%%%%%%%%%%%%%%%%%%%%%%%%%%%%%
 %%%%%%%%%%%%%%%%%%%%%%%%%%%%%%%%%%%%%%%%%%%%%%%%%%%%%%%%%
 %%%%%%%%%%%%%%%%%%%%%%%%%%%%%%%%%%%%%%%%%%%%%%%%%%%%%%%%%
 %%%%%%%%%%%%%%%%%%%%%%%%%%%%%%%%%%%%%%%%%%%%%%%%%%%%%%%%%
 %%%%%%%%%%%%%%%%%%%%%%%%%%%%%%%%%%%%%%%%%%%%%%%%%%%%%%%%%
 %%%%%%%%%%%%%%%%%%%%%%%%%%%%%%%%%%%%%%%%%%%%%%%%%%%%%%%%%
 
 \subsection{Allowed regions of $\tau$ at nearby fixed points}
 
 We have examined eighteen cases of leptons and quarks in above framework
 numerically as shown in Table 5. In this Table,
 the successful cases for the  mass matrix I and I\hspace{-.01em}I  at nearby fixed points are denoted by $\bigcirc$.
 On the other hand, $\times$ denotes a failure to reproduce
 observed mixing angles, and    $\bigotimes$ denotes the case in which 
 observed PMNS mixing angles are reproduced, but $\sum m_i\geq 120$\, meV.
 
  Among eighteen cases, seven cases of leptons and one case  of quarks
  are consistent with recent observed data.
  It is emphasized that the all cases of the mass matrix I work well
     at nearby $\tau=i$.
     %, but none of them works at $\tau=\omega$.
These results confirm our previous discussions.
 %%%%%%%%%%%%%%%%%%%%%%%%%%%%%%%%%%%%%%%%%%%%%%%%%%%%%%%%%%%%%%%%%%%%%
\begin{table}[h]
 \begin{small}
	\begin{center}
		\begin{tabular}{|c|c|c|c|}
			\hline 
			\rule[14pt]{0pt}{2pt} 
			{ Modulus } & { nearby $\tau=i$} & { nearby $\tau=\omega$}  & 
			{ towards $\tau=i\infty$  }\\ \hline
			\rule[14pt]{0pt}{2pt} 
			Lepton/Quark	&Lepton \quad\ Quark & Lepton \quad\ \ Quark& Lepton \quad\ \ Quark \\ \rule[14pt]{0pt}{2pt} 
			Neutrino mass hierarchy& \ \ NH \quad IH \qquad \qquad\qquad\quad
			&  \ \ NH \quad IH \qquad \qquad\qquad\quad 
			&   \ \ NH \quad IH \qquad \qquad\qquad\quad\\
			\hline \hline
			\rule[14pt]{0pt}{3pt}
			mass matrix I for $M_E$ and $M_q$ & $\bigcirc$ \ \ \quad  $\bigcirc$ \quad\quad  $\bigcirc$
			& $\bigotimes$ \ \quad  $\times$ \quad\quad  $\times$
			&  $\bigcirc$ \ \quad  $\times$ \quad\quad  $\times$\\
%			\rule[14pt]{0pt}{3pt}
%			$\langle  m_{ee} \rangle $\,[meV] & $15$--$31$ \  $17$--$31$\quad ---
%			&--- \qquad  ---\qquad ---& $16$--$18$ \ \ \ ---\qquad---\\
			\hline 
			\rule[14pt]{0pt}{3pt}
			mass matrix I\hspace{-.01em}I for $M_E$ and $M_q$ 
			& $\bigcirc$ \ \quad  $\bigotimes$ \quad\quad  $\times$
			& $\bigcirc$ \ \quad  $\bigcirc$ \quad\quad  $\times$
			& $\bigcirc$ \ \quad  $\bigotimes$ \quad\quad  $\times$\\
%			\rule[14pt]{0pt}{3pt}
%			$\langle  m_{ee} \rangle $\,[meV]& $1.4$--$27$  \ \ ---\qquad---
%			&$2.4$--$3.0$ \  $16$--$25$\quad ---
%			&  $8.8$--$14$  \ \ ---\qquad---\\
			\hline 
		\end{tabular}
	%\vskip -0.5 cm
		\caption{The successful cases for the  mass matrix I and I\hspace{-.01em}I  at nearby fixed points are denoted by
			$\bigcirc$.
			On the other hand, $\times$ denotes a failure to reproduce
			observed mixing angles, and    $\bigotimes$ denotes the case in which 
			observed mixing angles are reproduced, but $\sum m_i\geq 120$\, meV.
%The predicted ranges of the  effective mass for the $0\nu\beta\beta$ decay,
		%	$\langle  m_{ee} \rangle $\,[meV] are also presented.
	 }
		\label{summary}
	\end{center}
\end{small}
\end{table}
%%%%%%%%%%%%%%%%%%%%%%%%%%%%%%%%%%%%%%%%%%%%%%

We show 
allowed regions of $\tau$ at nearby $\tau=i$, $\tau=\omega$ 
and towards $\tau=i\infty$
 for eleven cases in Figs.\,1,\,2 and 3, respectively.
%Both cases of NH and IH of neutrinos  are examined numerically
%for each case of the  charged lepton mass matrix I and I\hspace{-.01em}I.
%Quark mass mass matrix I and I\hspace{-.01em}I are also examined numerically. 
In these figures, green points denote allowed ones by inputting masses and mixing angles  with the constraint $\sum m_i\leq 120$\,meV for leptons, but blue points denote the regions in which the sum of neutrino masses $\sum m_i$ is larger than $120$\,meV.  It is noted that blue points are hidden under  green points 
in the case of the charged lepton  I\hspace{-.01em}I (NH) of Fig.\,2
and charged lepton  I (NH) of Fig.\,3.
Green points for quarks  denote allowed region of $\tau$ by inputting masses, mixing angles and CP violating phase $\delta_{\rm CP}^{\rm CKM}$. 

As seen in Fig.\,1, the constraint $\sum m_i\leq 120$\,meV excludes
 the charged lepton  I\hspace{-.01em}I with IH of neutrinos. 
The allowed regions of $\tau$ (green points)  deviate from
the fixed point $\tau=i$ in magnitude of  $5$--$10$\%, which confirm the discussions in section 5. 
It is reasonable that the allowed points appear frequently at nearby $\tau=i$ since  one flavor mixing angle is generated
even at the fixed point $\tau=i$ as discussed in section 5.2.
In the quark sector, the mass matrix I works well, but the matrix I\hspace{-.01em}I does not because the mixing angles are canceled out each other
in the same type mass matrices of up-type  and down-type quarks.
It is emphasized that there is the common region of $\tau$
between charged lepton I (NH) and quark I.  Indeed,  
the region around {$\tau=\pm 0.04+ 1.05 \, i$} is common in quarks and leptons.
This common region has already discussed  in context with the quark-lepton unification in Ref.\cite{Okada:2019uoy}.

As seen in Fig.\,2, at nearby  $\tau=\omega$, 
the charged lepton mass matrix I with NH is excluded
by the constraint of $\sum m_i\leq 120$\,meV.
In the charged lepton mass matrix I with IH,  the PMNS mixing angles are not reproduced.
On the other hand, the allowed regions  are marginal
in the charged lepton  I\hspace{-.01em}I. Indeed, the green points are
 $0.1$ for NH and $0.15$ for IH away from $\tau=\omega$, respectively.
 The perturbative discussion of this IH case is  possibly broken.
Moreover, we cannot find allowed region of quarks  at nearby $\tau=\omega$.
That is expected in the discussion in section  6.3.

As seen in Fg.\,3, towards $\tau=i\infty$, both charged lepton mass matrix I and  I\hspace{-.01em}I reproduce
 the observed PMNS mixing angles for NH of neutrinos.
 In the charged lepton mass matrix I with IH,  the PMNS mixing angles are not reproduced.
 Although the charged lepton mass matrix  I\hspace{-.01em}I with IH reproduces
 three PMNS mixing angles, it is excluded
 by the constraint of $\sum m_i\leq 120$\,meV.
 We cannot  find allowed region for quarks.
These results are also consistent with discussions of section 5.4 and 6.4.

\newpage
%%%%%%%%%%%%%%%%%%%%%%%%%%%
\begin{figure}[H]
	\includegraphics[{width=0.47\linewidth}]{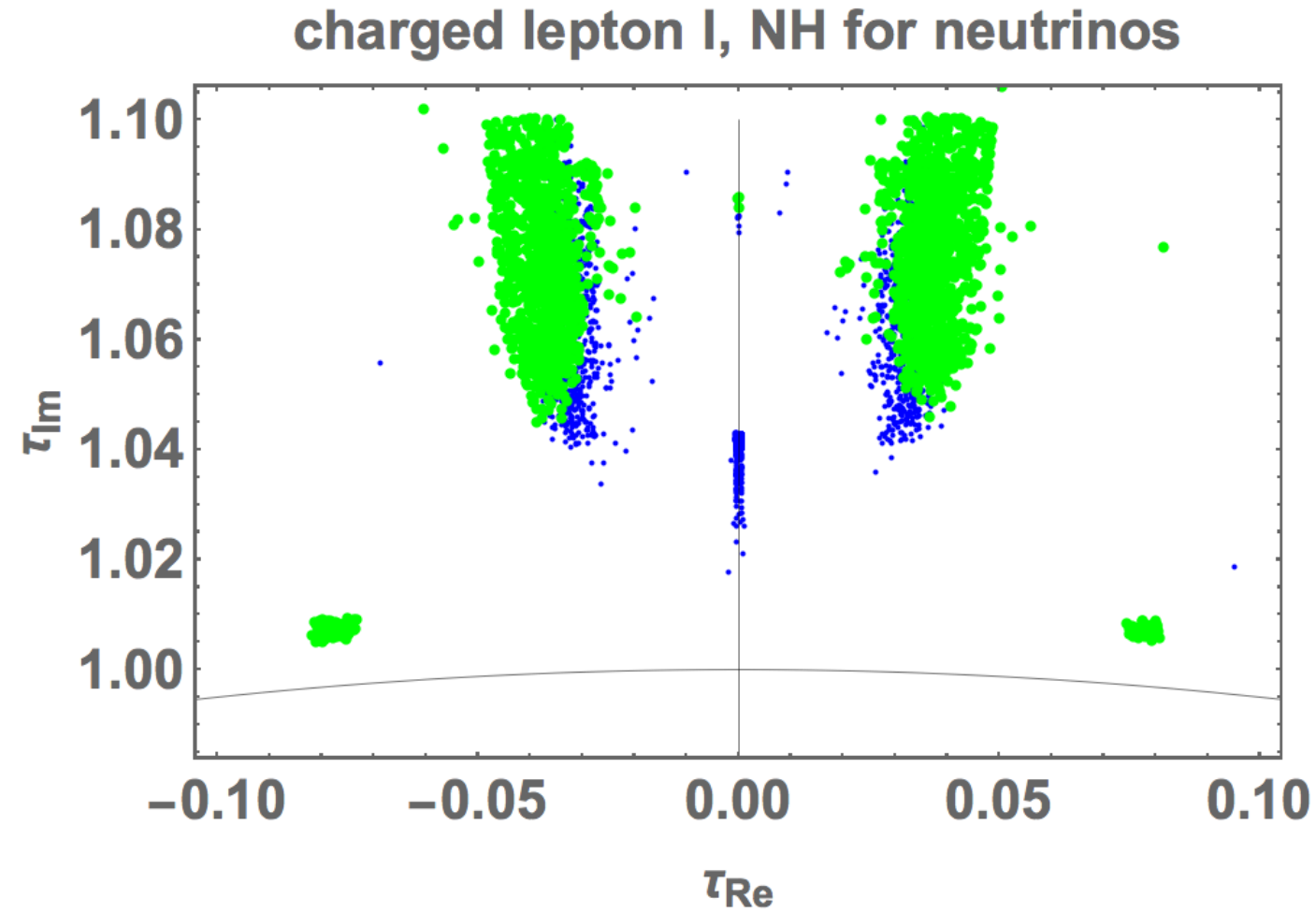}
	%		\caption{L-NH matrix I (222), $\tau=i$}
	\hskip 1 cm	\includegraphics[{width=0.47\linewidth}]{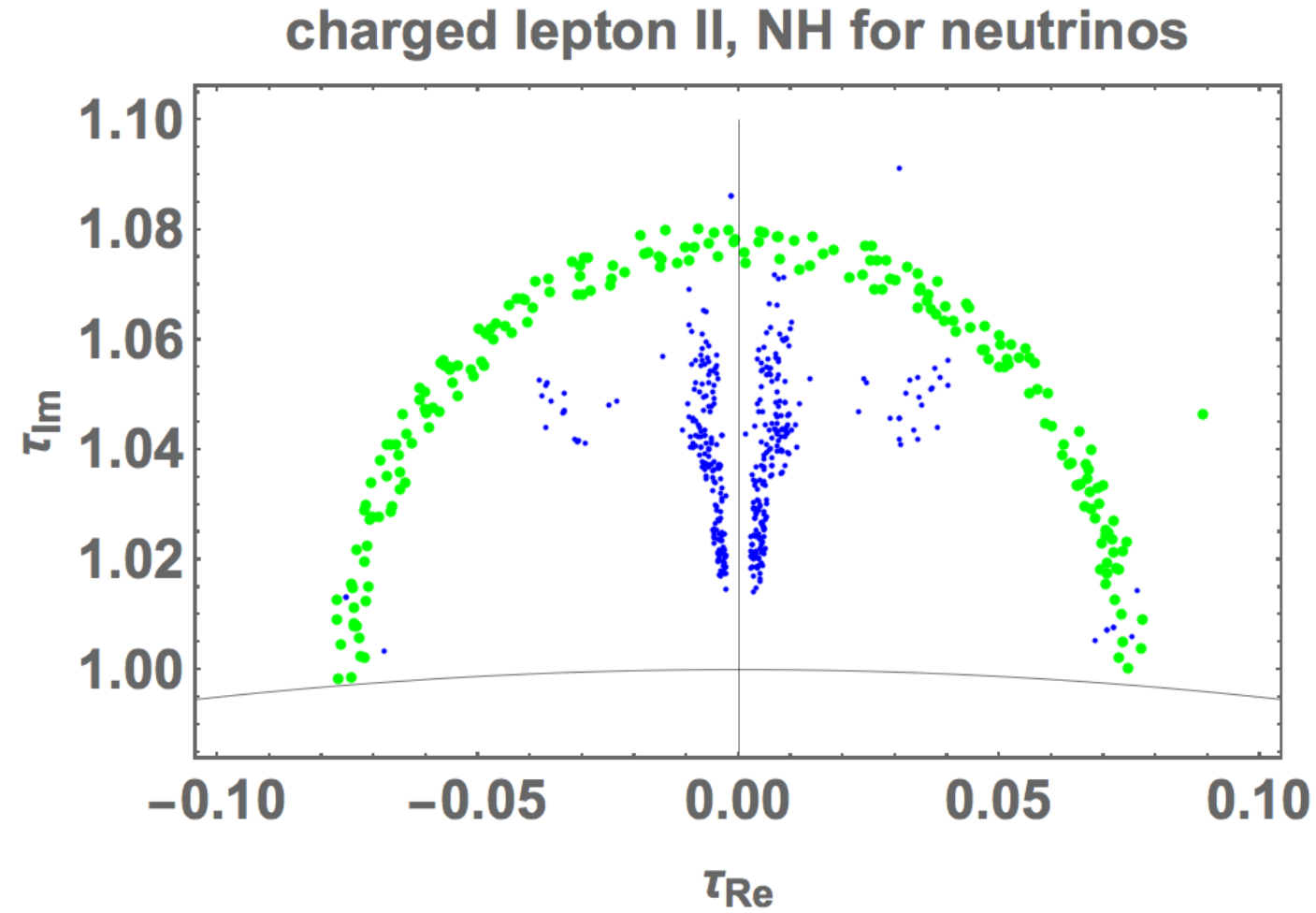}
	\vskip 1 cm
	%	\caption{L-NH  matrix  I\hspace{-.01em}I (642), $\tau=i$}
	\includegraphics[{width=0.47\linewidth}]{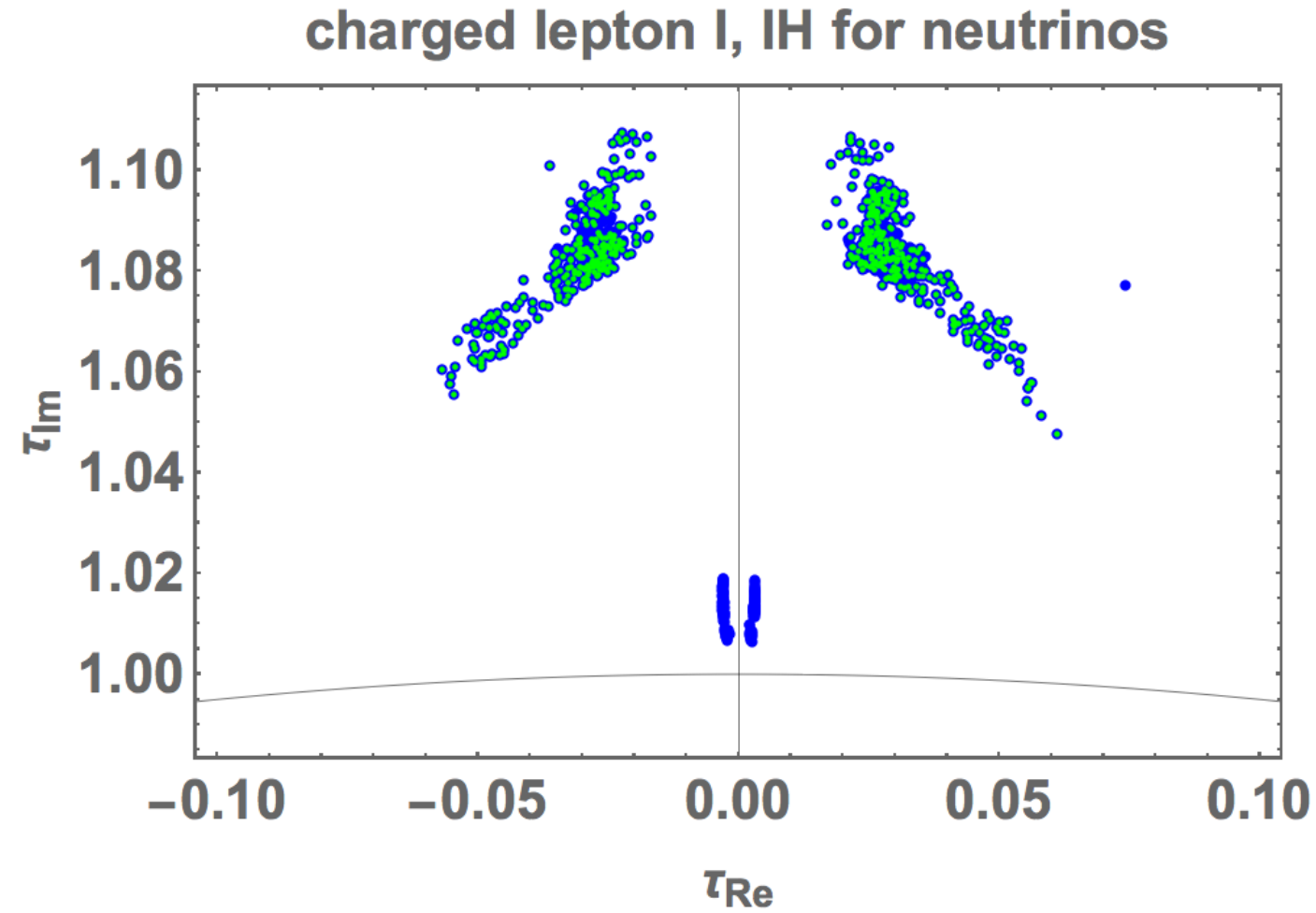}
	%		\caption{L-IH matrix I (222), $\tau=i$}
	\hskip 0.8 cm	\includegraphics[{width=0.47\linewidth}]{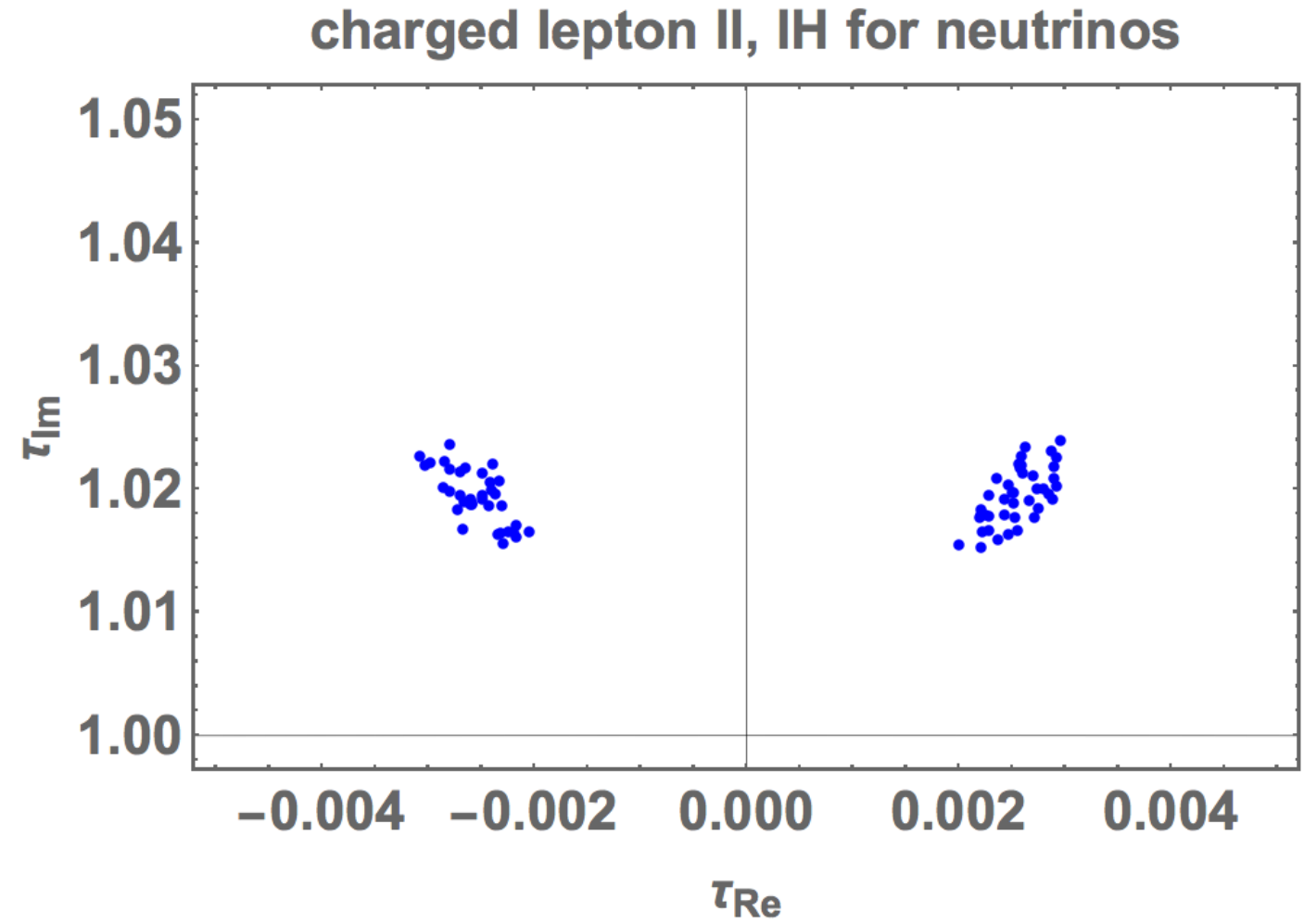}
	%	\vspace{-12mm}
	\includegraphics[{width=0.47\linewidth}]{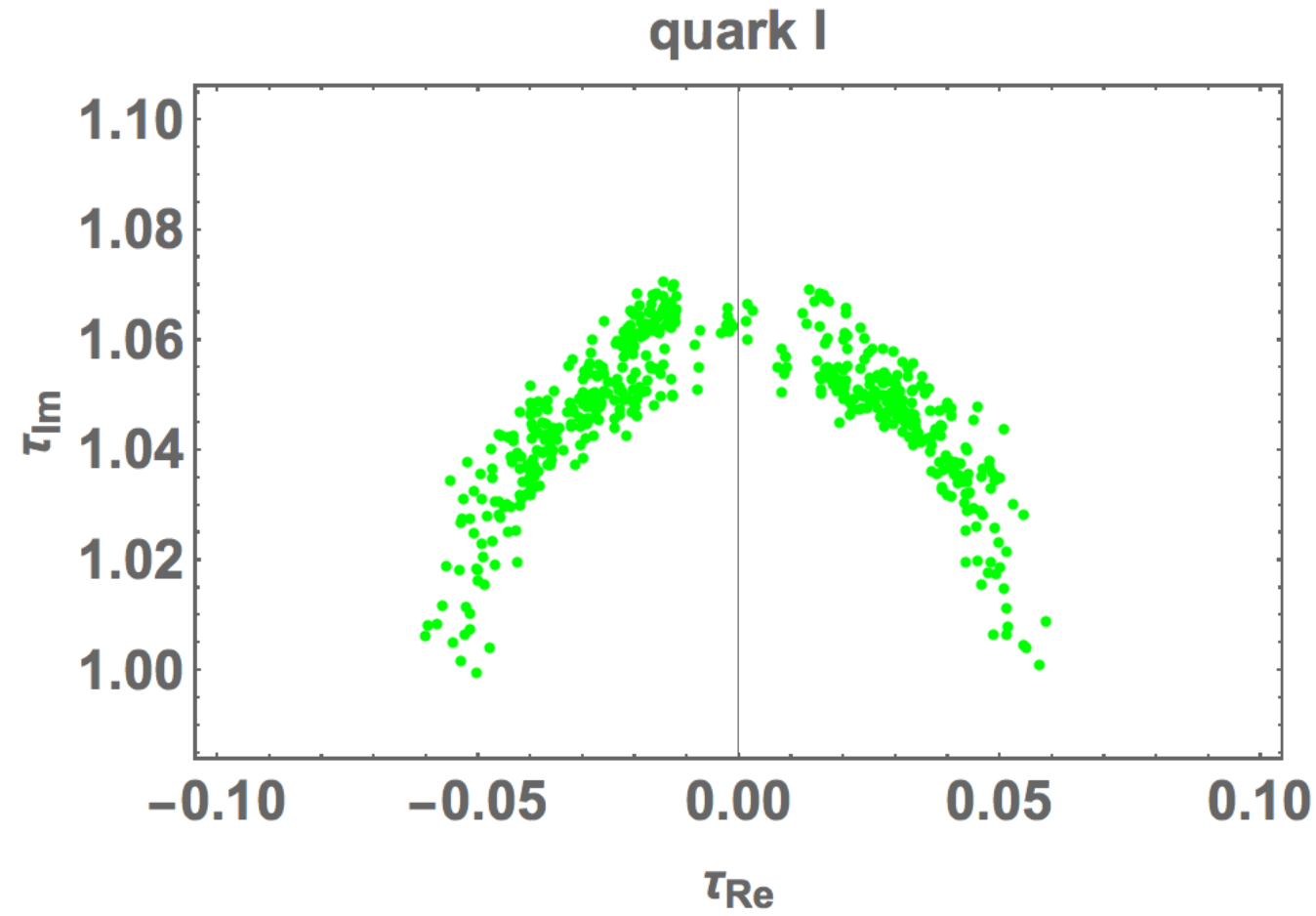}
	%	\caption{Quark mass  matrix I at nearby $\tau=i$}
	\hspace{1.5 cm}
	\includegraphics[bb=0 0 800 700,width=0.47\linewidth]{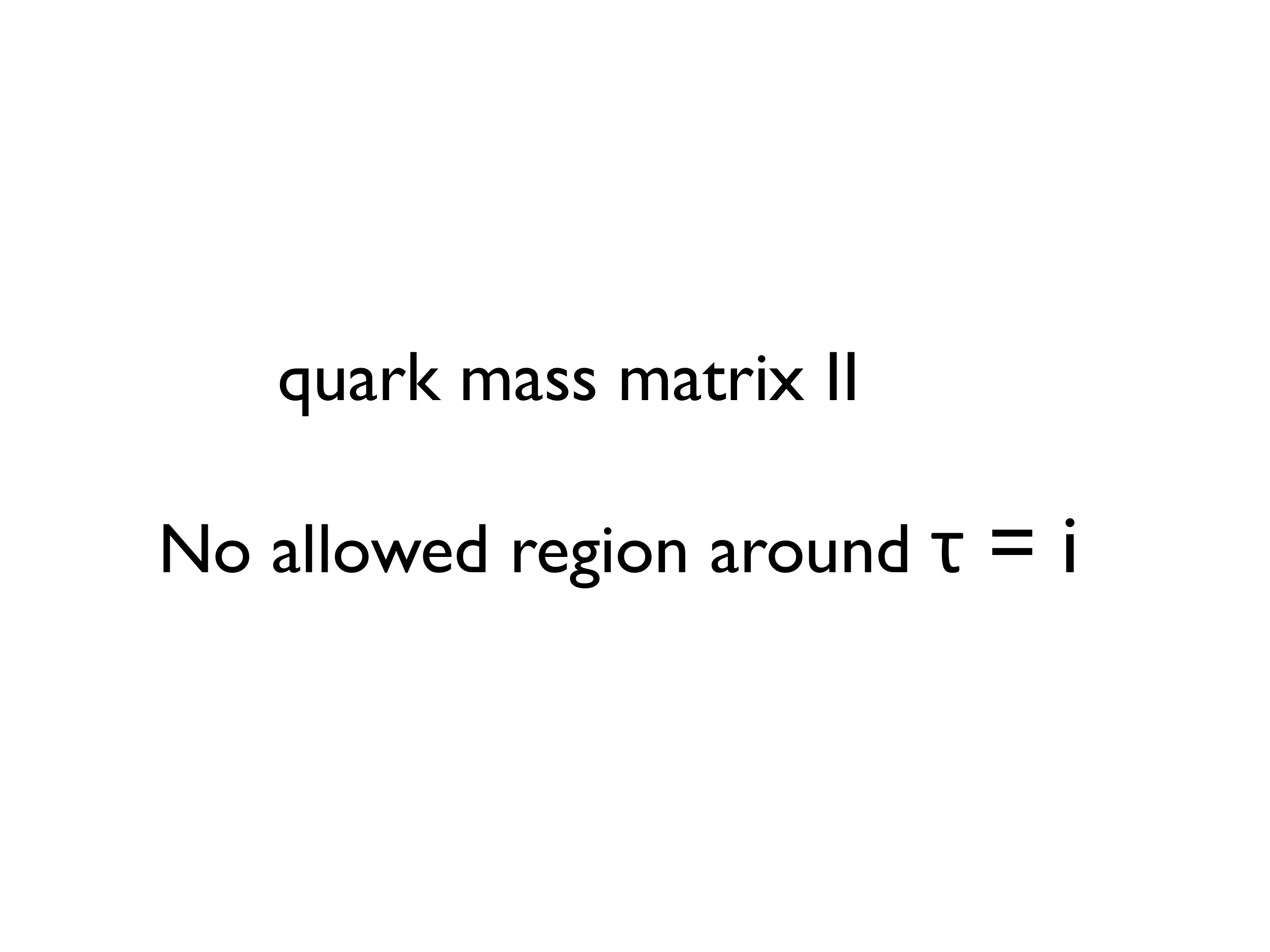}
	%			\includegraphics[{width=0.47\linewidth}]{}
	%		\caption{No solutions for Quark mass  matrix  I\hspace{-.01em}I at nearby $\tau=i$}
	%	\vspace{-12mm}
	\caption{Allowed regions of $\tau$ at nearby $\tau=i$ are shown by green points for
		charged lepton mass matrices I and I\hspace{-.01em}I with NH and IH of neutrinos,
	and  quark mass matrices  I, respectively. Blue points denote  regions in which the sum of neutrino masses $\sum m_i$ is larger than $120$\,meV.}
\end{figure}
%%%%%%%%%%%%%%%%%%%%%%%%%%% 
%%%%%%%%%%%%%%%%%%%%%%%%%%%
%%%%%%%%%%%%%%%%%%%%%%%%%%%
\newpage
%%%%%%%%%%%%%%%%%%%%%%%%%%%
%%%%%%%%%%%%%%%%%%%%%%%%%%%
%%%%%%%%%%%%%%%%%%%%%%%%%%%
\begin{figure}[H]
	\includegraphics[{width=0.47\linewidth}]{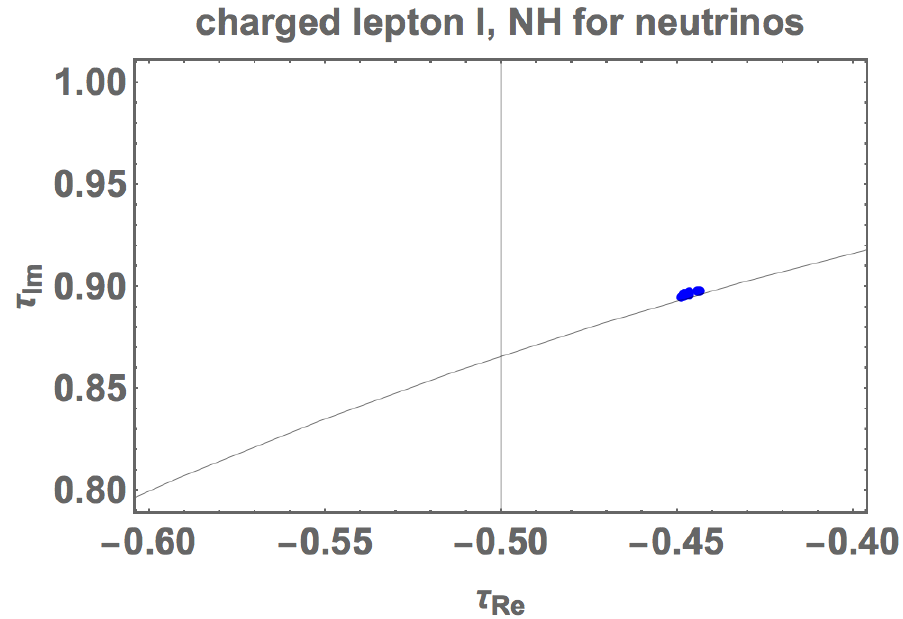}
	\hskip 0.7 cm\includegraphics[{width=0.47\linewidth}]{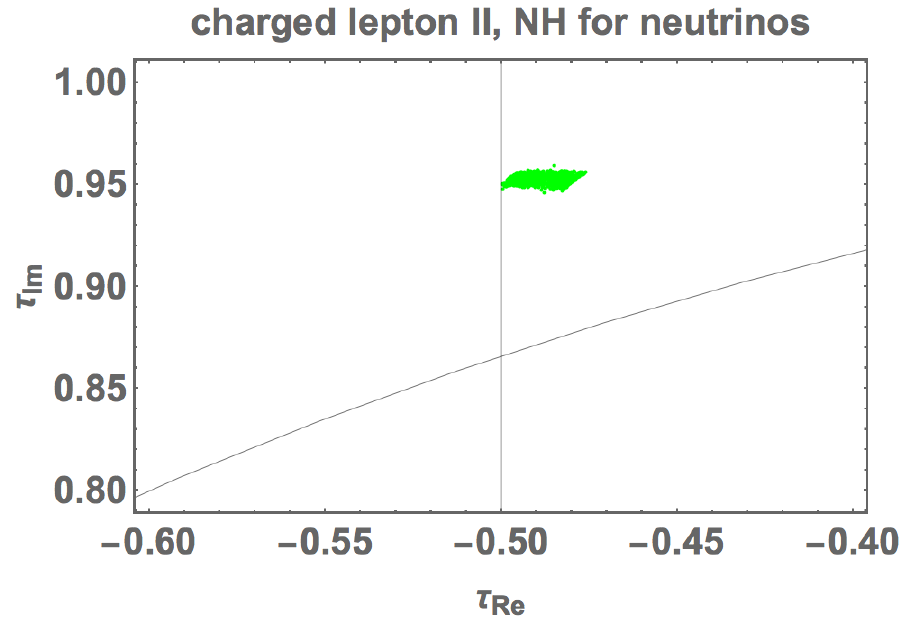}
	\vskip 0.5 cm
	\includegraphics[bb=0 0 800 700,width=0.47\linewidth]{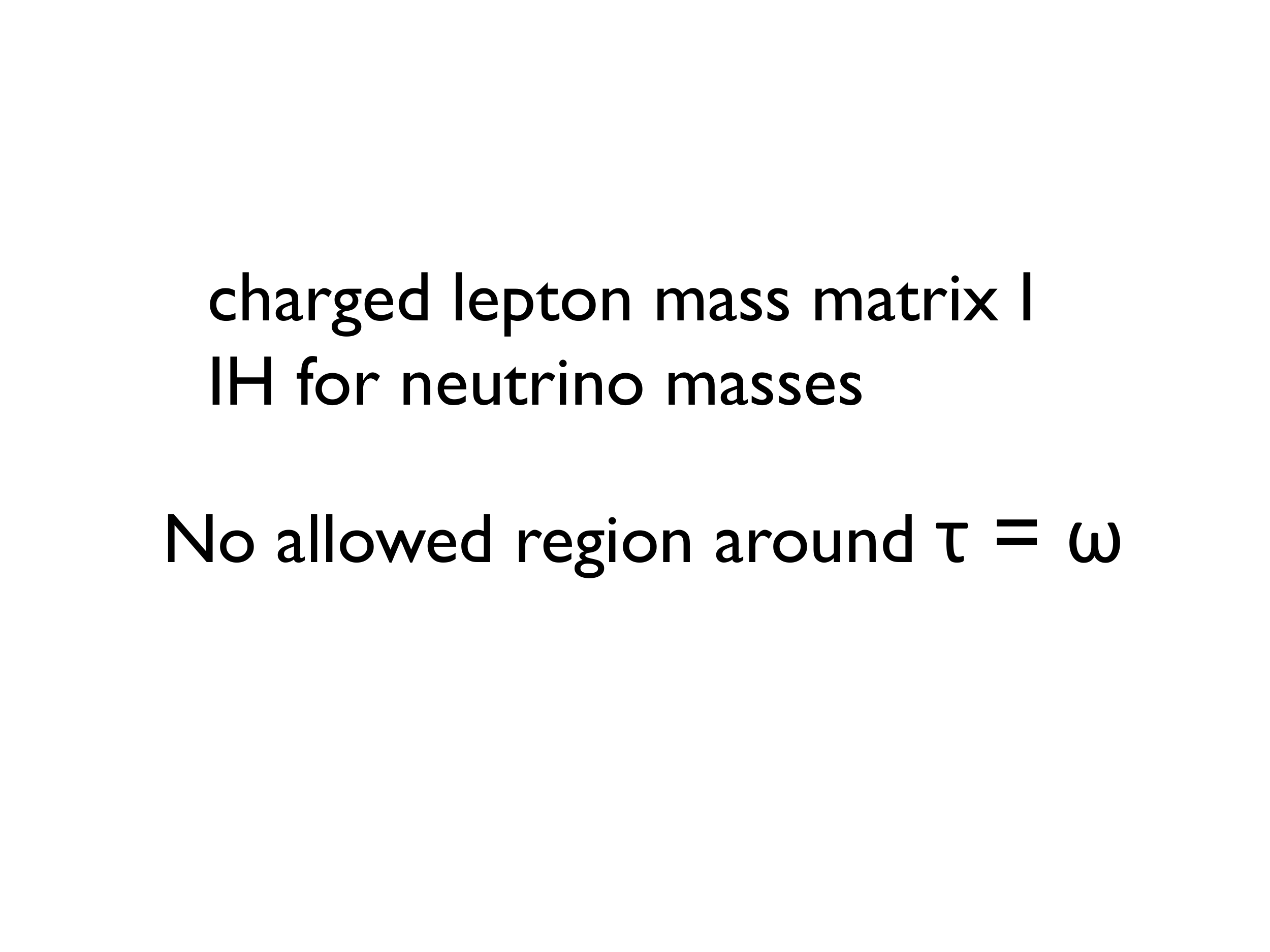}
	\hskip 1 cm\includegraphics[{width=0.47\linewidth}]{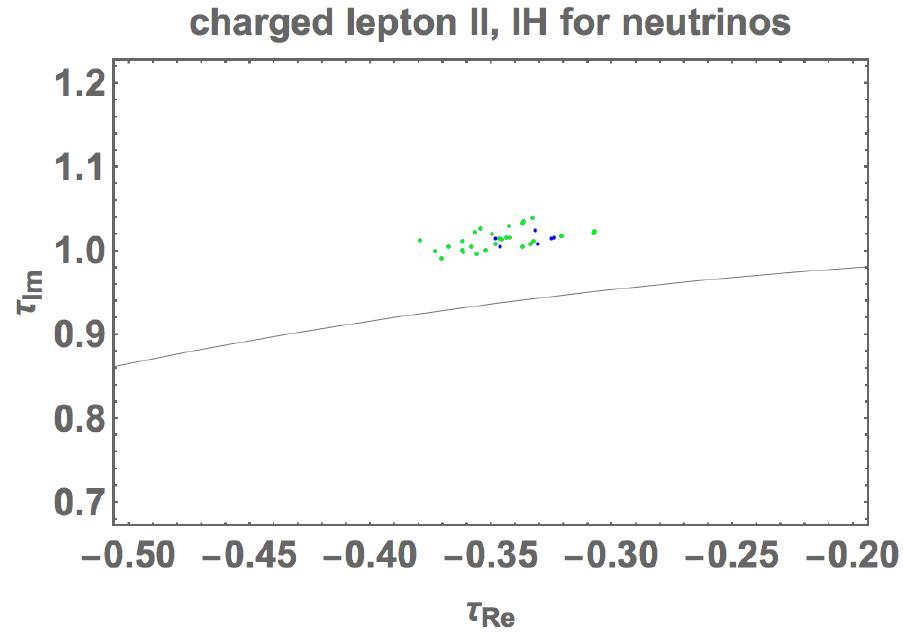}
	\includegraphics[bb=0 0 800 700,width=0.47\linewidth]{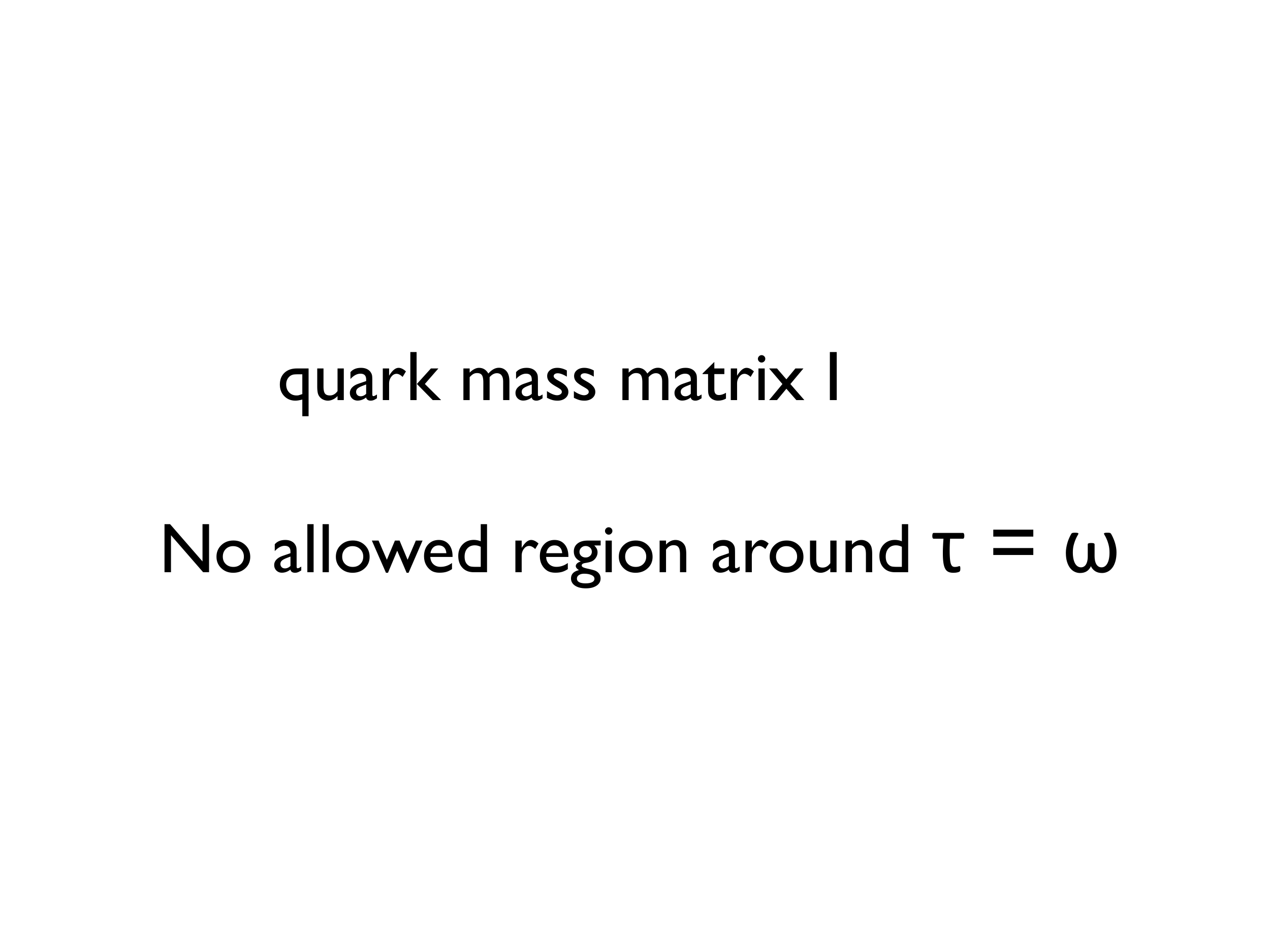}
	\hskip 2 cm
	\includegraphics[bb=0 0 800 700,width=0.47\linewidth]{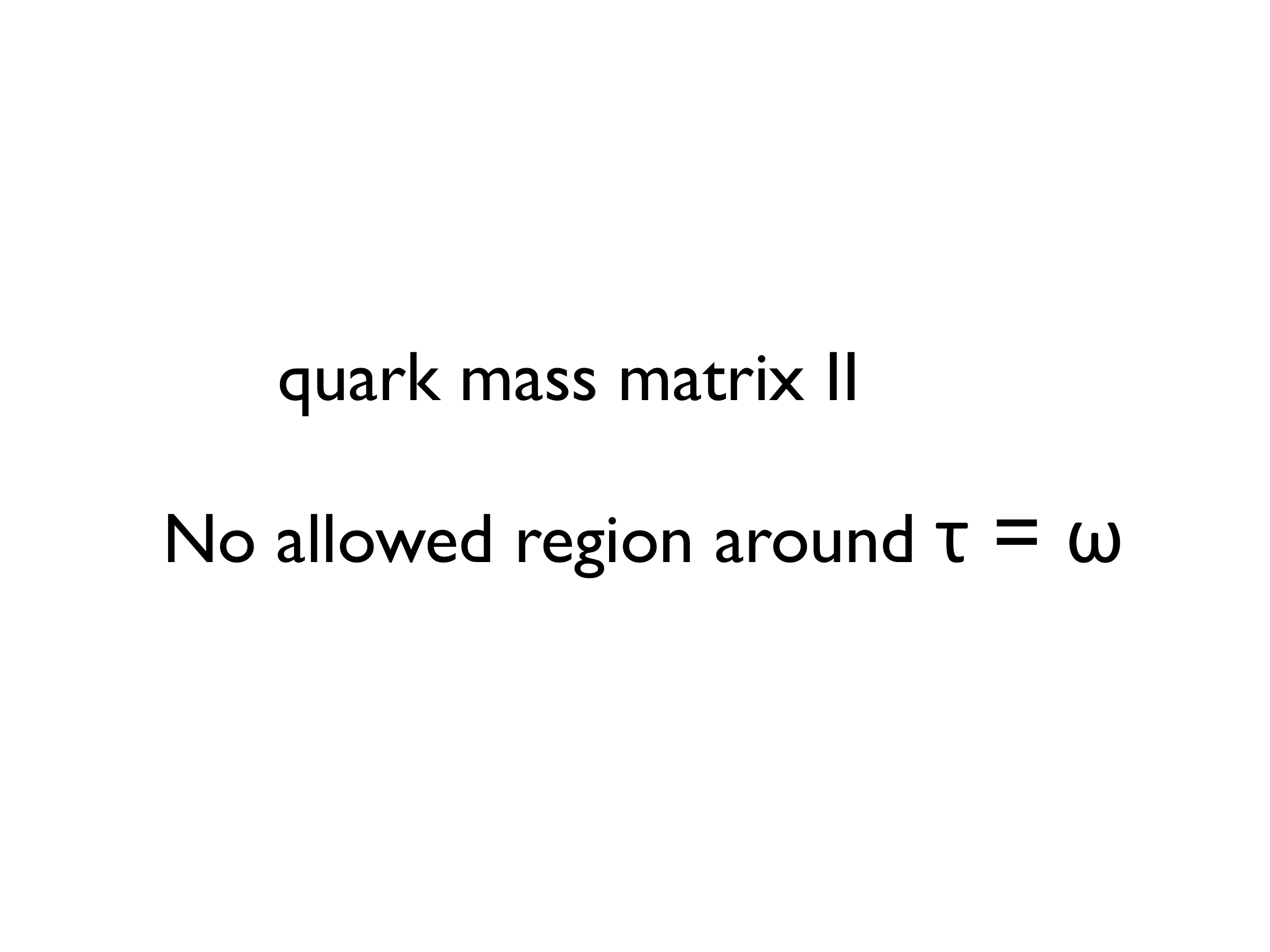}
	\vskip -1.5 cm
	\caption{Allowed regions of $\tau$ at nearby $\tau=\omega$ are shown by green points for the charged lepton mass matrix I and  I\hspace{-.01em}I with NH and IH of neutrinos, respectively. Blue points denote  regions in which the sum of neutrino masses $\sum m_i$ is larger than $120$\,meV.}
\end{figure}
%%%%%%%%%%%%%%%%%%%%%%%%%%% 
%%%%%%%%%%%%%%%%%%%
\newpage
%%%%%%%%%%%%%%%%%%%%%%%%%%%
%%%%%%%%%%%%%%%%%%%%%%%%%%%
%%%%%%%%%%%%%%%%%%%%%%%%%%%
%%%%%%%%%%%%%%%%%%%%%%%%%%%
%%%%%%%%%%%%%%%%%%%%%%%%%%%
\begin{figure}[H]
	\includegraphics[{width=0.47\linewidth}]{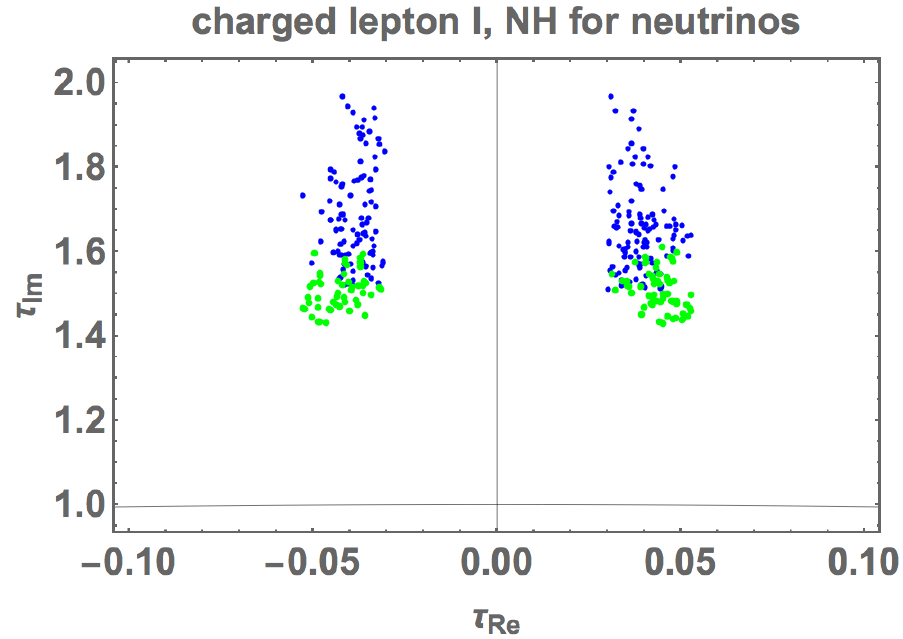}
	\hskip 1 cm
	\includegraphics[{width=0.47\linewidth}]{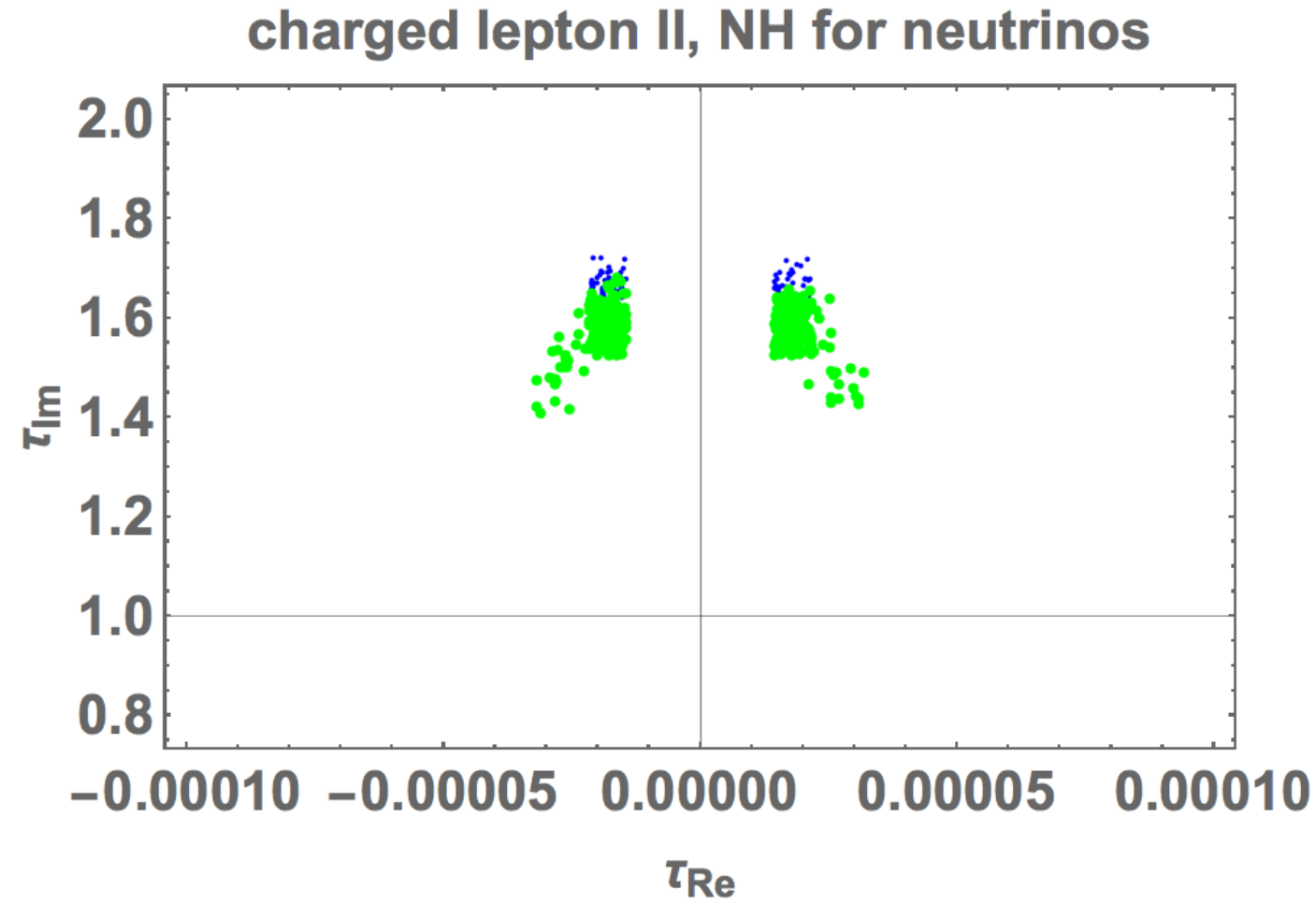}
	\vskip 1 cm
	\includegraphics[{bb=0 0 800 700,width=0.47\linewidth}]{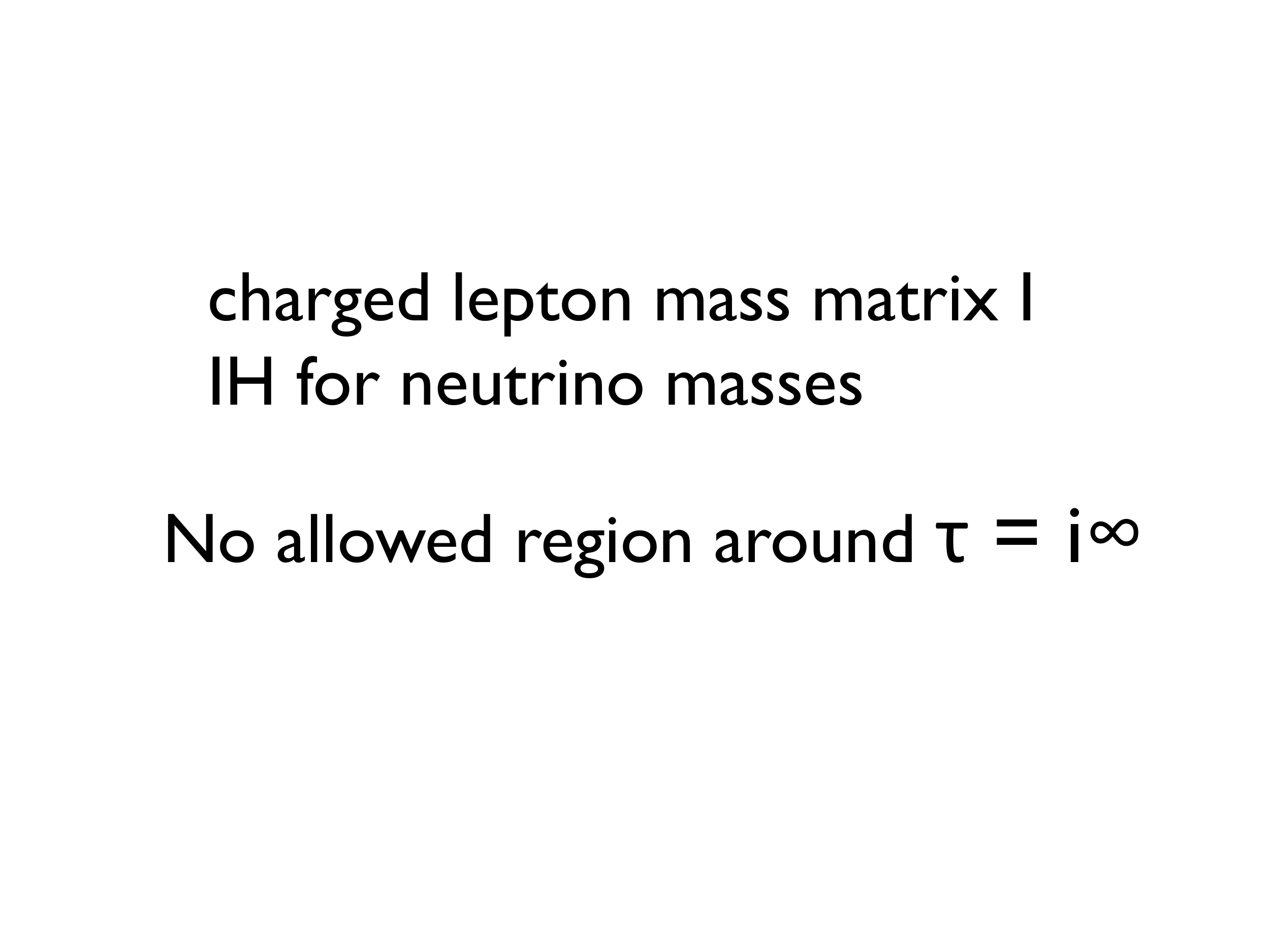}
	\hskip 1 cm	
	\includegraphics[{width=0.47\linewidth}]{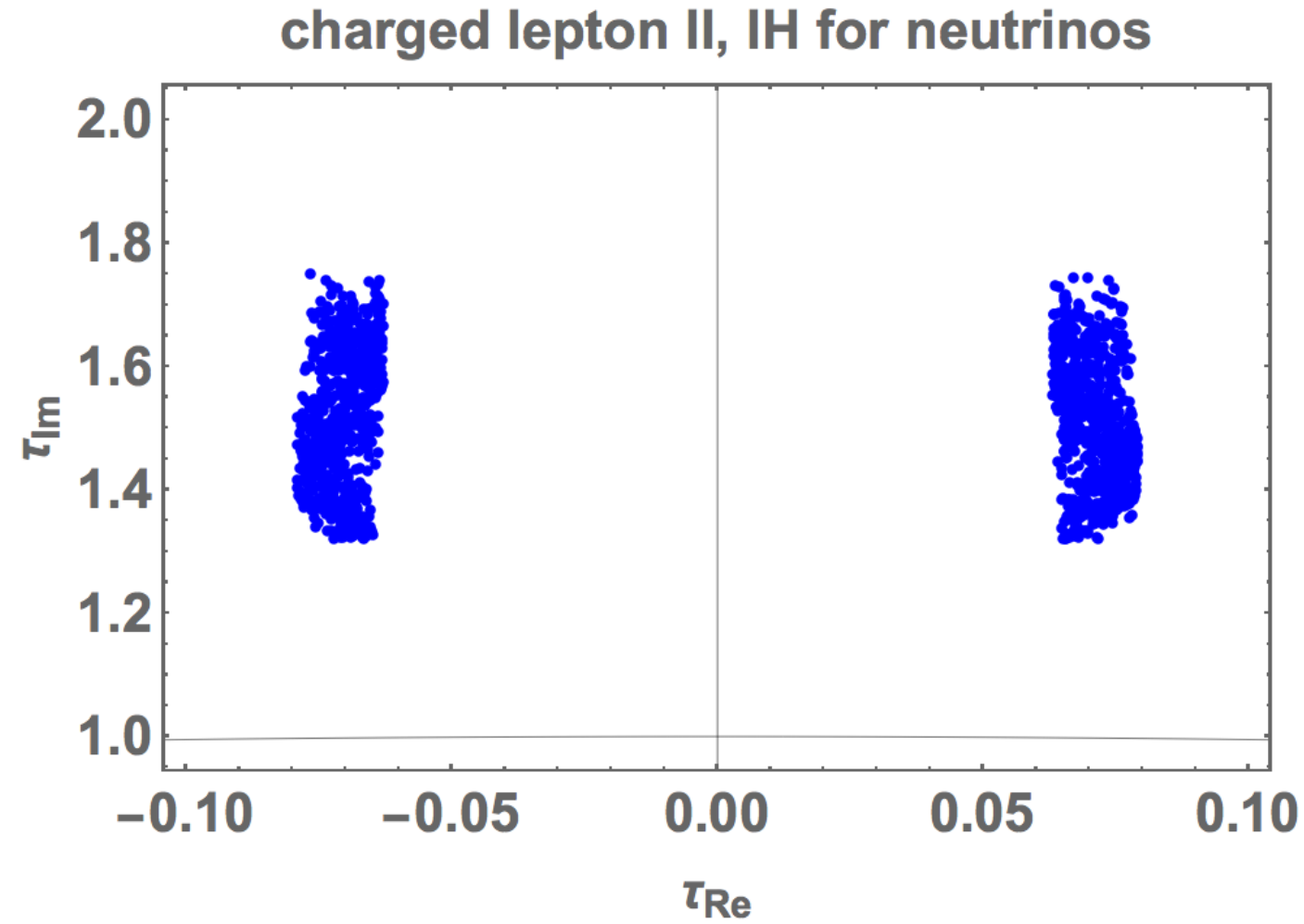}
	%	\hskip 1.5 cm
	\includegraphics[bb=0 0 800 700,width=0.47\linewidth]{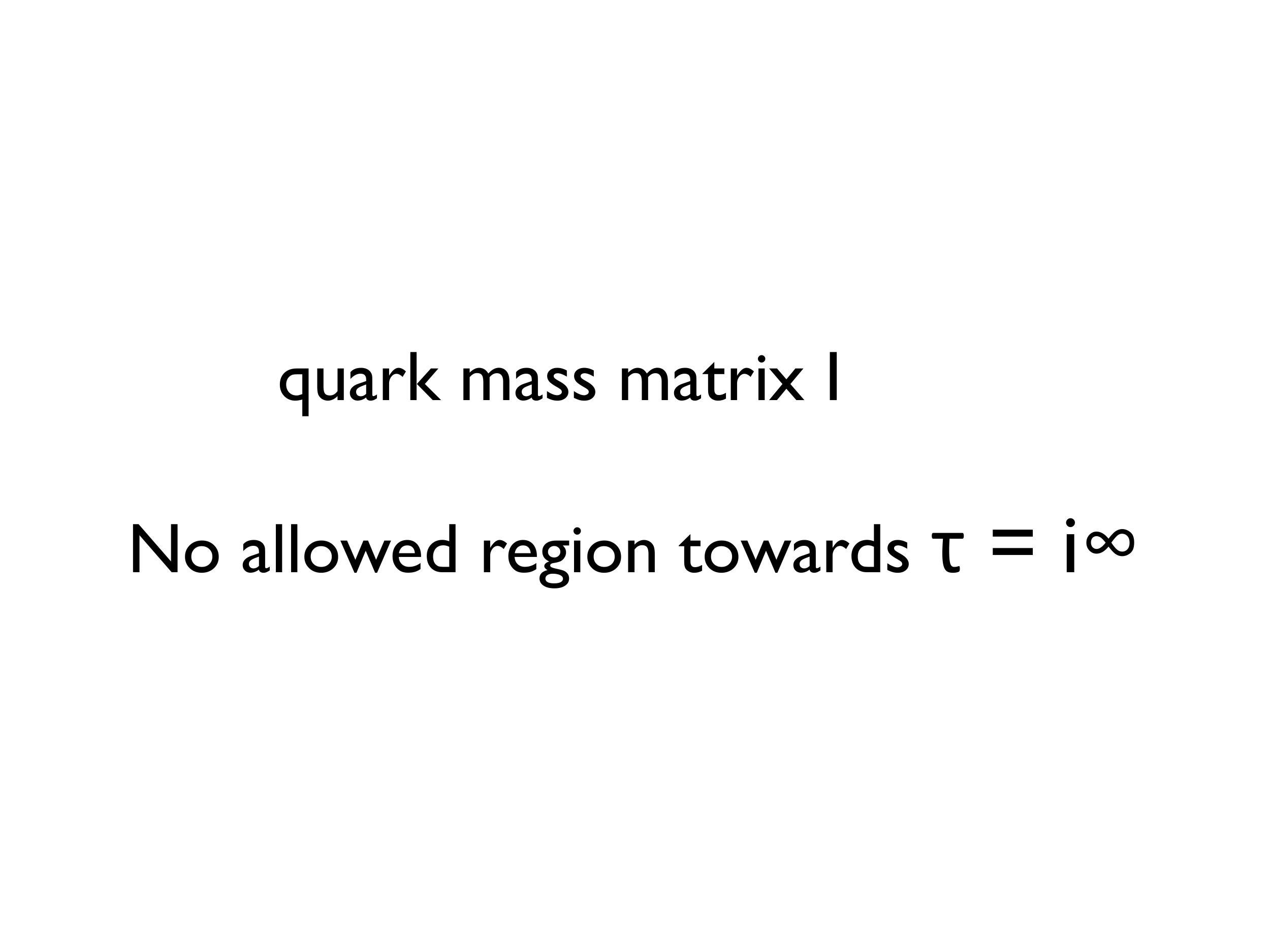}
	\hskip 2 cm
	\includegraphics[bb=0 0 800 700,width=0.47\linewidth]{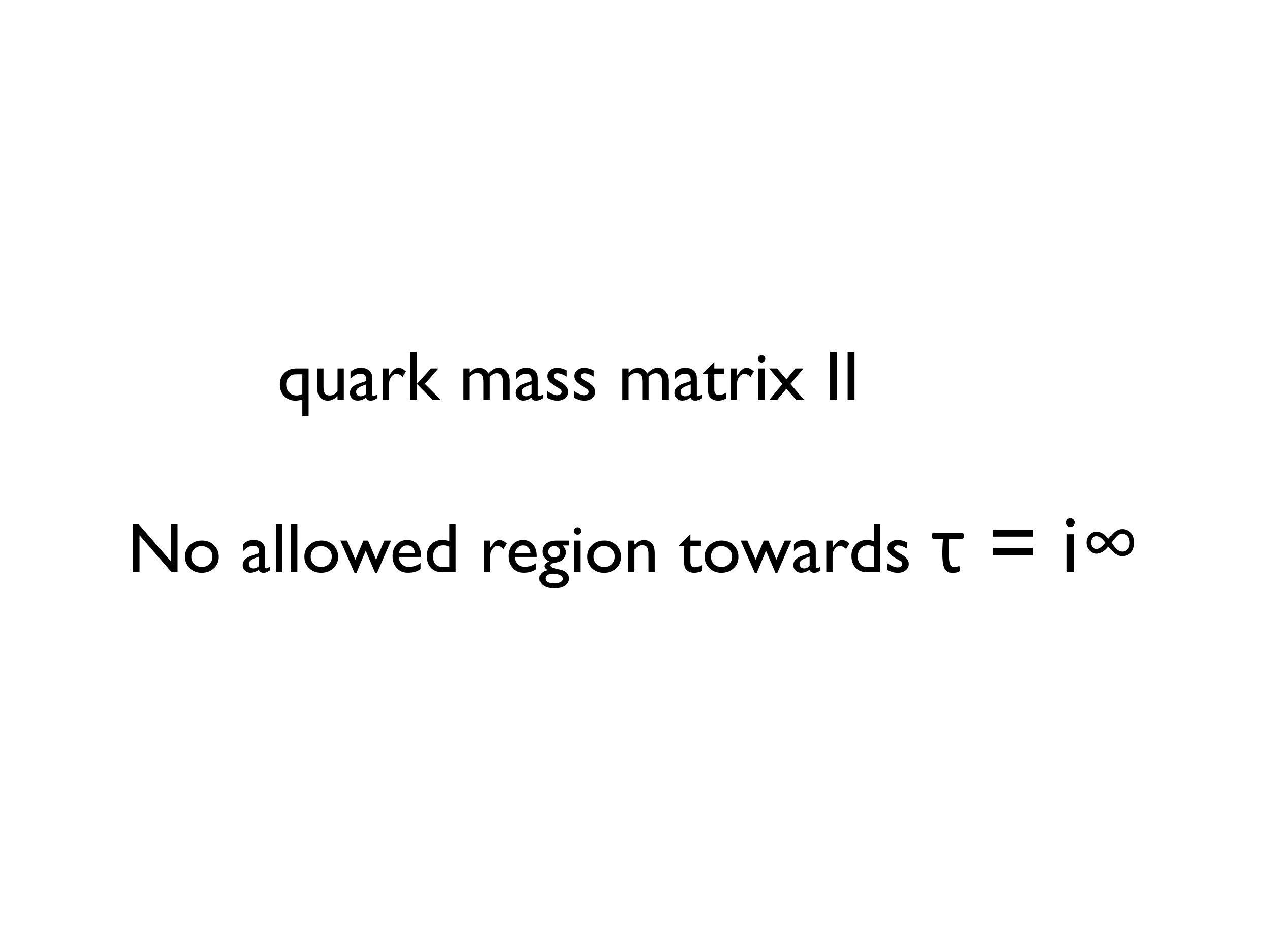}
	\vskip -1.5 cm
	\caption{Allowed regions of $\tau$  towards $\tau=i\infty$ are shown by green points
	for charged lepton mass matrices I and I\hspace{-.01em}I with NH and IH of neutrinos, respectively. Blue points denote  regions in which the sum of neutrino masses $\sum m_i$ is larger than $120$\,meV. }
\end{figure}
%%%%%%%%%%%%%%%%%%%%%%%%%%%
%%%%%%%%%%%%%%%%%%%%%%%%%%%
%%%%%%%%%%%%%%%%%%%%%%%%%%%
%%%%%%%%%%%%%%%%%%%%%%%%%%%
%%%%%%%%%%%%%%%%%%%%%%%%%%%
\newpage

\subsection{Predictions of CP violation and masses of neutrinos}

We predict  the leptonic CP violating phase 
$\delta_{\rm CP}^\ell$, 
the sum of neutrino masses $\sum m_i$ and
the  effective mass for the $0\nu\beta\beta$ decay 
 $|\langle m_{ee} \rangle |$ for each case of leptons
 since we input four observed quantities of neutrinos
 (three mixing angles of leptons
 and observed neutrino mass ratio $\Delta m_{\rm sol}^2/\Delta m_{\rm atm}^2$)
 and  three charged lepton masses.
 For quark sector, there is no prediction because ten observed quantities
 (quark masses and CKM elements) are put to obtain the region of
 the modulus $\tau$.

 In Table 6, the predicted ranges of the  effective mass for the $0\nu\beta\beta$ decay,
 $\langle  m_{ee} \rangle $ are presented for each case.
 We also summarize magnitudes of parameters $g_{\nu 1}$, 
 $g_{\nu 2}$, $g_{e}$ for leptons and $g_{u1}$, 
 $g_{u 2}$, $g_{u3}$ for quarks.
 Their phases are broad.
 We add
 hierarchies of $ \tilde\alpha_e^2,\,\tilde\beta_e^2, \,\tilde\gamma_e^2$
 and $ \tilde\alpha_q^2,\,\tilde\beta_q^2, \,\tilde\gamma_q^2$.
%to refer discussions of sections 5 and 6.

 %%%%%%%%%%%%%%%%%%%%%%%%%%%%%%%%%%%%%%%%%%%%%%%%%%%%%%%%%%%%%%%%%%%%%
\begin{table}[H]
	\begin{center}
		\begin{tabular}{|c|c|c|c|c|c|}
			\hline 
			\rule[14pt]{0pt}{3pt} 
			{  }&$\langle  m_{ee}\rangle  $\,[meV] & $|g_{\nu 1}|$&$|g_{\nu 2}|$ &$|g_e|$ &$ \tilde\alpha_e^2,\,\tilde\beta_e^2, \,\tilde\gamma_e^2$\\ 
			\hline
			\rule[14pt]{0pt}{3pt} 	
			NH,	charged lepton I,$\tau\simeq i$&$15$--$31$& $0.02$--$18$&$0.63$--$19$ &---
			&$ \tilde\gamma_e^2\gg\tilde\alpha_e^2\gg\tilde\beta_e^2$\\
			\hline 
			\rule[14pt]{0pt}{3pt} 	
			IH,	\, charged lepton I, $\tau\simeq i$	&$17$--$31$ & $0.56$--$3.9$& $1.6$--$4.9$&---
			&$ \tilde\gamma_e^2\gg\tilde\alpha_e^2\gg\tilde\beta_e^2$\\
			\hline 
			\rule[14pt]{0pt}{3pt} 	
			NH,	 charged lepton I\hspace{-.01em}I, $\tau\simeq i$&$1.4$--$27$ & $0.53$-- $7.0$& $0.56$--$6.9$ & $0.63$--$8.9$
			&$ \tilde\alpha_e^2\gg\tilde\gamma_e^2\gg\tilde\beta_e^2$\\
			\hline 
			\rule[14pt]{0pt}{3pt} 	
			NH, charged lepton I\hspace{-.01em}I, $\tau\simeq \omega$	&$2.4$--$3.0$ &
			$0.03$--$0.05$ & $0.53$--$0.65$& $0.22$--$0.28$
			&$ \tilde\alpha_e^2\gg\tilde\beta_e^2\gg\tilde\gamma_e^2$\\
			\hline 
			\rule[14pt]{0pt}{3pt} 	
			IH, charged lepton I\hspace{-.01em}I, $\tau\simeq \omega$	& $16$--$25$
			&  $1.2$--$1.8$& $1.1$--$1.5$&$5.5$--$9.8$
			&$ \tilde\alpha_e^2\gg\tilde\beta_e^2\gg\tilde\gamma_e^2$\\
			\hline 
			\rule[14pt]{0pt}{3pt} 	
			NH,	charged lepton I, $\tau\simeq i\infty$	&$16$--$18$
			&$0.25$--$0.53$ &$1.0$--$1.2$ & ---
			&$ \tilde\gamma_e^2\gg\tilde\beta_e^2\gg\tilde\alpha_e^2$\\
			\hline 
			\rule[14pt]{0pt}{3pt} 	
			NH, charged lepton I\hspace{-.01em}I, $\tau\simeq i\infty$	&$8.8$--$14$ 
			& $0.13$--$0.33$&$0.76$--$0.87$ &$3.1$--$5.6$
			&$ \tilde\alpha_e^2\gg\tilde\gamma_e^2\gg\tilde\beta_e^2$\\
			\hline 
			\hline
			\rule[14pt]{0pt}{3pt} 
			{  }& & $|g_{u1}|$&$|g_{u2}|$ &$|g_{u3}|$
			&$ \tilde\alpha_q^2,\,\tilde\beta_q^2, \,\tilde\gamma_q^2$\\ 
			\hline
			\rule[14pt]{0pt}{2pt} 	
			quark mass matrices I, $\tau\simeq i$	& --- & $0.01$--$0.86$ &$0.14$--$1.29$ &$0.02$-$0.07$
			&$ \tilde\gamma_u^2\gg\tilde\beta_u^2\gg\tilde\alpha_u^2$\\
	 & & & & &$ \tilde\gamma_d^2\gg\tilde\alpha_d^2\gg\tilde\beta_d^2$\\
			\hline
		\end{tabular}
		\caption{Magnitudes of parameters $g_{\nu 1}$, 
			$g_{\nu 2}$, $g_{e}$ for leptons
			and $g_{u1}$, 
			$g_{u 2}$, $g_{u3}$ for quarks are shown.
			Predicted ranges of the  effective mass for the $0\nu\beta\beta$ decay,
			$\langle  m_{ee} \rangle $\,[meV] are also given.
			In addition, hierarchies of $ \tilde\alpha_e^2,\,\tilde\beta_e^2, \,\tilde\gamma_e^2$
			and $ \tilde\alpha_q^2,\,\tilde\beta_q^2, \,\tilde\gamma_q^2$
			are  presented.}
		\label{g-parameters}
	\end{center}
\end{table}
%%%%%%%%%%%%%%%%%%%%%%%%%%%%%%%%%%%%%%%%%%%%%%%%%%%%%%%%%%%%%%%
%%%%%%%%%%%%%%%%%%%%%%%%%%%%%%%%%%%%%%%%%%%%%%%%%%%%%%%%%%%%%%%
%%%%%%%%%%%%%%%%%%%%%%%%%%%%%%%%%%%%%%%%%%%%%%%%%%%%%%%%%%%%%%%
We present numerical predictions on   $\sum m_i$--$\delta_{\rm CP}^\ell$
and $\delta_{\rm CP}^\ell$--$\sin^2\theta_{23}$ planes for successful seven cases  in Figs.\,4--10.
In Fig.\,4,  we show them at nearby $\tau=i$	
for the charged lepton mass matrix  I with  NH of neutrinos. 
The predicted range of the sum of  neutrino masses is
$\sum m_i=86$--$120$\,meV.
%where $120$\,meV is the cosmological upper-bound.
The  predicted $\delta_{\rm CP}^\ell$ depends on $\sum m_i$.
A crucial test will be presented in the near future
by  cosmological observations.
The correlation between
$\sin^2\theta_{23}$ and  $\delta_{\rm CP}^\ell$ is also helpful to test  
this case.

%%%%%%%%%%%%%%%%%%%%%%%%%%%%%%%%%%%%%%%%%%%%%%%%%%%%%%%%%%%%%%%%%%    

In Fig.\,5, we show them at nearby $\tau=i$	for the charged lepton mass matrix  I with IH of neutrinos. 
The predicted range of the sum of  neutrino masses is
$\sum m_i=90$--$120$\,meV.
The  prediction of  $\delta_{\rm CP}^\ell$ is clearly given versus $\sum m_i$.
On the other hand, $\sin^2\theta_{23}$ is predicted  to be smaller than $0.52$.
Crucial test will be available  
by  cosmological observations  and  neutrino oscillation experiments
in the near future.

 %%%%%%%%%%%%%%%%%%%%%%%%%%%%%%%%%%%%%%%%%%%%%%%%%%%%%%%%%%%%%%%%%%

In Fig.\,6, 
we show them at nearby $\tau=i$
for the charged lepton mass matrix  I\hspace{-.01em}I with NH of neutrinos. 
The predicted range of the sum of  neutrino masses is
$\sum m_i=58$--$83$\,meV
while  $\delta_{\rm CP}^\ell$ is allowed  in $[-\pi,\pi]$.
There is no correlation between
$\sin^2\theta_{23}$ and  $\delta_{\rm CP}^\ell$.
The rather small value  of the sum of neutrino masses is a characteristic
prediction in this case.

%%%%%%%%%%%%%%%%%%%%%%%%%%%%%%%%%%%%%%%%%%%%%%%%%%%%%%%%%%%%%%%
%%%%%%%%%%%%%%%%%%%%%%%%%%%%%%%%%%%%%%%%%%%%%%%%%%%%%%%%%%%%%%%
%%%%%%%%%%%%%%%%%%  Figures  %%%%%%%%%%%%%%%%%%%%%%%%%%%%%%%%%%
%%%%%%%%%%%%%%%%%%%%%%%%%%%%%%%%%%%%%%%%%%%%%%%%%%%%%%%%%%%%%%%
%%%%%%%%%%%%%%%%%%%%%%%%%%%%%%%%%%%%%%%%%%%%%%%%%%%%%%%%%%%%%%%
%%%%%%%%%%%%%%%%%%%%%%%%%  tau=i  %%%%%%%%%%%%%%%%%%%%%%%%%%%%%
%%%%%%%%%%%%%%%%%%%%%%%%%%%%%%%%%%%%%%%%%%%%%%%%%%%%%%%%%%%%%%%
\begin{figure}[H]
	\centering
	\includegraphics[{width=0.42\linewidth}]{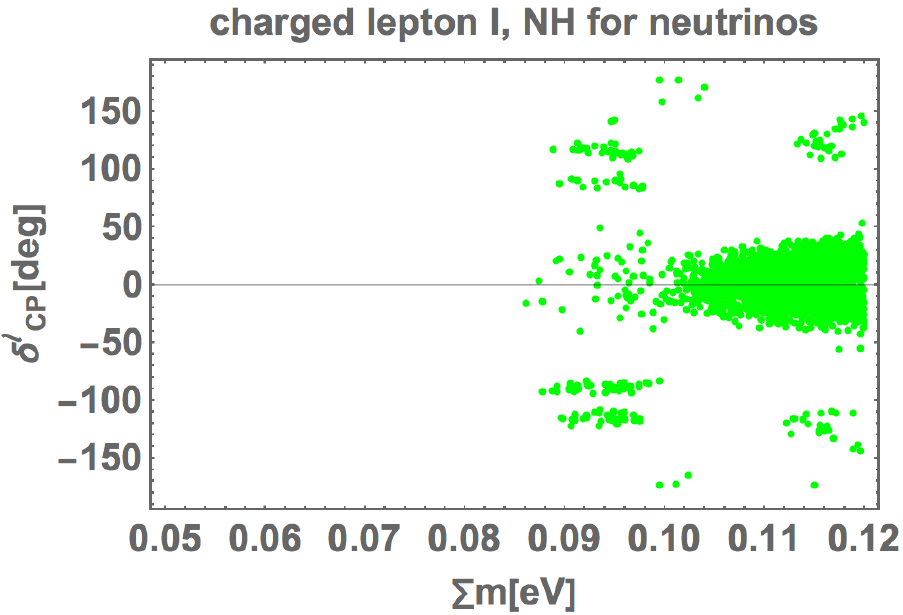}
	\hspace{10mm}
	\includegraphics[{width=0.42\linewidth}]{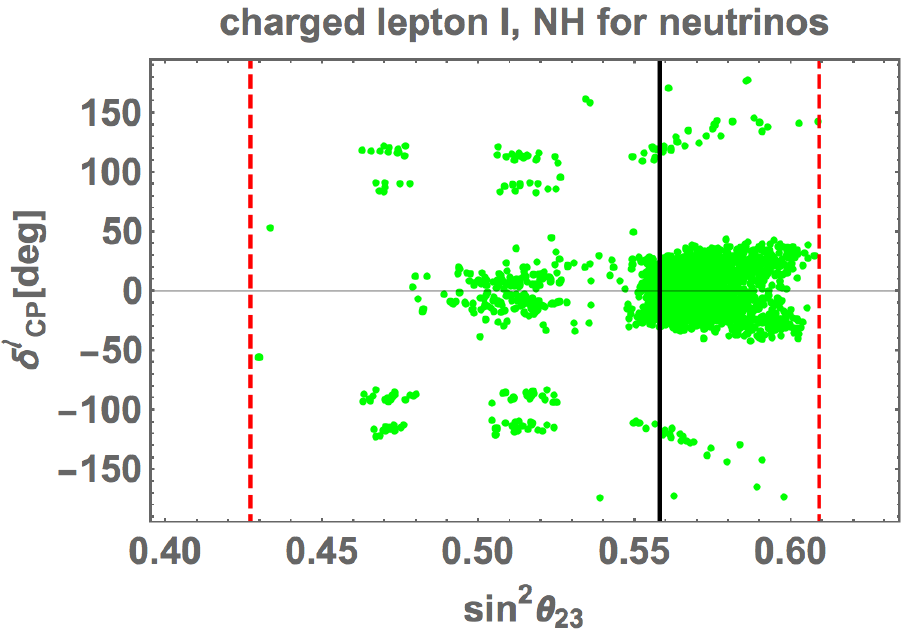}
	\caption{Allowed regions on   $\sum m_i$--$\delta_{\rm CP}^\ell$
		and $\delta_{\rm CP}^\ell$--$\sin^2\theta_{23}$ planes
		at nearby $\tau=i$	for the charged lepton mass matrix  I with NH of neutrinos.
		The solid black line denotes observed best-fit value of 
		$\sin^2\theta_{23}$,   and  red dashed-lines denote
		its upper(lower)-bound  of  $3\sigma$ interval.
}
\end{figure}
%%%%%%%%%%%%%%%%%%%%%%%%%%% 
\begin{figure}[H]
	\centering
	\includegraphics[{width=0.42\linewidth}]{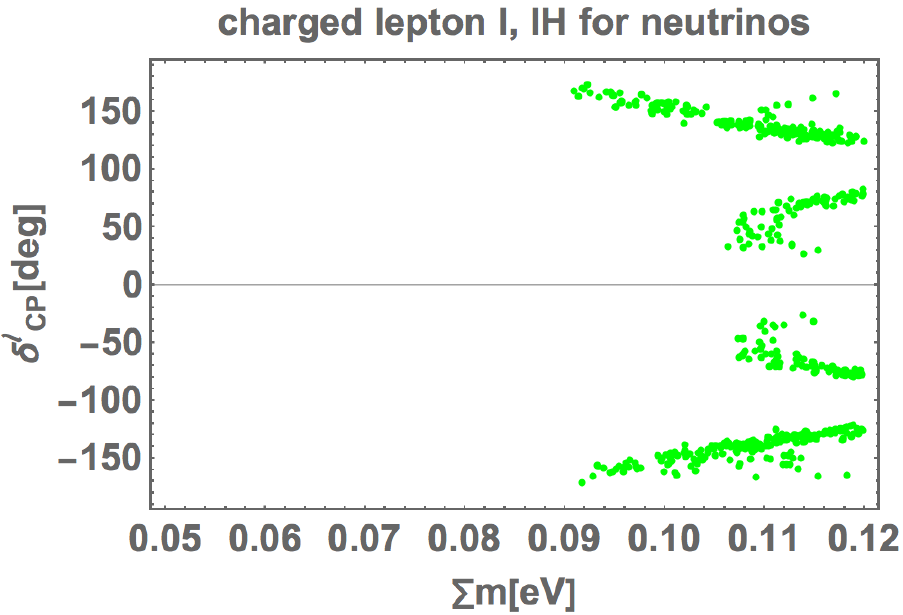}
	\hspace{10mm}
	\includegraphics[{width=0.42\linewidth}]{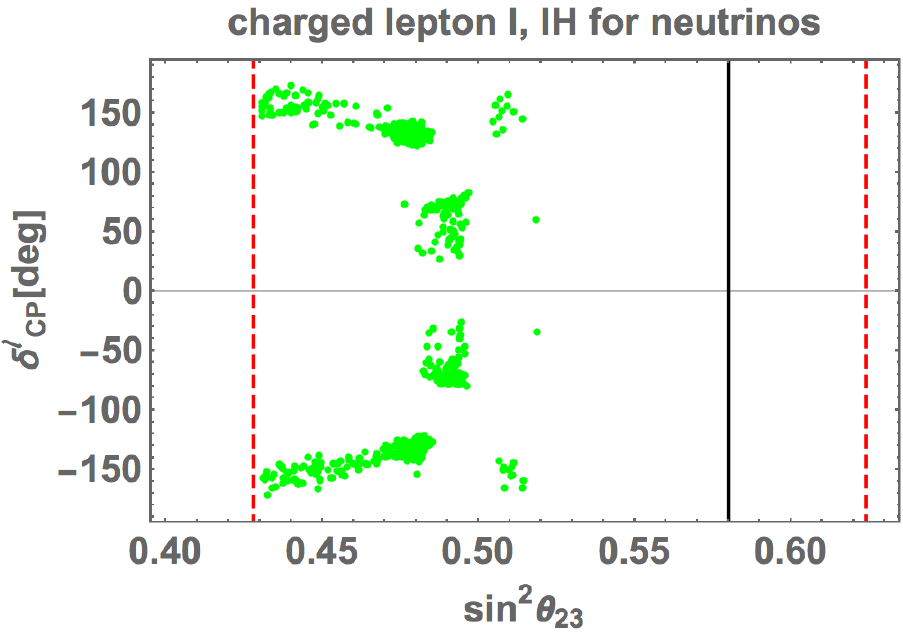}
	\caption{Allowed regions on   $\sum m_i$--$\delta_{\rm CP}^\ell$
		and $\delta_{\rm CP}^\ell$--$\sin^2\theta_{23}$ planes
		at nearby $\tau=i$	for the charged lepton mass matrix I with IH
		of neutrinos. 
	%	The solid black line denotes observed best-fit value of 
	%	$\sin^2\theta_{23}$,   and  red dashed-lines denote
	%	its upper(lower)-bound  of  $3\sigma$ interval.
	}
\end{figure}
%%%%%%%%%%%%%%%%%%%%%%%%%%% 
\begin{figure}[H]
	\centering
	\includegraphics[{width=0.42\linewidth}]{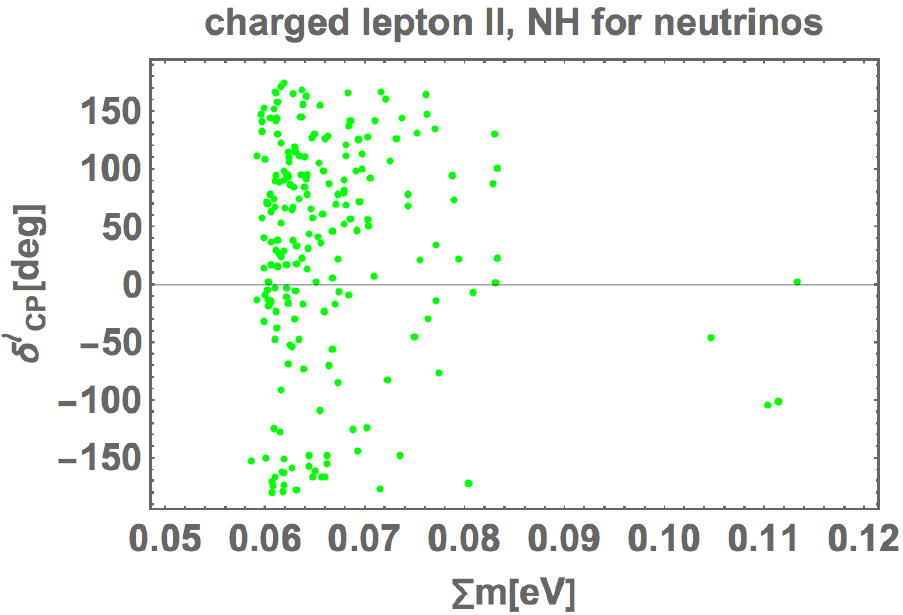}
	\hspace{10mm}
	\includegraphics[{width=0.42\linewidth}]{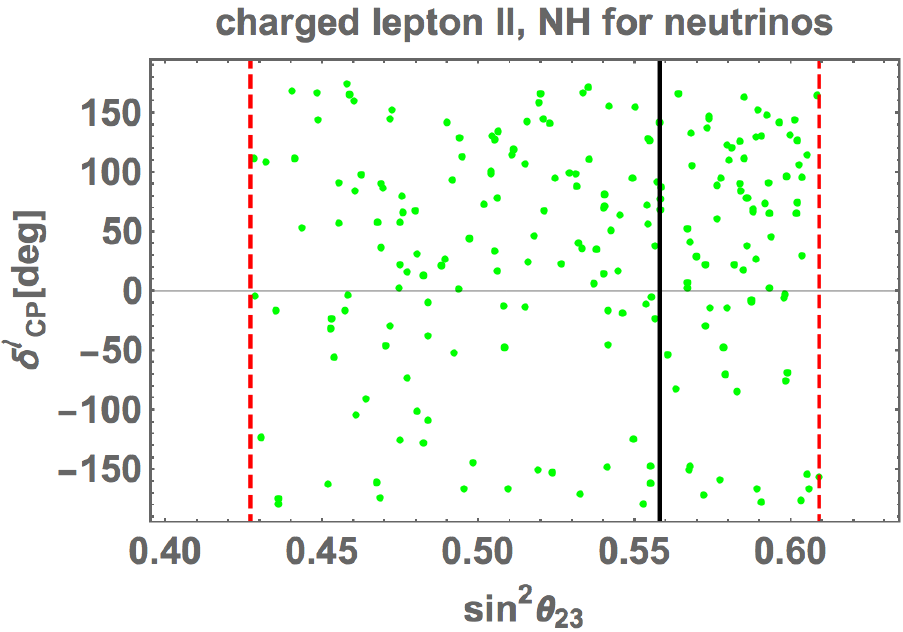}
	\caption{Allowed regions on   $\sum m_i$--$\delta_{\rm CP}^\ell$
		and $\delta_{\rm CP}^\ell$--$\sin^2\theta_{23}$ planes
		at nearby $\tau=i$	for the charged lepton mass matrix  I\hspace{-.01em}I with NH	of neutrinos. 
		%The solid black line denotes observed best-fit value of 
	%	$\sin^2\theta_{23}$,   and  red dashed-lines denote
	%	its upper(lower)-bound  of  $3\sigma$ interval.
}
\end{figure}
%%%%%%%%%%%%%%%%%%%%%%%%%%%%%%%%%%%%%%%%%%%%%%%%%%%%%%%%%%%%%%%
%%%%%%%%%%%%%%%%%%%%%%%%%  tau=omega  %%%%%%%%%%%%%%%%%%%%%%%%%
%%%%%%%%%%%%%%%%%%%%%%%%%%%%%%%%%%%%%%%%%%%%%%%%%%%%%%%%%%%%%%%
%%%%%%%%%%%%%%%%%%%%%%%%%%% 
\begin{figure}[H]
	\centering
	\includegraphics[{width=0.42\linewidth}]{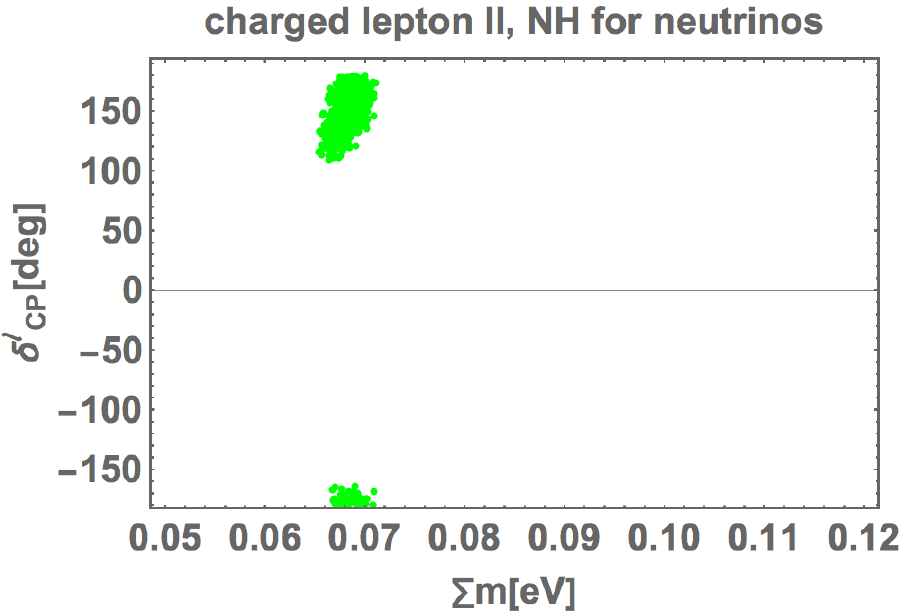}
	\hspace{10mm}
	\includegraphics[{width=0.42\linewidth}]{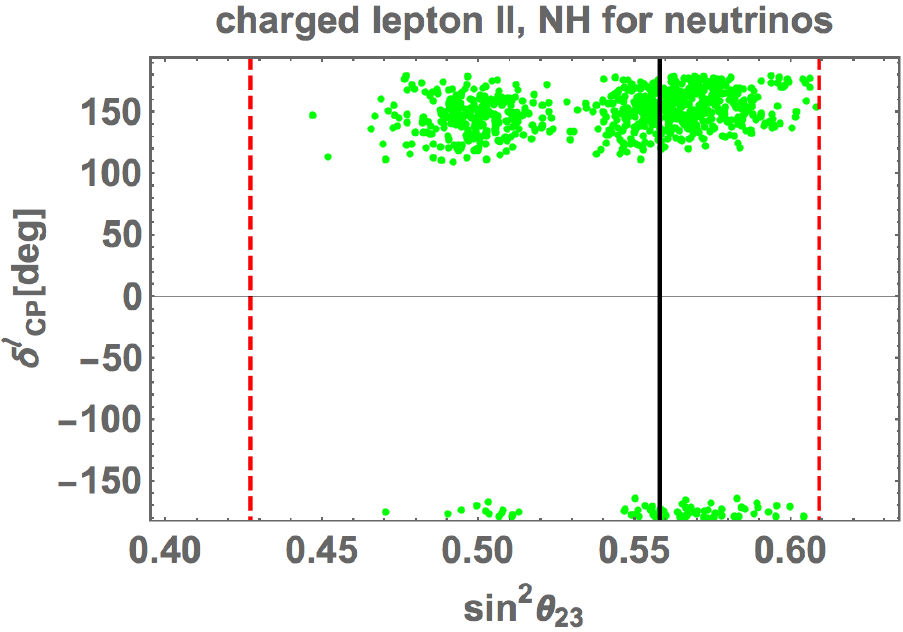}
	\caption{Allowed regions on   $\sum m_i$--$\delta_{\rm CP}^\ell$
		and $\delta_{\rm CP}^\ell$--$\sin^2\theta_{23}$ planes
		at nearby $\tau=\omega$	for the charged lepton mass matrix  I\hspace{-.01em}I with NH	of neutrinos. 
		%The solid black line denotes observed best-fit value of 
		%$\sin^2\theta_{23}$,   and  red dashed-lines denote
		%its upper(lower)-bound  of  $3\sigma$ interval.
	}
\end{figure}
%%%%%%%%%%%%%%%%%%%%%%%%%%%
%%%%%%%%%%%%%%%%%%%%%%%%%%% 
\begin{figure}[H]
	\centering
	\includegraphics[{width=0.42\linewidth}]{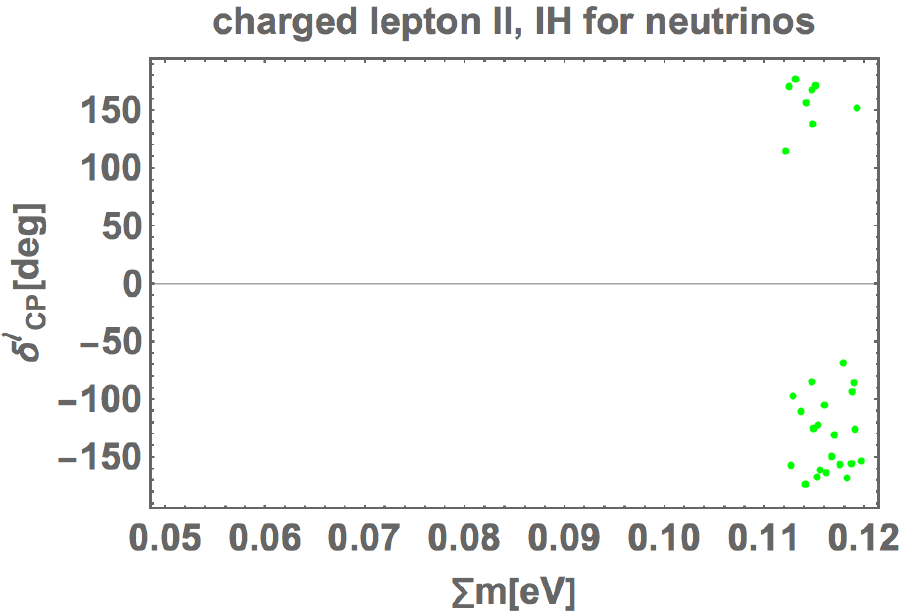}
	\hspace{10mm}
	\includegraphics[{width=0.42\linewidth}]{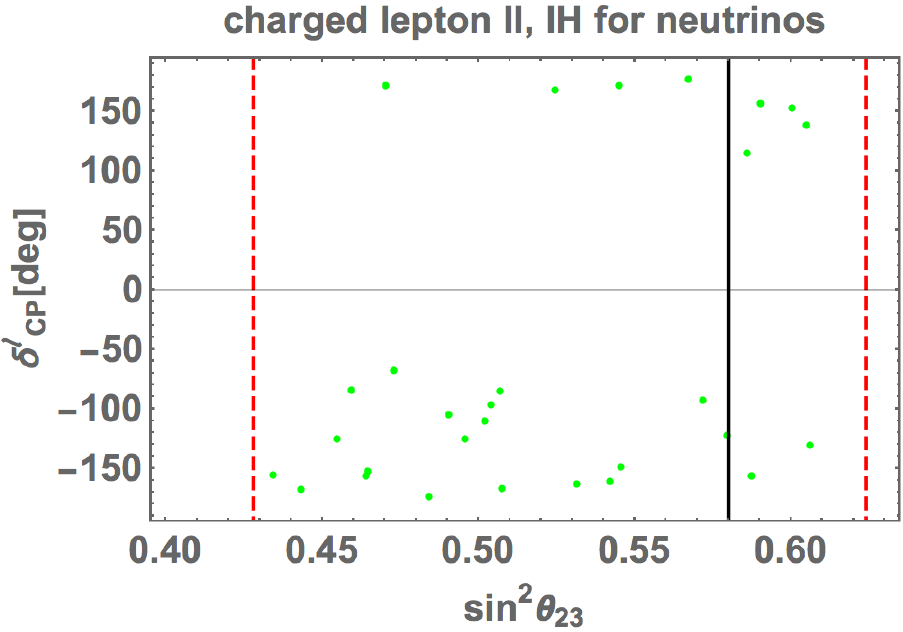}
	\caption{Allowed regions on   $\sum m_i$--$\delta_{\rm CP}^\ell$
		and $\delta_{\rm CP}^\ell$--$\sin^2\theta_{23}$ planes
		at nearby $\tau=\omega$	for the charged lepton mass matrix  I\hspace{-.01em}I with IH of neutrinos. 
		%The solid black line denotes observed best-fit value of 
		%$\sin^2\theta_{23}$,   and  red dashed-lines denote
		%its upper(lower)-bound  of  $3\sigma$ interval.
	}
\end{figure}
%%%%%%%%%%%%%%%%%%%%%%%%%%%%%%%%%%%%%%%%%%%%%%%%%%%%%%%%%%%%%%%
%%%%%%%%%%%%%%%%%%%%%%%%%  tau=i infty  %%%%%%%%%%%%%%%%%%%%%%%
%%%%%%%%%%%%%%%%%%%%%%%%%%%%%%%%%%%%%%%%%%%%%%%%%%%%%%%%%%%%%%%
%%%%%%%%%%%%%%%%%%%%%%%%%%%%%%%%%%%%%%%%%%%%%%%%%%%%%%%%%%%%%%%
\begin{figure}[H]
	\centering
	\includegraphics[{width=0.42\linewidth}]{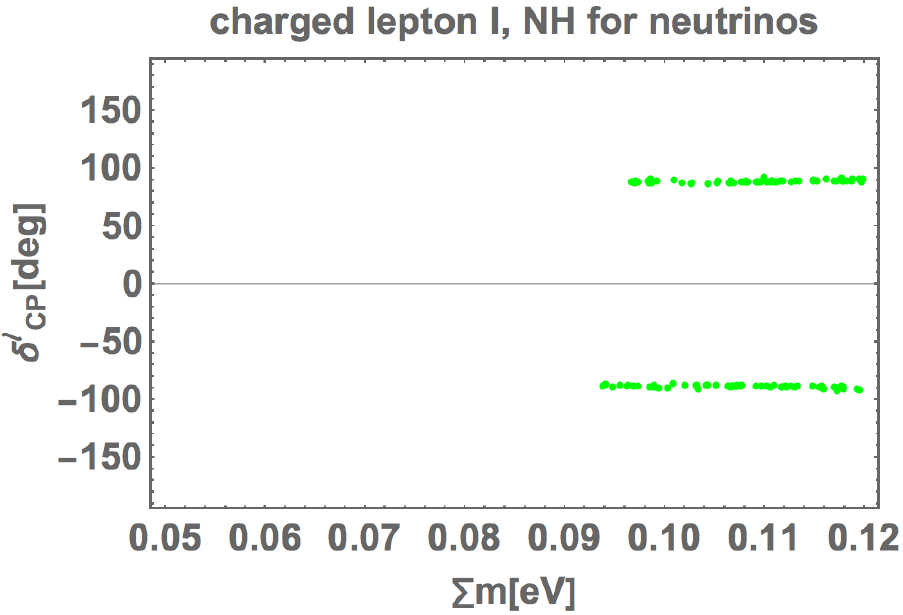}
	\hspace{10mm}
	\includegraphics[{width=0.42\linewidth}]{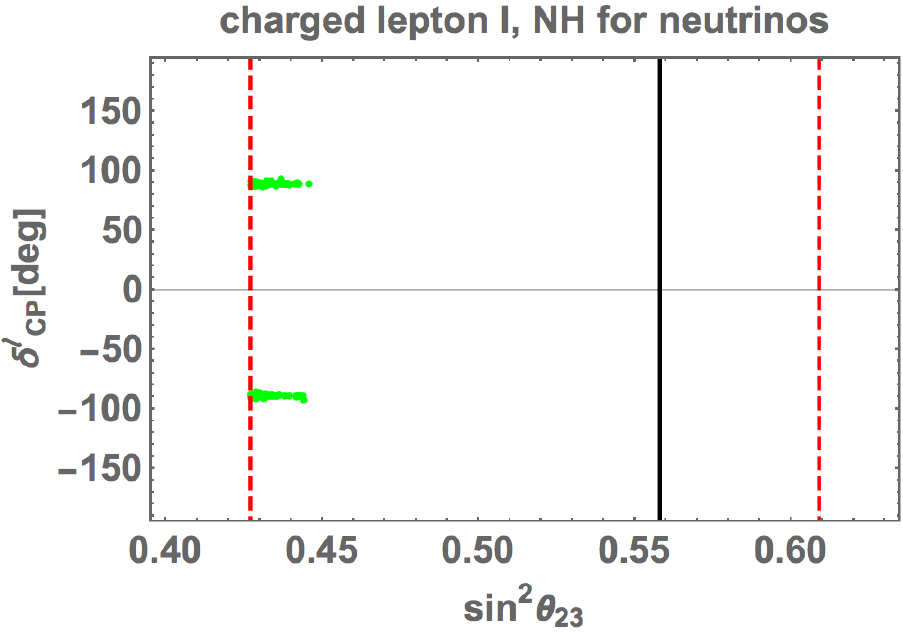}
	\caption{Allowed regions on   $\sum m_i$--$\delta_{\rm CP}^\ell$
		and $\delta_{\rm CP}^\ell$--$\sin^2\theta_{23}$ planes towards 
$\tau=i\infty$ for the charged lepton mass matrix I with NH of neutrinos. 
		%The solid black line denotes observed best-fit value of 
	%	$\sin^2\theta_{23}$,   and  red dashed-lines denote
	%	its upper(lower)-bound  of  $3\sigma$ interval.
}
\end{figure}
%%%%%%%%%%%%%%%%%%%%%%%%%%% 
%%%%%%%%%%%%%%%%%%%%%%%%%%% 
\begin{figure}[H]
	\centering
	\includegraphics[{width=0.42\linewidth}]{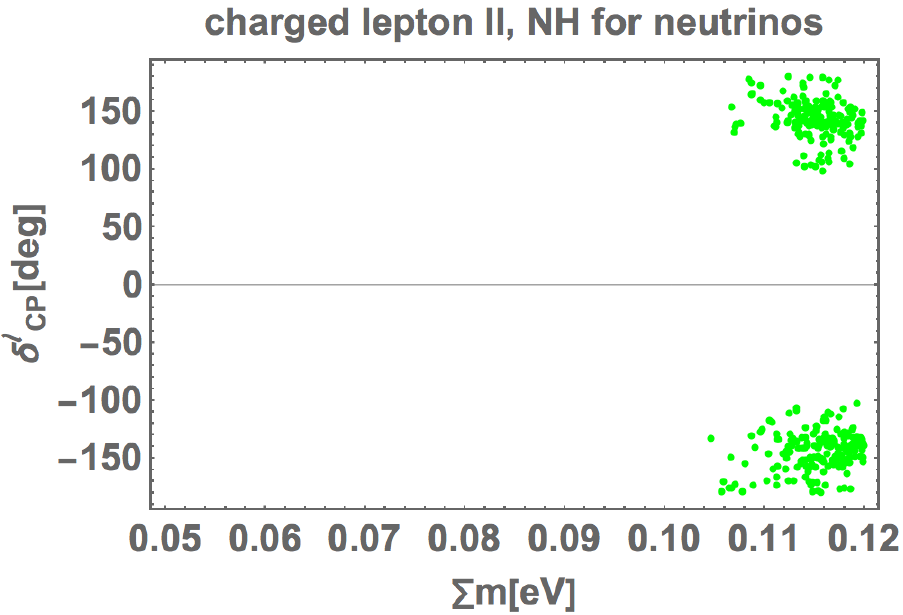}
	\hspace{10mm}
	\includegraphics[{width=0.42\linewidth}]{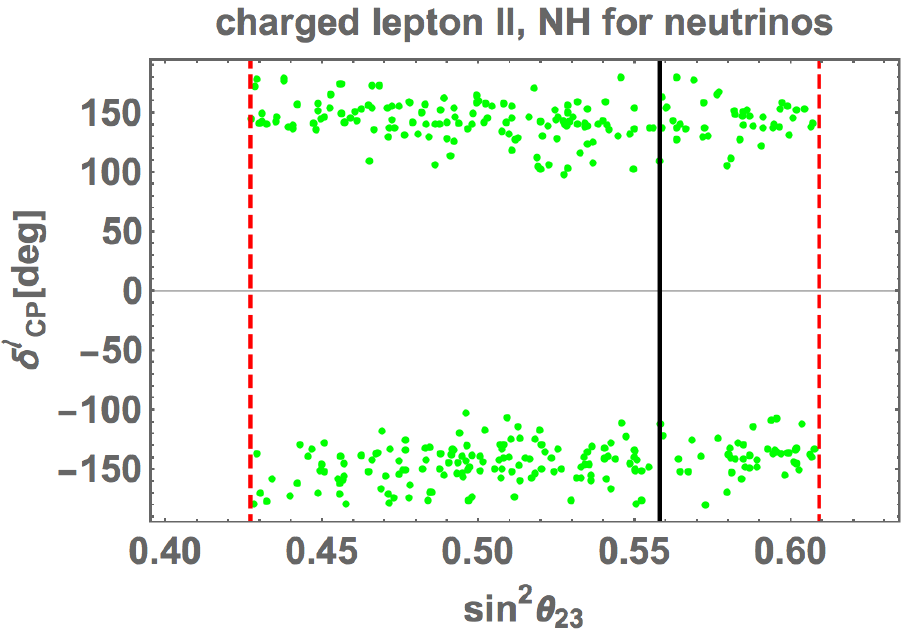}
	\caption{Allowed regions on   $\sum m_i$--$\delta_{\rm CP}^\ell$
		and $\delta_{\rm CP}^\ell$--$\sin^2\theta_{23}$ planes
		towards $\tau=i\infty$	for the charged lepton mass matrix  I\hspace{-.01em}I with NH	of neutrinos.
		% The solid black line denotes observed best-fit value of 
		%$\sin^2\theta_{23}$,   and  red dashed-lines denote
	%	its upper(lower)-bound  of  $3\sigma$ interval.
}
\end{figure}
%%%%%%%%%%%%%%%%%%%%%%%%%%%%%%%%%%%%%%%%%%%%%%%%%%%%%%%%%%%%%%%%%%

 %%%%%%%%%%%%%%%%%%%%%%%%%%%%%%%%%%%%%%%%%%%%%%%%%%%%%%%%%%%%%%%%%%% 
 %%%%%%%%%%%%%%%%%%%%%%%%%%%%%%%%%%%%%%%%%%%%%%%%%%%%%%%%%%%%%%%%%%%
 %%%%%%%%%%%%%%%%%%%%%%%%%%%%%%%%%%%%%%%%%%%%%%%%%%%%%%%%%%%%%%%%%%%  
     Let us give our predictions
     on   $\sum m_i$--$\delta_{\rm CP}^\ell$
     and $\delta_{\rm CP}^\ell$--$\sin^2\theta_{23}$ planes at nearby $\tau=\omega$.
  In Fig.\,7, we show them for the charged lepton mass matrix I\hspace{-.01em}I with NH of neutrinos.
     The predicted range of the sum of  neutrino masses is
     $\sum m_i=65$--$71$\,meV.  
     The  ranges of    $\delta_{\rm CP}^\ell$ is clearly given
     in [$110^\circ$,$180^\circ$] and [$-180^\circ$,$-160^\circ$].
     On the other hand, $\sin^2\theta_{23}$ is predicted 
      in both 1st- and 2nd-octant.

%%%%%%%%%%%%%%%%%%%%%%%%%%%%%%%%%%%%%%%%%%%%%%%%%%%%%%%%%%%   
     
       In Fig.\,8, 
       we show them for the charged lepton mass matrix I\hspace{-.01em}I 
       with IH of neutrinos  at nearby $\tau=\omega$.
       The predicted range of the sum of  neutrino masses is 
     $\sum m_i=112$--$120$\,meV, which may be excluded 
     in the near future due to the cosmological observations.
      The  predicted CP violating phase is
        $\delta_{\rm CP}^\ell =[-180^\circ,-60^\circ]$
      and $[110^\circ,180^\circ]$.
    There is no clear correlation between  $\sin^2\theta_{23}$  and  
    $\delta_{\rm CP}^\ell$. 
    
   It is noticed that the predicted CP violating phase $\delta_{\rm CP}^\ell$
    is asymmetric for plus and minus signs  in both Figs.\,7 and 8.
   That is due to excluding the $\tau$ region at nearby $\tau=\omega$ outside the fundamental domain of  $\rm PSL(2,\mathbb{Z})$.
   Indeed, the excluded region
   corresponds to the other region inside at nearby  the fixed point $\tau=-\omega^2$, where we obtain   $\delta_{\rm CP}^\ell$
   with the reversed sign  of Figs.\,7 and 8.
     
%%%%%%%%%%%%%%%%%%%%%%%%%%%%%%%%%%%%%%%%%%%%%%%%%%%%%%%%%%%     
Finally,   we  show predictions
on   $\sum m_i$--$\delta_{\rm CP}^\ell$
and $\delta_{\rm CP}^\ell$--$\sin^2\theta_{23}$ planes towards $\tau=i\infty$.     
     In Fig.\,9,  we show them 
     for the charged lepton mass matrix  I with  NH of neutrinos. 
     The predicted range of the sum of  neutrino masses is
      in the narrow range  of
     $\sum m_i=94$--$120$\,meV.
     The  predicted $\delta_{\rm CP}^\ell$ is close to $\pm\pi/2$.
   On the other hand, $\sin^2\theta_{23}$ is predicted to be smaller than $0.45$.
    The predicted CP violation is favored 
    by the  T2K  experiment \cite{T2K:2020},
    however the predicted  $\sin^2\theta_{23}$ may be excluded
     in the near future since  it is far from the best fit value.
     
%%%%%%%%%%%%%%%%%%%%%%%%%%%%%%%%%%%%%%%%%%%%%%%%%%%%%%%%%%%%
     
         In Fig.\,10, we show them 
     for the charged lepton mass matrix  I\hspace{-.01em}I with NH of neutrinos. 
     The predicted range of the sum of neutrino masses is
   in  $\sum m_i=105$--$120$\,meV.
     The  predicted $\delta_{\rm CP}^\ell$ is
      is clearly given  in $\pm (100^\circ$--$180^\circ)$. 
     On the other hand, $\sin^2\theta_{23}$ is allowed
      in full range of $3\sigma$ error-bar.
       Crucial test will be available  
      by  cosmological observations  and 
      CP violation experiments of neutrinos in the future.
      
 Thus, lepton mass matrices at nearby fixed points provide
   characteristic predictions for  $\sum m_i$ and  $\delta_{\rm CP}^\ell$.
   On the other hand, there is no prediction for the quark sector. 
     
%%%%%%%%%%%%%%%%%%%%%%%%%%%%%%%%%%%%%%%%%%%%%%%%%%%%%%%%%%
%%%%%%%%%%%%%%%%%%%%%%%%%%%%%%%%%%%%%%%%%%%%%%%%%%%%%%%%%%
\section{Summary}

In the  modular invariant flavor model of $\rm A_4$, we have studied the hierarchical structure of lepton/quark flavors at the nearby fixed points of the  modulus. There are only two inequivalent fixed  points in the fundamental domain
of  $\rm PSL(2,\mathbb{Z})$, $\tau= i$ and  $ \tau =\omega$.
  These fixed points  correspond to   the  residual symmetries 
  $\mathbb{Z}_2^{S}=\{I, S \}$
  and   $\mathbb{Z}_3^{ST}=\{ I, ST, (ST)^2 \}$ of  $\rm A_4$, respectively.
There is also infinite point $\tau = i \infty$,
in which  the subgroup  $\mathbb{Z}^T_3=\{ I,T,T^2 \}$ of $\rm A_4$
is preserved.
We have examined  typical two-type mass matrices for  charged leptons and quarks
by using modular forms of weights $2$, $4$ and $6$ while
the neutrino mass matrix  with the modular forms of weight $4$ 
through the Weinberg operator.  
By performing Taylor expansion of modular forms around fixed points, 
we have obtained  linear modular forms in good approximations.
By using those explicit modular forms, we have found 
the hierarchical structure of these mass matrices  in the diagonal base of $S$, $T$ and $ST$, in which the flavor mixing angles are easily estimated.
The observed PMNS  mixing angles  are reproduced  
at the nearby fixed point in ten cases of lepton mass matrices.
Among them, seven cases satisfy  the cosmological bound $\sum m_i\leq 120$\,meV.
On the other hand, only  one case of  quark mass matrices 
is consistent with the observed CKM matrix.
Our results have  been  confirmed  by scanning model parameters numerically
as seen in $\tau$ regions of  Figs.\,1,\,2 and 3.

We have also presented  predictions for  $\sum m_i$ and  $\delta_{\rm CP}^\ell$
for seven cases. Some cases will be tested in the near future.
Although there is no prediction for the quark sector,
 the  obtained  $\tau$  provides an interesting subject,
 the possibility of  the common $\tau$ between quarks and leptons.
 Indeed,  there exists  the common region around {$\tau=\pm 0.04+ 1.05 \, i$}
   for the charged lepton mass matrix  I with NH of neutrinos
   as seen Fig.\,1.
 
 We have worked by using two-type specific mass matrices for charged leptons and quarks while one Majorana neutrino mass matrix
 in order to clarify the behavior  at nearby fixed points.
 More studies including other mass matrices are necessary to understand the phenomenology of fixed points completely.
 The modular symmetry provides a good outlook for the flavor structure
 of leptons and quarks  at nearby fixed points.
 We also should pay attention to the recent theoretical work: the spontaneous CP violation in Type IIB string theory  is possibly realized at nearby  fixed points, where the moduli stabilization is performed in a controlled way \cite{Abe:2020vmv,Kobayashi:2020uaj}.
 Thus, the modular symmetry at nearby fixed points gives  us an attractive approach to flavors.

%%%%%%%%%%%%%%%%%%%%%%%%%%%%%%%%%%%%%%%%%%%%%%%%%%%
%%%%%%%%%%%%%%%%%%%%%%%%%%%%%%%%%%%%%%%%%%%%%%%%%%%

%%%%%%%%%%%%%%%%%%%%%%%%%%%%%%%%%%%%%%%%%%%%%%%%%%%%%%%%%%%%%
\section*{Acknowledgments}
This research was supported by an appointment to the JRG Program at the APCTP through the Science and Technology Promotion Fund and Lottery Fund of the Korean Government. This was also supported by the Korean Local Governments - Gyeongsangbuk-do Province and Pohang City (H.O.). 
H. O. is sincerely grateful for the KIAS member. 

\appendix
\section*{Appendix}

%%%%%%%%%%%%%%%%%%%%%%%%%%%%%%%%%%%%%%%%%%%%%%%%%%%%%%%%%%%
%%%%%%%%%%%%%%%%%%%%%%%%%%%%%%%%%%%%%%%%%%%%%%%%%%%%%%%%%%%
\section{Tensor product of  $\rm A_4$ group}
%%%%%%%%%%%%%%%%%%%%%%%%%%%%%%%%%%%%%%%%%%%%%%%%%%%%%%%%%%%%%%%%%%%%%%%%

We take the generators of $A_4$ group for the triplet as follows:
\begin{align}
\begin{aligned}
S=\frac{1}{3}
\begin{pmatrix}
-1 & 2 & 2 \\
2 &-1 & 2 \\
2 & 2 &-1
\end{pmatrix},
\end{aligned}
\qquad 
\begin{aligned}
T=
\begin{pmatrix}
1 & 0& 0 \\
0 &\omega& 0 \\
0 & 0 & \omega^2
\end{pmatrix}, 
\end{aligned}
\end{align}
where $\omega=e^{i\frac{2}{3}\pi}$ for a triplet.
In this base,
the multiplication rule is
\begin{align}
\begin{pmatrix}
a_1\\
a_2\\
a_3
\end{pmatrix}_{\bf 3}
\otimes 
\begin{pmatrix}
b_1\\
b_2\\
b_3
\end{pmatrix}_{\bf 3}
&=\left (a_1b_1+a_2b_3+a_3b_2\right )_{\bf 1} 
\oplus \left (a_3b_3+a_1b_2+a_2b_1\right )_{{\bf 1}'} \nonumber \\
& \oplus \left (a_2b_2+a_1b_3+a_3b_1\right )_{{\bf 1}''} \nonumber \\
&\oplus \frac13
\begin{pmatrix}
2a_1b_1-a_2b_3-a_3b_2 \\
2a_3b_3-a_1b_2-a_2b_1 \\
2a_2b_2-a_1b_3-a_3b_1
\end{pmatrix}_{{\bf 3}}
\oplus \frac12
\begin{pmatrix}
a_2b_3-a_3b_2 \\
a_1b_2-a_2b_1 \\
a_3b_1-a_1b_3
\end{pmatrix}_{{\bf 3}\  } \ , \nonumber \\
\nonumber \\
{\bf 1} \otimes {\bf 1} = {\bf 1} \ , \qquad &
{\bf 1'} \otimes {\bf 1'} = {\bf 1''} \ , \qquad
{\bf 1''} \otimes {\bf 1''} = {\bf 1'} \ , \qquad
{\bf 1'} \otimes {\bf 1''} = {\bf 1} \  ,
\end{align}
where
\begin{align}
T({\bf 1')}=\omega\,,\qquad T({\bf 1''})=\omega^2. 
\end{align}
More details are shown in the review~\cite{Ishimori:2010au,Ishimori:2012zz}.

\section{Mass matrix in arbitrary base of  $S$ and  $T$ }
%%%%%%%%%%%%%%%%%%%%%%%%%%%
Define the new basis of generators, $\hat S$ and $\hat T$
by a unirary transformation as:
\begin{align}
\begin{aligned}
\hat S= U S U^\dagger , \qquad \hat T= U T U^\dagger \  ,
\end{aligned}
\end{align}
where $\hat S$, $S$, $\hat T$, $T$ and $U$ are $3\times 3$ matrices.
Since the $\rm A_4$ triplet transforms under the $S$  ($T$)  transformation as:
\begin{align}
\begin{pmatrix}
a_1\\
a_2\\
a_3
\end{pmatrix}_{\bf 3}
\to S \  (T)
\begin{pmatrix}
a_1\\
a_2\\
a_3
\end{pmatrix}_{\bf 3}
= U^\dagger \hat S \ (\hat T) U
\begin{pmatrix}
a_1\\
a_2\\
a_3
\end{pmatrix}_{\bf 3} \ .
\end{align}
Thus, in the new base,  the $\rm A_4$ triplet transforms as:
\begin{align} 
\begin{pmatrix}
\hat a_1\\
\hat a_2\\
\hat a_3
\end{pmatrix}_{\bf 3}
\to \hat S \  (\hat T)\  
\begin{pmatrix}
\hat a_1\\
\hat a_2\\
\hat a_3
\end{pmatrix}_{\bf 3},
\end{align}
where 
\begin{align} 
\begin{pmatrix}
\hat a_1\\
\hat a_2\\
\hat a_3
\end{pmatrix}_{\bf 3}
= U
\begin{pmatrix}
a_1\\
a_2\\
a_3
\end{pmatrix}_{\bf 3}.
\end{align}
%%%%%%%%%%%%%%%%%%%%%%%%%%%%%%%%%
%Modular forms with weight $k$ are also transformed as:
%\begin{align}
%\begin{aligned}
%\hat Y^{(k)}_{\bf 3} = U Y^{(k)}_{\bf 3} .
%\end{aligned}
%\end{align}

%%%%%%%%%%%%%%%%%%%%%%%%%%%%%%%%%
Let us rewrite the Dirac mass matrix $M_{RL}$
in the new base ($\hat S$, $\hat T$) of
the triplet left-handed fields.
Denoting $L$ and $\hat L$ to be  triplets of the left-handed fields in the basse  of $S$ and $\hat S$, respectively, and $R$ to be right-handed singlets,
 the Dirac mass matrix is written as:
\begin{align}
\begin{aligned}
\bar R M_{RL} L=\bar R M_{RL} U^\dagger  \hat L  \,
%R^c_{\bf 1(1")(1')}  Y^{(k)}_{\bf 3}  L_{\bf 3} =
%R^c_{\bf 1(1")(1')}   Y^{(k)}_{\bf 3}  U^\dagger \hat L_{\bf 3}
%= R^c_{\bf 1(1")(1')}  M_{RL} U^\dagger\hat L_{\bf 3}\,,
\end{aligned}
\end{align}
where 
\begin{align}
\hat L=U L\, .
\end{align}
Then, the Dirac mass matrix $\hat M_{RL}$ in the new base is given as:
\begin{align}
\hat M_{RL}=  M_{RL}U^\dagger \, .
\label{newmassmatrix}
\end{align}

%%%%%%%%%%%%%%%%%%%%%%%%%%%%%%%%%%
On the other hand, the Majorana mass matrix $M_{LL}$
in the new base ($\hat S$, $\hat T$) is written as
% of the triplet left-handed fields, $\hat L$ as:
\begin{align}
\begin{aligned}
L^c M_{LL} L= \hat L^c U M_{LL} U^\dagger  L 
%L^c_{\bf 3}  Y^{(k)}_{\bf 3}  L_{\bf 3} =
%\hat  L^c_{\bf 3} U  Y^{(k)}_{\bf 3}  U^\dagger \hat L_{\bf 3}
%= \hat L^c U M_{LL} U^\dagger\hat L_{\bf 3}
\end{aligned} \, .
\end{align}
Therefore, the Majorana mass matrix $\hat M_{LL}$ is given as:
\begin{align}
\hat M_{LL}= U M_{LL}U^\dagger \, .
\label{newmassmatrixLL}
\end{align}
%%%%%%%%%%%%%%%%%%%%%%%%%%%%%%%%%%%%%%%%%%%%%%%%%%%%%%%%%%%%%%   
%%%%%%%%%%%%%%%%%%%%%%%%%%%%%%%%%%%%%%%%%%%%%%%%%%%%%%%%%%%
\section{Modular forms  at nearby  fixed points }
%%%%%%%%%%%%%%%%%%%%%%%%%%%%%%%%%%%%%%%%%%%%%%%%%%%%%%%%%%%%%%%%% 
\subsection{Modular forms at nearby   $\tau=i$}

Let us present the behavior of modular forms at nearby $\tau=i$.
We obtain  approximate linear forms  of  $Y_1(\tau)$, $Y_2(\tau)$ and $Y_3(\tau)$ by performing Taylor expansion of modular forms around $\tau=i$.
We parametrize $\tau$ as:
\begin{align}
\begin{aligned}
\tau=i +\epsilon \ , \qquad {\rm with } \qquad \epsilon=\epsilon_R+i\,\epsilon_I
\ , 
\end{aligned}
\label{epsilonS}
\end{align}
where $|\epsilon|$ is supposed to be enough small $|\epsilon|\ll 1$.
For the case of the pure imaginary number of  $\epsilon$, that is  $\epsilon=i\,\epsilon_I$ ($\epsilon_I$ is real), 
 we obtain the linear fit of $\epsilon$ by
\begin{align}
\begin{aligned}
\frac{Y_2(\tau)}{Y_1(\tau)}\simeq (1-2.05\,  \epsilon_I )\, (1-\sqrt{3}) \ , \qquad 
\frac{Y_3(\tau)}{Y_1(\tau)}\simeq (1-4.1\,  \epsilon_I )\, (-2+\sqrt{3}) \ ,
\end{aligned}
\end{align}
where coefficients are obtained  by numerical fittings.
These ratios decrease linearly for $\epsilon_I\geq 0$.

On the other hand, for the case of the real number of  $\epsilon$, that is $\epsilon=\epsilon_R$, ($\epsilon_R$ is real), we obtain   as:
\begin{align}
& \begin{aligned}
{\rm Re }\, \frac{Y_2(\tau)}{Y_1(\tau)}\simeq (1-1.9\,  \epsilon_R^2 )\, (1-\sqrt{3}) \ , \qquad 
{\rm Re}\,\frac{Y_3(\tau)}{Y_1(\tau)}\simeq (1-8\,  \epsilon_R^2 )\, (-2+\sqrt{3}) \, ,
\end{aligned}
\nonumber\\
& \begin{aligned}
{\rm Im }\, \frac{Y_2(\tau)}{Y_1(\tau)}\simeq 2.05\, \epsilon_R \,  (1-\sqrt{3}) \ , \qquad\qquad\, 
{\rm Im}\, \frac{Y_3(\tau)}{Y_1(\tau)}\simeq 4.1\,  \epsilon_R \, (-2+\sqrt{3})
 \, ,
\end{aligned}
\end{align}
where the liner terms of $\epsilon$ disappear in the real parts.
Finally, after neglecting ${\cal O}(\epsilon_R^2)$, we obtain approximately
\begin{align}
\begin{aligned}
\frac{Y_2(\tau)}{Y_1(\tau)}\simeq (1+\epsilon_1)\, (1-\sqrt{3}) \, , \quad 
\frac{Y_3(\tau)}{Y_1(\tau)}\simeq (1+\epsilon_2)\, (-2+\sqrt{3}) \, ,
\quad \epsilon_1=\frac{1}{2} \epsilon_2=2.05\,i\,\epsilon\,.
\end{aligned}
\label{epS12}
\end{align}
These approximate  forms are  agreement with exact numerical values within  $0.1\,\%$
for $|\epsilon|\leq 0.05$.

%%%%%%%%%%%%%%%%%%%%%%%%%%%%%%%%%%%%%%%%%%%%%%%
%%%%%%%%%%%%  Higher weights  %%%%%%%%%%%%%%%%%
%%%%%%%%%%%%%%%%%%%%%%%%%%%%%%%%%%%%%%%%%%%%%%%
We have also  higher weight modular forms $ Y_i^{(k)}$
in Eqs.\,(\ref{weight4}) and (\ref{weight6})
in terms of $\epsilon_1$ and  $\epsilon_2$. 
 For weight $4$, they are
\begin{align}
&  \begin{aligned}
\frac{Y_1^{(4)}(\tau)}{Y_1^2(\tau)}
\simeq 6-3\sqrt{3}+(5-3\sqrt{3})(\epsilon_1+\epsilon_2) \ , \qquad 
\frac{Y_2^{(4)}(\tau)}{Y_1^2(\tau)} \simeq 6-3\sqrt{3}+(\sqrt{3}-1)\epsilon_1+(14-8\sqrt{3})\epsilon_2\ ,
\end{aligned}
\nonumber\\
&  \begin{aligned}
\frac{Y_3^{(4)}(\tau)}{Y_1^2(\tau)} \simeq 6-3\sqrt{3}+(8-4\sqrt{3})\epsilon_1+(2-\sqrt{3})\epsilon_2\ ,
\end{aligned}
\\
&  \begin{aligned}
\frac{Y^{(4)}_{\mathbf 1}(\tau)}{Y_1^2(\tau)}  \simeq -9+6\sqrt{3}+(6\sqrt{3}-10)(\epsilon_1+\epsilon_2)\ ,
\quad
\frac{Y^{(4)}_{\mathbf 1'}(\tau)}{Y_1^2(\tau)}  \simeq 9-6\sqrt{3}+(2-2\sqrt{3})\, \epsilon_1+
(14-8\sqrt{3})\,\epsilon_2\,. \nonumber
\end{aligned}
%\label{epS4A}
\end{align}
For weight $6$, they are
\begin{align}
&  \begin{aligned}
\frac{Y_1^{(6)}(\tau)}{3Y_1^3(\tau)} \simeq 2\sqrt{3}-3+
\left (2\sqrt{3}-\frac{10}{3}\right )(\epsilon_1+\epsilon_2) \, ,  \quad
\end{aligned} \nonumber \\
&  \begin{aligned}
\frac{Y_2^{(6)}(\tau)}{3Y_1^3(\tau)} \simeq 5\sqrt{3}-9+\left (\frac{31}{\sqrt{3}}-\frac{55}{3}\right )\epsilon_1+
\left (\frac{16}{\sqrt{3}}-\frac{28}{3} \right )\epsilon_2\, ,
\end{aligned}
\nonumber\\
&  \begin{aligned}
\frac{Y_3^{(6)}(\tau)}{3Y_1^3(\tau)} \simeq 12-7\sqrt{3}+\left (\frac{38}{3}-\frac{22}{\sqrt{3}}
\right )\epsilon_1+
\left (\frac{74}{3}- \frac{43}{\sqrt{3}}\right )\epsilon_2\, ,
\end{aligned}
\nonumber\\
&  \begin{aligned}
\frac{Y_1^{'(6)}(\tau)}{3Y_1^3(\tau)} \simeq 7\sqrt{3}-12+
\left (2\sqrt{3}-\frac{10}{3}\right )\epsilon_1+
\left (17\sqrt{3}-\frac{88}{3}\right )\epsilon_2 \, , \end{aligned} \nonumber\\
&\begin{aligned}
\frac{Y_2^{'(6)}(\tau)}{3Y_1^3(\tau)}  \simeq 3-2\sqrt{3}+\left (\frac{2}{3}- \frac{2}{\sqrt{3}}\right )
\epsilon_1+
\left (\frac{14}{3}-\frac{8}{\sqrt{3}}- \right )\epsilon_2\, ,
\end{aligned}
\nonumber\\
&  \begin{aligned}
\frac{Y_3^{'(6)}(\tau)}{3Y_1^3(\tau)}  \simeq9-5\sqrt{3}+\left (\frac{35}{3}-\frac{19}{\sqrt{3}}
\right )\epsilon_1+
\left (\frac{38}{3}- \frac{22}{\sqrt{3}}\right )\epsilon_2\, ,
\end{aligned}
\nonumber\\
&  \begin{aligned}
\frac{Y_{\mathbf  1}^{(6)}(\tau)}{3Y_1^3(\tau)}  \simeq
 (15-9\sqrt{3})\epsilon_1 + (12\sqrt{3}-21)\epsilon_2
\, .
\end{aligned}
\label{epS666}
\end{align}
%where the last one denotes for the $A_4$ singlet.

%%%%%%%%%%%%%%%%%%%%%%%%%%%%%%%%%%%%%%%%%%%%%%%%%%%%%%%%%%%%%%%%%%%%%%
%%%%%%%%%%%%%%%%%%%%%%%%%%%%%%%%%%%%%%%%%%%%%%%%%%%%%%%%%%%%%%%%%%%%%%
%%%%%%%%%%%%%%%%%%%%%%%%%%%%%%%%%%%%%%%%%%%%%%%%%%%%%%%%%%%%%%%%%%%%%%
\subsection{Modular forms at nearby  $\tau=\omega$}

Let us present the behavior of modular forms at nearby
$\tau=\omega$.
We perform linear approximation of the modular forms $Y_1(\tau)$, $Y_2(\tau)$ and $Y_3(\tau)$ by performing Taylor expansion around $\tau=\omega$.
We parametrize $\tau$ as:
\begin{align}
\begin{aligned}
\tau= \omega+\epsilon \, , \qquad {\rm with } \qquad \epsilon=\epsilon_R+i\,\epsilon_I
\ , 
\end{aligned}
\label{epST}
\end{align}
where we suppose $|\epsilon|\ll 1$.
For the case of $\epsilon=i\,\epsilon_I$, which is a pure imaginary number, we obtain the linear fit of $\epsilon$ as:
\begin{align}
\begin{aligned}
\frac{Y_2(\tau)}{Y_1(\tau)}\simeq \omega \,(1-2.1\,  \epsilon_I ) \ , \qquad 
\frac{Y_3(\tau)}{Y_1(\tau)}\simeq -\frac{1}{2}\omega^2\, (1-4.2\,  \epsilon_I ) \ ,
\end{aligned}
\label{}
\end{align}
where coefficients are obtained  by numerical fittings.
These ratios decrease linearly for $\epsilon_I\geq 0$.
On the other hand, for the case of $\epsilon=\epsilon_R$,
which  is a real number, we obtain as:
\begin{align}
& \begin{aligned}
{\rm Re }\, \frac{Y_2(\tau)}{Y_1(\tau)}\simeq\omega \,(1-3\,  \epsilon_R^2 )\ , \qquad \quad
{\rm Re}\,\frac{Y_3(\tau)}{Y_1(\tau)}\simeq
-\frac{1}{2}\omega^2\,(1-11\,  \epsilon_R^2 ) \ .
\end{aligned}
\nonumber\\
& \begin{aligned}
{\rm Im }\, \frac{Y_2(\tau)}{Y_1(\tau)}\simeq \omega\,(2.1\, \epsilon_R)  \ , \qquad\qquad\, 
{\rm Im}\, \frac{Y_3(\tau)}{Y_1(\tau)}\simeq -\frac{1}{2}\omega^2\, 
(4.2\,  \epsilon_R ) \, ,
\end{aligned}
\end{align}
where the linear terms of $\epsilon$ disappear in the real parts.
After neglecting ${\cal O}(\epsilon_R^2)$, we obtain approximately
\begin{align}
\begin{aligned}
\frac{Y_2(\tau)}{Y_1(\tau)}\simeq \omega\,(1+\,\epsilon_1) \, , \quad 
\frac{Y_3(\tau)}{Y_1(\tau)}\simeq -\frac{1}{2}\omega^2 \, 
(1+\, \epsilon_2)\,  ,
\quad \epsilon_1=\frac{1}{2} \epsilon_2=2.1\,i\,\epsilon\,,
\end{aligned}
\label{epST12}
\end{align}
where $|\epsilon|\ll 1$.
These approximate  forms are  agreement with exact numerical values within  $1\,\%$
 for $|\epsilon|\leq 0.05$.

%%%%%%%%%%%%%%%%%%%%%%%%%%%%%%%%%%%%%%%%%%%%%%%
%%%%%%%%%%%%  Higher weights  %%%%%%%%%%%%%%%%%
%%%%%%%%%%%%%%%%%%%%%%%%%%%%%%%%%%%%%%%%%%%%%%%
We have also  higher weight modular forms $ Y_i^{(k)}$
in Eqs.\,(\ref{weight4}) and (\ref{weight6})
in terms of $\epsilon_1$ and  $\epsilon_2$. 
 For weight $4$, they are
\begin{align}
&  \begin{aligned}
\frac{Y_1^{(4)}(\tau)}{Y_1^2(\tau)}
\simeq \frac{3}{2}(1+\epsilon_1+\epsilon_2) \, , \quad 
\frac{Y_2^{(4)}(\tau)}{Y_1^2(\tau)} \simeq -\frac{3}{2}\,\omega\,
\left( \frac{1}{2}+\frac{2}{3}\epsilon_1+\frac{1}{6}\epsilon_2\right )\, ,
\quad
\frac{Y_3^{(4)}(\tau)}{Y_1^2(\tau)} \simeq \frac{3}{2}\,\omega^2\,   \left( 1-\frac{4}{3}\epsilon_1-\frac{2}{3}\epsilon_2\right )\, ,
\end{aligned}
\nonumber \\
&  \begin{aligned}
\frac{Y^{(4)}_{\mathbf 1}(\tau)}{Y_1^2(\tau)}  \simeq 
-(\epsilon_1+\epsilon_2)\, ,
\qquad\ \ 
\frac{Y^{(4)}_{\mathbf 1'}(\tau)}{Y_1^2(\tau)}  \simeq  
\frac{9}{4}\omega\left(1+\frac{8}{9}\epsilon_1+\frac{2}{9}\epsilon_2\right )\,.
\end{aligned}
\label{epST4}
\end{align}
For weight $6$, they are
\begin{align}
&  \begin{aligned}
\frac{Y_1^{(6)}(\tau)}{Y_1^3(\tau)} \simeq -(\epsilon_1+\epsilon_2)\, ,  \quad
\end{aligned} \nonumber \\
&  \begin{aligned}
\frac{Y_2^{(6)}(\tau)}{Y_1^3(\tau)} \simeq
-\omega\, (\epsilon_1+\epsilon_2)\,  ,
\end{aligned}
\nonumber\\
&  \begin{aligned}
\frac{Y_3^{(6)}(\tau)}{Y_1^3(\tau)} \simeq
\frac{1}{2}\,\omega^2\, (\epsilon_1+\epsilon_2) ,
\end{aligned}
\nonumber\\
&  \begin{aligned}
\frac{Y_1^{'(6)}(\tau)}{Y_1^3(\tau)} \simeq -\frac{9}{8}
\left (1+\frac{8}{9}\epsilon_1+\frac{11}{9}\epsilon_2\right )\, , \end{aligned} \nonumber\\
&\begin{aligned}
\frac{Y_2^{'(6)}(\tau)}{Y_1^3(\tau)}  \simeq
\frac{9}{4}\,\omega\,
\left (1+\frac{8}{9}\epsilon_1+\frac{2}{9}\epsilon_2\right ) \, ,
\end{aligned}
\nonumber\\
&  \begin{aligned}
\frac{Y_3^{'(6)}(\tau)}{Y_1^3(\tau)}  \simeq
\frac{9}{4}\,\omega^2\,
\left (1+\frac{17}{9}\epsilon_1+\frac{2}{9}\epsilon_2\right )\, ,
\end{aligned}
\nonumber\\
&  \begin{aligned}
\frac{Y_{\mathbf 1}^{(6)}(\tau)}{Y_1^3(\tau)}  \simeq
\frac{27}{8}\left(1+\frac{4}{3}\epsilon_1+\frac{1}{3}\epsilon_2\right ) \, .
\end{aligned}
\label{epST666}
\end{align}

%%%%%%%%%%%%%%%%%%%%%%%%%%%%%%%%%%%%%%%%%%%%%%%%%%%%%%%%%%%%%%%%%%%%%%
\subsection{Modular forms  towards  $\tau=i\infty$}

We show the behavior of modular forms at large ${\rm Im}\tau$, where
$q=\exp{(2\pi i\tau)}$ is suppressed.
Taking leading terms of  Eq.\,(\ref{Y(2)}), we can express modular forms 
approximately  as:
\begin{align}
Y_1(\tau)\simeq 1+ 12 p\,\epsilon\,,\quad  
Y_2(\tau)\simeq -6 p^{\frac{1}{3}}\,\epsilon^{\frac{1}{3}}\,,
\quad  Y_3(\tau)\simeq -18 p^{\frac{2}{3}}\,\epsilon^{\frac{2}{3}}\,, \quad
p=e^{2\pi i\, {\rm Re}\, \tau}\,,\quad 
\epsilon=e^{-2\pi\, {\rm Im}\, \tau}\,.
\label{epT12}
\end{align}

%%%%%%%%%%%%%%%%%%%%%%%%%%%%%%%%%%%%%%%%%%%%%%%%%%%%%

%%%%%%%%%%%%%%%%%%%%%%%%%%%%%%%%%%%%%%%%%%%%%%%%%%%%%%%%
%%%%%%%%%%%%%%%%%%%  weight 4 and 6  %%%%%%%%%%%%%%%%%%%
%%%%%%%%%%%%%%%%%%%%%%%%%%%%%%%%%%%%%%%%%%%%%%%%%%%%%%%% 
Higher weight modular forms $ Y_i^{(k)}$
in Eqs.\,(\ref{weight4}) and (\ref{weight6})
are obtained 
in terms of $p$ and $\epsilon$ approximately.
For weight $4$, they are
\begin{align}
&  \begin{aligned}
Y_1^{(4)}(\tau)
\simeq 1-84 p \, \epsilon \ , \qquad 
Y_2^{(4)}(\tau) \simeq 6 p^{\frac{1}{3}}\,\epsilon^{\frac{1}{3}}\,,
\qquad 
Y_3^{(4)}(\tau) \simeq 54 p^{\frac{2}{3}}\,\epsilon^{\frac{2}{3}}\,,
\end{aligned}
\nonumber\\
&  \begin{aligned}
Y^{(4)}_{\mathbf 1}(\tau)  \simeq 1+240\, p \, \epsilon \,, 
\quad\ \ 
Y^{(4)}_{\mathbf 1'}(\tau)\simeq -12 p^{\frac{1}{3}}\,\epsilon^{\frac{1}{3}}\,.
\end{aligned}
\label{epT4}
\end{align}
Weight $6$ modular forms are given:
\begin{align}
&  \begin{aligned}
Y_1^{(6)}(\tau)
\simeq 1+252\, p \, \epsilon \, , \qquad 
Y_2^{(6)}(\tau) \simeq -6\, p^{\frac{1}{3}}\,\epsilon^{\frac{1}{3}}\,,
\qquad 
Y_3^{(6)}(\tau) \simeq -18\, p^{\frac{2}{3}}\,\epsilon^{\frac{2}{3}}\,,
\end{aligned}
\nonumber\\
&  \begin{aligned}
Y_1^{'(6)}(\tau)
\simeq 216\,p \, \epsilon \, , \qquad\quad\  
Y_2^{'(6)}(\tau) \simeq -12\, p^{\frac{1}{3}}\,\epsilon^{\frac{1}{3}}\,,
\quad \ \
Y_3^{'(6)}(\tau) \simeq 72\, p^{\frac{2}{3}}\,\epsilon^{\frac{2}{3}}\,,
\end{aligned}
\nonumber\\
&  \begin{aligned}
Y_{\mathbf 1}^{(6)}(\tau)
\simeq 1-504\, p \, \epsilon \, .  
\end{aligned} 
\label{epT666}
\end{align}

%%%%%%%%%%%%%%%%%%%%%%%%%%%%%%%%%%%%%%%%%%%%%%%%%%%%%%%%%
\section{Majorana and Dirac phases and $\langle m_{ee}	\rangle $
	in  $0\nu\beta\beta$ decay }

Supposing neutrinos to be Majorana particles, 
the PMNS matrix $U_{\text{PMNS}}$~\cite{Maki:1962mu,Pontecorvo:1967fh} 
is parametrized in terms of the three mixing angles $\theta _{ij}$ $(i,j=1,2,3;~i<j)$,
one CP violating Dirac phase $\delta _\text{CP}^\ell$ and two Majorana phases 
$\alpha_{21}$, $\alpha_{31}$  as follows:
\begin{align}
U_\text{PMNS} =
\begin{pmatrix}
c_{12} c_{13} & s_{12} c_{13} & s_{13}e^{-i\delta^\ell_\text{CP}} \\
-s_{12} c_{23} - c_{12} s_{23} s_{13}e^{i\delta^\ell_\text{CP}} &
c_{12} c_{23} - s_{12} s_{23} s_{13}e^{i\delta^\ell_\text{CP}} & s_{23} c_{13} \\
s_{12} s_{23} - c_{12} c_{23} s_{13}e^{i\delta^\ell_\text{CP}} &
-c_{12} s_{23} - s_{12} c_{23} s_{13}e^{i\delta^\ell_\text{CP}} & c_{23} c_{13}
\end{pmatrix}
%\times
\begin{pmatrix}
1&0 &0 \\
0 & e^{i\frac{\alpha_{21}}{2}} & 0 \\
0 & 0 & e^{i\frac{\alpha_{31}}{2}}
\end{pmatrix},
\label{UPMNS}
\end{align}
where $c_{ij}$ and $s_{ij}$ denote $\cos\theta_{ij}$ and $\sin\theta_{ij}$, respectively.

The rephasing invariant CP violating measure of leptons \cite{Jarlskog:1985ht,Krastev:1988yu}
is defined by the PMNS matrix elements $U_{\alpha i}$. 
It is written in terms of the mixing angles and the CP violating phase as:
\begin{equation}
J_{CP}=\text{Im}\left [U_{e1}U_{\mu 2}U_{e2}^\ast U_{\mu 1}^\ast \right ]
=s_{23}c_{23}s_{12}c_{12}s_{13}c_{13}^2\sin \delta^\ell_\text{CP}~ ,
\label{Jcp}
\end{equation}
where $U_{\alpha i}$ denotes the each component of the PMNS matrix.

There are also other invariants $I_1$ and $I_2$ associated with Majorana phases
%\cite{Bilenky:2001rz}-\cite{Girardi:2016zwz},
\begin{equation}
I_1=\text{Im}\left [U_{e1}^\ast U_{e2} \right ]
=c_{12}s_{12}c_{13}^2\sin \left (\frac{\alpha_{21}}{2}\right )~, \quad
I_2=\text{Im}\left [U_{e1}^\ast U_{e3} \right ]
=c_{12}s_{13}c_{13}\sin \left (\frac{\alpha_{31}}{2}-\delta^\ell_\text{CP}\right )~.
\label{Jcp}
\end{equation}
We can calculate $\delta^\ell_\text{CP}$, $\alpha_{21}$ and $\alpha_{31}$ with these relations by taking account of 
\begin{eqnarray}
&&\cos\delta^\ell_{\rm CP}=\frac{|U_{\tau 1}|^2-
	s_{12}^2 s_{23}^2 -c_{12}^2c_{23}^2s_{13}^2}
{2 c_{12}s_{12}c_{23}s_{23}s_{13}}~ , \nonumber \\
&&\text{Re}\left [U_{e1}^\ast U_{e2} \right ]
=c_{12}s_{12}c_{13}^2\cos \left (\frac{\alpha_{21}}{2}\right )~, \qquad
\text{Re}\left [U_{e1}^\ast U_{e3} \right ]
=c_{12}s_{13}c_{13}\cos\left(\frac{\alpha_{31}}{2}-\delta^\ell_\text{CP}\right )~.
\end{eqnarray}
In terms of these parameters, the effective mass for the $0\nu\beta\beta$ decay is given as follows:
\begin{align}
\langle m_{ee}	\rangle=\left| m_1 c_{12}^2 c_{13}^2+ m_2s_{12}^2 c_{13}^2 e^{i\alpha_{21}}+
m_3 s_{13}^2 e^{i(\alpha_{31}-2\delta^\ell_{\rm CP})}\right|  \, .
\end{align}
%%%%%%%%%%%%%%%%%%%%%%%%%%%%%%%%%%%%%%%%%%%%%%%%%%%%%%%%%%
%%%%%%%%%%%%%%%%%%%%%%%%%%%%%%%%%%%%%%%%%%%%%%%%%%%%%%%%%%
%%%%%%%%%%%%%%%%%%%%%   References   %%%%%%%%%%%%%%%%%%%%% %%%%%%%%%%%%%%%%%%%%%%%%%%%%%%%%%%%%%%%%%%%%%%%%%%%%%%%%%%
%\newpage

\end{document}